\documentclass{emulateapj} 
\usepackage{apjfonts}
\usepackage{color}
\usepackage{hyperref}
\usepackage{graphicx}

\hypersetup{
    colorlinks,%
    citecolor=blue,%
    linkcolor=blue,%
    urlcolor=blue
}

\begin{document}

\title{THREE-DIMENSIONAL RELATIVISTIC PAIR PLASMA RECONNECTION WITH RADIATIVE FEEDBACK IN THE CRAB NEBULA}

\shorttitle{Three-dimensional relativistic pair plasma reconnection with radiative feedback in the Crab Nebula}

\author{B.~Cerutti$^{1,2,3}$, G.~R.~Werner$^{3}$, D.~A.~Uzdensky$^{3}$ \& M.~C.~Begelman$^{4,5}$} \shortauthors{Cerutti, Werner, Uzdensky, \& Begelman}

\affil{$^1$Department of Astrophysical Sciences, Princeton University, Princeton, NJ 08544, USA; bcerutti@astro.princeton.edu}

\affil{$^2$Lyman Spitzer Jr. Fellow}

\affil{$^3$ Center for Integrated Plasma Studies, Physics Department, University of Colorado, UCB 390, Boulder, CO 80309-0390, USA; greg.werner@colorado.edu, uzdensky@colorado.edu}

\affil{$^4$ JILA, University of Colorado and National Institute of Standards and Technology, UCB 440, Boulder, CO 80309-0440, USA; mitch@jila.colorado.edu}

\affil{$^5$ Department of Astrophysical and Planetary Sciences, University of Colorado, UCB 391, Boulder, CO 80309-0391, USA}

\begin{abstract}
The discovery of rapid synchrotron gamma-ray flares above $100~$MeV from the Crab Nebula has attracted new interest in alternative particle acceleration mechanisms in pulsar wind nebulae. Diffuse shock-acceleration fails to explain the flares because particle acceleration and emission occur during a single or even sub-Larmor timescale. In this regime, the synchrotron energy losses induce a drag force on the particle motion that balances the electric acceleration and prevents the emission of synchrotron radiation above $160~$MeV. Previous analytical studies and 2D particle-in-cell (PIC) simulations indicate that relativistic reconnection is a viable mechanism to circumvent the above difficulties. The reconnection electric field localized at X-points linearly accelerates particles with little radiative energy losses. In this paper, we check whether this mechanism survives in 3D, using a set of large PIC simulations with radiation reaction force and with a guide field. In agreement with earlier works, we find that the relativistic drift kink instability deforms and then disrupts the layer, resulting in significant plasma heating but few non-thermal particles. A moderate guide field stabilizes the layer and enables particle acceleration. We report that 3D magnetic reconnection can accelerate particles above the standard radiation reaction limit, although the effect is less pronounced than in 2D with no guide field. We confirm that the highest energy particles form compact bunches within magnetic flux ropes, and a beam tightly confined within the reconnection layer, which could result in the observed Crab flares when, by chance, the beam crosses our line of sight.
\end{abstract}

\keywords{Acceleration of particles --- Magnetic reconnection --- Radiation mechanisms: non-thermal --- ISM: individual (Crab Nebula)}

\section{INTRODUCTION}\label{intro}

The non-thermal radiation emitted in pulsar wind nebulae is commonly associated with ultra-relativistic electron-positron pairs injected by the pulsar and accelerated at the termination shock. In the Crab Nebula, the particle spectrum above $\sim 1~$TeV responsible for the X-ray to gamma-ray synchrotron emission is well modeled by a single power-law distribution of index $-2.2$, which is usually associated with first-order Fermi acceleration at the shock front (see e.g., \citealt{2009ASSL..357..421K}). Since the detections of the first flares of high-energy gamma rays in 2010 \citep{2011Sci...331..739A, 2011Sci...331..736T,2011A&A...527L...4B} and the following ones detected since then \citep{2011ApJ...741L...5S, 2012ApJ...749...26B, 2013ApJ...765...52S, 2013ApJ...775L...37S, 2013ATel.5485....1B}, we know that the Crab Nebula occasionally accelerates particles up to a few $10^{15}$~eV (see reviews by \citealt{2012SSRv..173..341A} and \citealt{2013arXiv1309.7046B}). This discovery is very puzzling because the particles are accelerated to such energies within a few days, which corresponds to their Larmor gyration time in the Nebula. This is far too fast for Fermi-type acceleration mechanisms which operate over multiple crossings of the particles through the shock (e.g., \citealt{1987PhR...154....1B}). In addition, the observed particle spectrum is very hard, which is not compatible with the steep power-law $\gtrsim 2$ expected with diffuse shock-acceleration \citep{2012ApJ...749...26B}. Even more surprising, the particles emit synchrotron radiation above the well-established radiation reaction limit photon energy of $160~$MeV \citep{1983MNRAS.205..593G, 1996ApJ...457..253D, 2010MNRAS.405.1809L, 2011ApJ...737L..40U}. It implies that the particles must be subject to extreme synchrotron cooling over a sub-Larmor timescale. Hence, in principle, synchrotron cooling should prevent the acceleration of pairs to such high energies in the first place.

Fortunately, there is a simple way to circumvent these tight constraints on particle acceleration if there is a region of strong coherent electric field associated with a low magnetic field perpendicular to the particle motion, i.e., if $E>B_{\perp}$. This supposes that a non-ideal, dissipative magnetohydrodynamic process is at work somewhere in the Nebula. Using a simple semi-analytical approach, \citet{2011ApJ...737L..40U} and \citet{2012ApJ...746..148C} showed that such extreme particle acceleration can occur within a Sweet-Parker-like reconnection layer \citep{1957JGR....62..509P, 1958IAUS....6..123S, 2009ARA&A..47..291Z}, where the reversing reconnecting magnetic field traps and confines the highest energy particles deep inside the layer where $E>B_{\perp}$ \citep{1965JGR....70.4219S, 2004PhRvL..92r1101K, 2007A&A...472..219C}. The reconnection electric field accelerates the particles almost linearly along a few light-day long layer. This solution solves the sub-Larmor acceleration problem at the same time. Two-dimensional (2D) particle-in-cell (PIC) simulations of relativistic pair plasma reconnection with radiation reaction force have confirmed and strengthened the viability of this scenario \citep{2013ApJ...770..147C}. These simulations can also explain the observed rapid intra-flare time variability of the $>160~$MeV synchrotron flux, the apparent photon spectral shape, as well as the flux/cutoff energy correlation \citep{2012ApJ...749...26B}.

Although these 2D PIC simulations provide a fairly complete assessment of extreme particle acceleration in reconnection layers, it is still a simplified picture of a truly three-dimensional process. We know from previous 3D reconnection studies \citep{2005PhRvL..95i5001Z, 2008ApJ...677..530Z, 2011NatPh...7..539D, 2011PhPl...18e2105L, 2011ApJ...741...39S, 2013ApJ...774...41K, 2013PhPl...20h2105M} that the reconnection layer is unstable to the relativistic tearing and kink modes, and a combination of these two into oblique modes. The kink (and oblique) instabilities, which cannot arise in the 2D simulations of \citet{2013ApJ...770..147C}, can lead to significant deformation or even disruption of the reconnection layer in 3D simulations, subsequently suppressing particle acceleration.  However, a moderate guide magnetic field can stabilize the layer \citep{2005PhRvL..95i5001Z, 2007ApJ...670..702Z, 2008ApJ...677..530Z, 2011ApJ...741...39S}.

In this work, we extend the previous 2D study of \citet{2013ApJ...770..147C} by performing large 3D PIC simulations of pair plasma reconnection with radiative feedback and with guide field, in the context of the Crab flares. In the next section, we first present the numerical techniques and the setup of the simulations chosen for this study. Then, we investigate separately the effect of the tearing and the kink instabilities on the efficiency of particle acceleration, using a set of 2D simulations in Section~\ref{results2d}. In Section~\ref{results3d}, we establish the conditions for particle acceleration above the radiation reaction limit and emission of $>160~$MeV synchrotron radiation in 3D reconnection. In addition, we report in this section on strong anisotropy and inhomogeneity of the highest-energy particles in 3D reconnection consistent with 2D results, and their important role in explaining the observed {\em Fermi}-LAT gamma-ray flux of the Crab flares. We summarize and discuss the results of this work in Section~\ref{ccl}.

\section{NUMERICAL APPROACH AND SIMULATION SETUP}\label{setup}

\subsection{Numerical techniques}

All the simulations presented in this work were performed with {\tt Zeltron}\footnote{\url{http://benoit.cerutti.free.fr/zeltron.html}}, a parallel three-dimensional electromagnetic PIC code \citep{2013ApJ...770..147C}. {\tt Zeltron} solves self-consistently Maxwell's equations using the Yee finite-difference time-domain (FDTD) algorithm \citep{1966ITAP...14..302Y}, and Newton's equation following the Boris FDTD algorithm \citep{Birdsall...Langdon}. Unlike most PIC codes, {\tt Zeltron} includes the effect of the radiation reaction force in Newton's equation (or the so-called ``Lorentz-Abraham-Dirac equation'') induced by the emission of radiation by the particles (see also, e.g., \citealt{2009PhRvL.103g5002J,2010NJPh...12l3005T,2012PhRvE..86c6401C}). In the ultra-relativistic regime, the radiation reaction force, $\mathbf{g}$, is akin to a continuous friction force, proportional to the radiative power and opposite to the particle's direction of motion (e.g., \citealt{1975ctf..book.....L,2010NJPh...12l3005T,2012ApJ...746..148C}). The expression of the radiation reaction force used in {\tt Zeltron} is given by
\begin{equation}
\mathbf{g}=-\frac{2}{3}r_{\rm e}^2\gamma\left[\left(\mathbf{E}+\frac{\mathbf{u}\times\mathbf{B}}{\gamma}\right)^2-\left(\frac{\mathbf{u}\cdot\mathbf{E}}{\gamma}\right)^2\right]\mathbf{u},
\label{gforce}
\end{equation}
where $r_{\rm e}\approx 2.82\times 10^{-13}~$cm is the classical radius of the electron, $\gamma$ is the Lorentz factor of the particle, $\mathbf{E}$ and $\mathbf{B}$ are the electric and magnetic fields, and $\mathbf{u}=\gamma\mathbf{v}/c$ is the four-velocity divided by the speed of light. This formulation is valid if $\gamma B/B_{\rm QED}\ll 1$, where $B_{\rm QED}=4.4\times 10^{13}~$G is the quantum critical magnetic field. Because of the relativistic effects, the typical frequency of the expected radiation is $\sim\gamma^3\gg 1$ times the relativistic cyclotron frequency. Hence, the radiation is not resolved by the grid and time step of the simulation. It must be calculated separately. {\tt Zeltron} computes the emitted optically thin radiation (spectrum, and angular distributions) assuming it is pure synchrotron radiation. This is valid if the change of the particle energy is small, $\Delta\gamma/\gamma\ll1$, during the formation length of a synchrotron photon, given by the relativistic Larmor radius divided by $\gamma$. We checked {\em a posteriori} that this assumption is indeed correct. The code also models the inverse Compton drag force on the particle motion in an imposed photon field, but this effect is negligible in the context of the Crab flares \citep{2012ApJ...746..148C}, hence this capability will not be utilized in the following. To perform the large 3D simulations presented in this paper, {\tt Zeltron} ran on 97,200 cores on the Kraken supercomputer\footnote{National Institute for Computational Sciences (\url{www.nics.tennessee.edu/}).} with nearly perfect scaling.

\begin{figure}
\centering
\includegraphics[width=9cm]{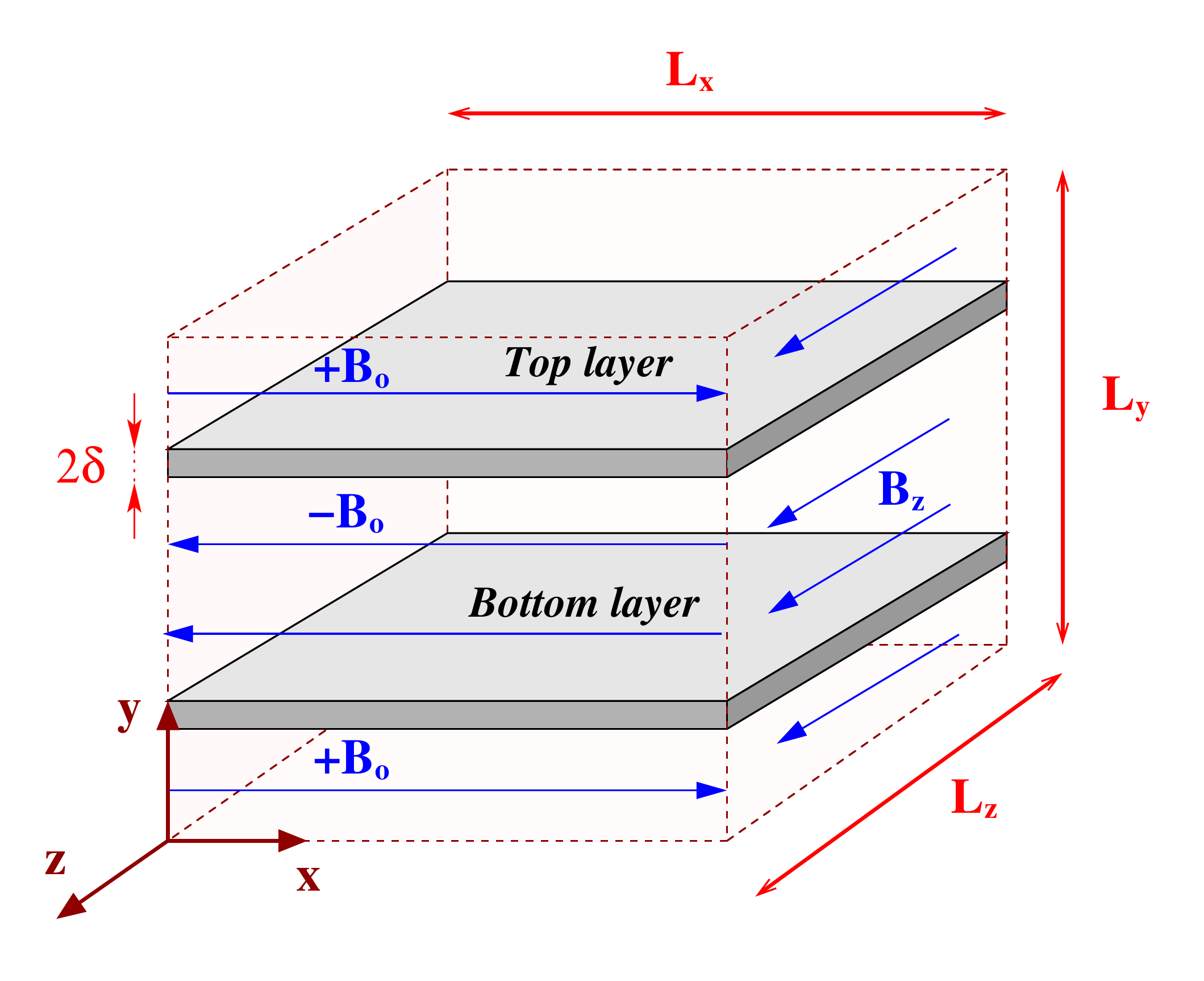}
\caption{Initial simulation setup and geometry. The computational domain is a rectangular box of volume $L_{\rm x}\times L_{\rm y}\times L_{\rm z}$ with periodic boundary conditions in all directions. The box initially contains two flat, anti-parallel, relativistic Harris layers in the $xz$-plane centered at $y=L_{\rm y}/4$ and $y=3L_{\rm y}/4$, of some thickness $2\delta$. The magnetic field structure is composed of the reconnecting field, $B_0$, along the $x$-direction, which reverses across the layers, and a uniform guide field, $B_{\rm z}=\alpha B_0$, along the $z$-direction.}
\label{fig_config}
\end{figure}

\begin{deluxetable*}{ccccccc}
\tablecaption{Complete list of the numerical simulations presented in this paper\label{tab_simu}}
\tablewidth{500pt}
\tablehead{
\colhead{Name simulation} & \colhead{$L_{\rm x}/\rho_0$} & \colhead{$L_{\rm y}/\rho_0$} & \colhead{$L_{\rm z}/\rho_0$} & \colhead{Grid cells} & \colhead{\# particles} & \colhead{$\alpha$}
}
\startdata

{\tt 2DXY0} & $200$ & $200$ & --- & $1440^2$ & $3.32\times 10^7$ & $0$ \\
{\tt 2DXY025} & $200$ & $200$ & --- & $1440^2$ & $3.32\times 10^7$ & $0.25$ \\
{\tt 2DXY050} & $200$ & $200$ & --- & $1440^2$ & $3.32\times 10^7$ & $0.5$ \\
{\tt 2DXY075} & $200$ & $200$ & --- & $1440^2$ & $3.32\times 10^7$ & $0.75$ \\
{\tt 2DXY1} & $200$ & $200$ & --- & $1440^2$ & $3.32\times 10^7$ & $1$ \\
\hline
{\tt 2DYZ0} & --- & $200$ & $200$ & $1440^2$ & $3.32\times 10^7$ & $0$ \\
{\tt 2DYZ025} & --- & $200$ & $200$ & $1440^2$ & $3.32\times 10^7$ & $0.25$ \\
{\tt 2DYZ040} & --- & $200$ & $200$ & $1440^2$ & $3.32\times 10^7$ & $0.4$ \\
{\tt 2DYZ050} & --- & $200$ & $200$ & $1440^2$ & $3.32\times 10^7$ & $0.5$ \\
{\tt 2DYZ060} & --- & $200$ & $200$ & $1440^2$ & $3.32\times 10^7$ & $0.6$ \\
{\tt 2DYZ075} & --- & $200$ & $200$ & $1440^2$ & $3.32\times 10^7$ & $0.75$ \\
{\tt 2DYZ1} & --- & $200$ & $200$ & $1440^2$ & $3.32\times 10^7$ & $1$ \\
\hline
{\tt 3D0} & $200$ & $200$ & $200$ & $1440^3$ & $4.78\times 10^{10}$ & $0$ \\
{\tt 3D050} & $200$ & $200$ & $200$ & $1440^3$ & $4.78\times 10^{10}$ & $0.5$ \\
\enddata
\tablecomments{There are three distinct subsets of simulations. The first subset comprises $5$ 2D simulations of reconnection in the $xy$-plane, designed to study the effect of the guide field strength $\alpha$ on the dynamics of reconnection and particle acceleration. The second subset comprises $7$ 2D simulations of the reconnection layer in the $yz$-plane, in order to study the development of the kink instability as a function of the guide field strength $\alpha$. The last set of simulations is chosen to test particle acceleration beyond the radiation reaction limit in 3D, in the best and in the worst cases identified in the 2D subsets.}
\end{deluxetable*}

\subsection{Simulation setup}

The simulation setup chosen here is almost identical to our previous two-dimensional pair plasma reconnection simulations with radiation reaction force in \citet{2013ApJ...770..147C}. The computational domain is a rectangular box of dimensions $L_{\rm x}$, $L_{\rm y}$ and $L_{\rm z}$, respectively along the $x$-, $y$- and $z$-directions, with periodic boundary conditions in all directions. We set up the simulation with two flat anti-parallel relativistic Harris current layers \citep{2003ApJ...591..366K} in the $xz$-plane located at $y=L_{\rm y}/4$ and $y=3L_{\rm y}/4$ (Figure~\ref{fig_config}). Having two current sheets is only a convenient numerical artifact that allows us to use periodic boundary conditions along the $y$-direction, but it does not have a physical meaning in the context of our model of the Crab flares where only one reconnection layer is involved. The electric current, $J_{\rm z}$, flows in the $\pm z$-directions, and is supported by electrons counter-streaming with positrons at a mildly relativistic drift velocity (relative to the speed of light) $\beta_{\rm drift}=0.6$. The plasma (electrons and positrons) is spatially distributed throughout the domain, with the following density profile
\begin{equation}
n = \left\{ \begin{array}{lcl} n_0\left[\cosh\left(\frac{y-L_{\rm y}/4}{\delta}\right)\right]^{-2}+0.1n_0 &\mbox{if} & y<L_{\rm y}/2\\
n_0\left[\cosh\left(\frac{y-3L_{\rm y}/4}{\delta}\right)\right]^{-2}+0.1n_0 & \mbox{if} & y>L_{\rm y}/2 \end{array} \right. .
\label{plasma_density}
\end{equation}
The first term is the density of the drifting pairs carrying the initial current, concentrated within the layer half-thickness $\delta=\lambda_{\rm D}/\beta_{\rm drift}$, where $\lambda_{\rm D}$ is the relativistic Debye length \citep{2003ApJ...591..366K}. This population is modeled with a uniform and isotropic (in the co-moving frame) distribution of macro-particles with variable weights to account for the density profile and to decrease the numerical noise in low-density regions. The second term is a uniform and isotropic background pair plasma at rest in the laboratory frame with a density chosen to be 10 times lower than at the center of the layers (i.e., $0.1 n_0$). The drifting and the background particles are distributed in energy according to a relativistic Maxwellian with the same temperature $\theta_0\equiv kT/m_{\rm e}c^2=10^8$, where $k$ is the Boltzmann constant and $m_{\rm e}$ is the rest mass of the electron. The temperature of the drifting particles is defined in the co-moving frame. This temperature models the ultra-relativistic plasma already present in the Crab Nebula, prior to reconnection, whose particles could have been accelerated at the wind termination shock or even by other reconnection events throughout the nebula. However, observations show that in reality the background plasma is distributed according to a broad and steep power-law, extending roughly between $\gamma_{\rm min}=10^6$ and $\gamma_{\rm max}=10^9$ (responsible for the UV to 100 MeV synchrotron spectrum). This large dynamic range of particle energies translates directly into an equally large dynamic range of relativistic Larmor radii and hence of length scales that must be resolved in the simulation, which is beyond the reach of our numerical capabilities.

\begin{figure*}
\centering
\includegraphics[width=8.5cm]{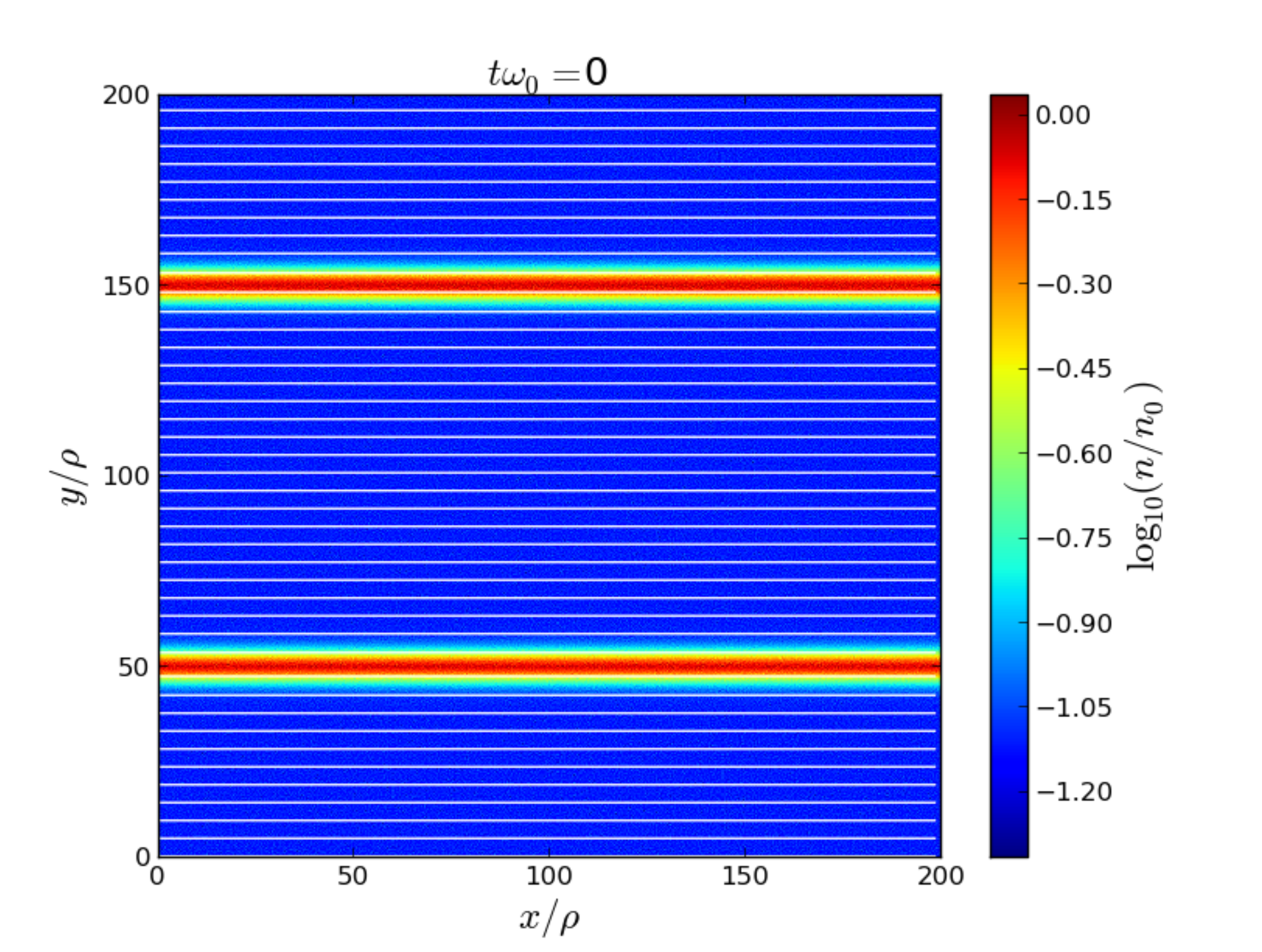}
\includegraphics[width=8.5cm]{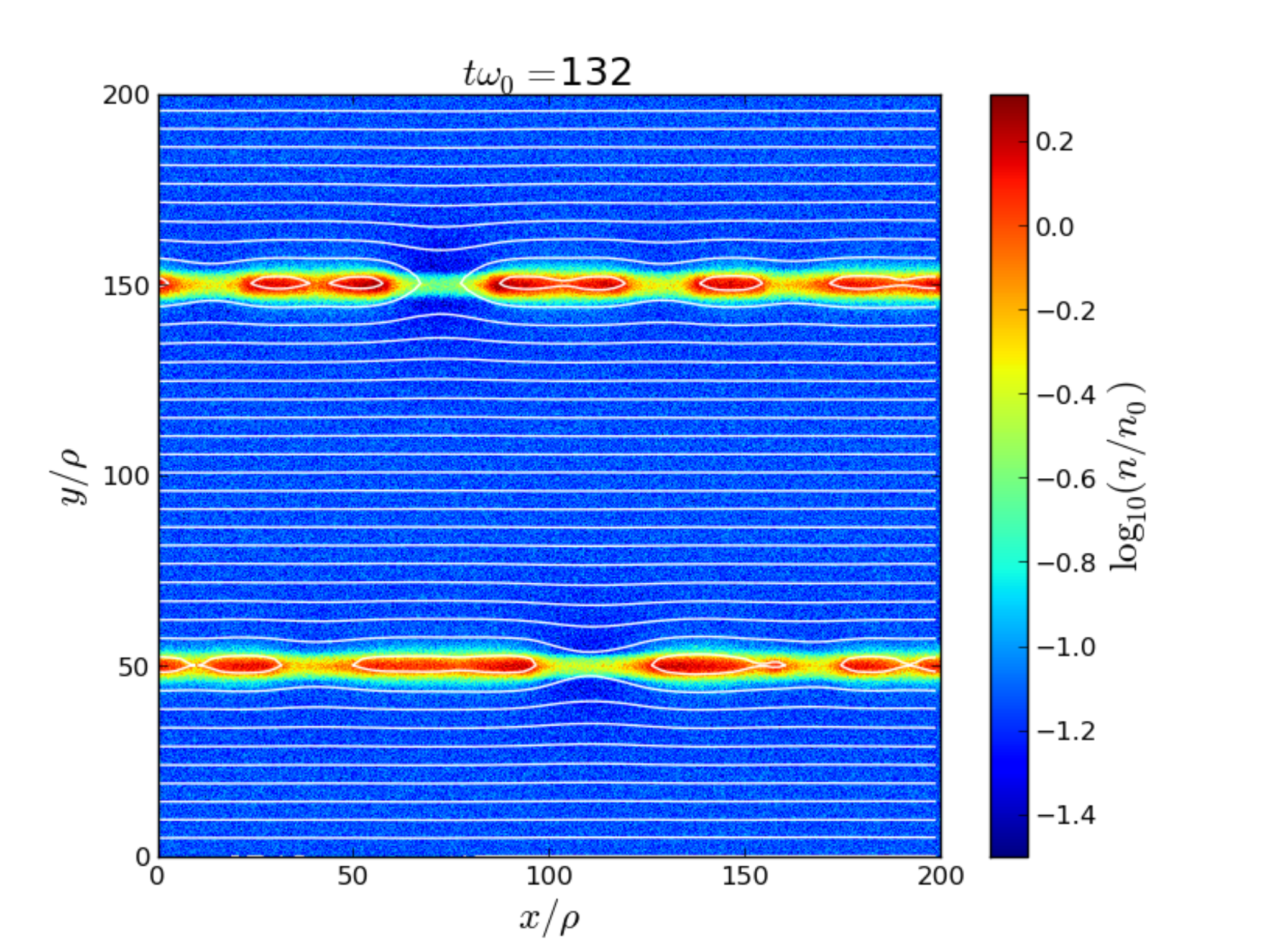}
\includegraphics[width=8.5cm]{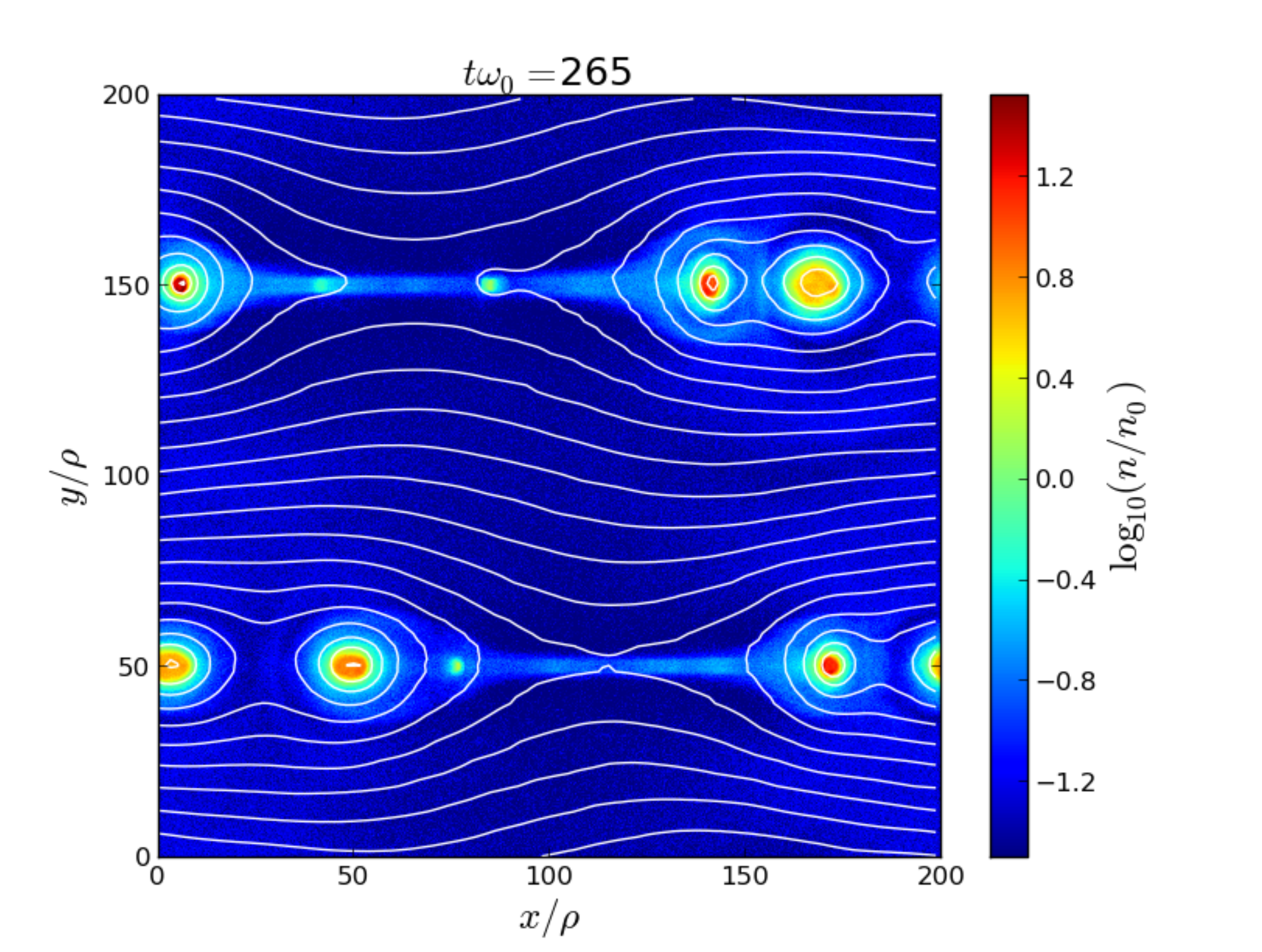}
\includegraphics[width=8.5cm]{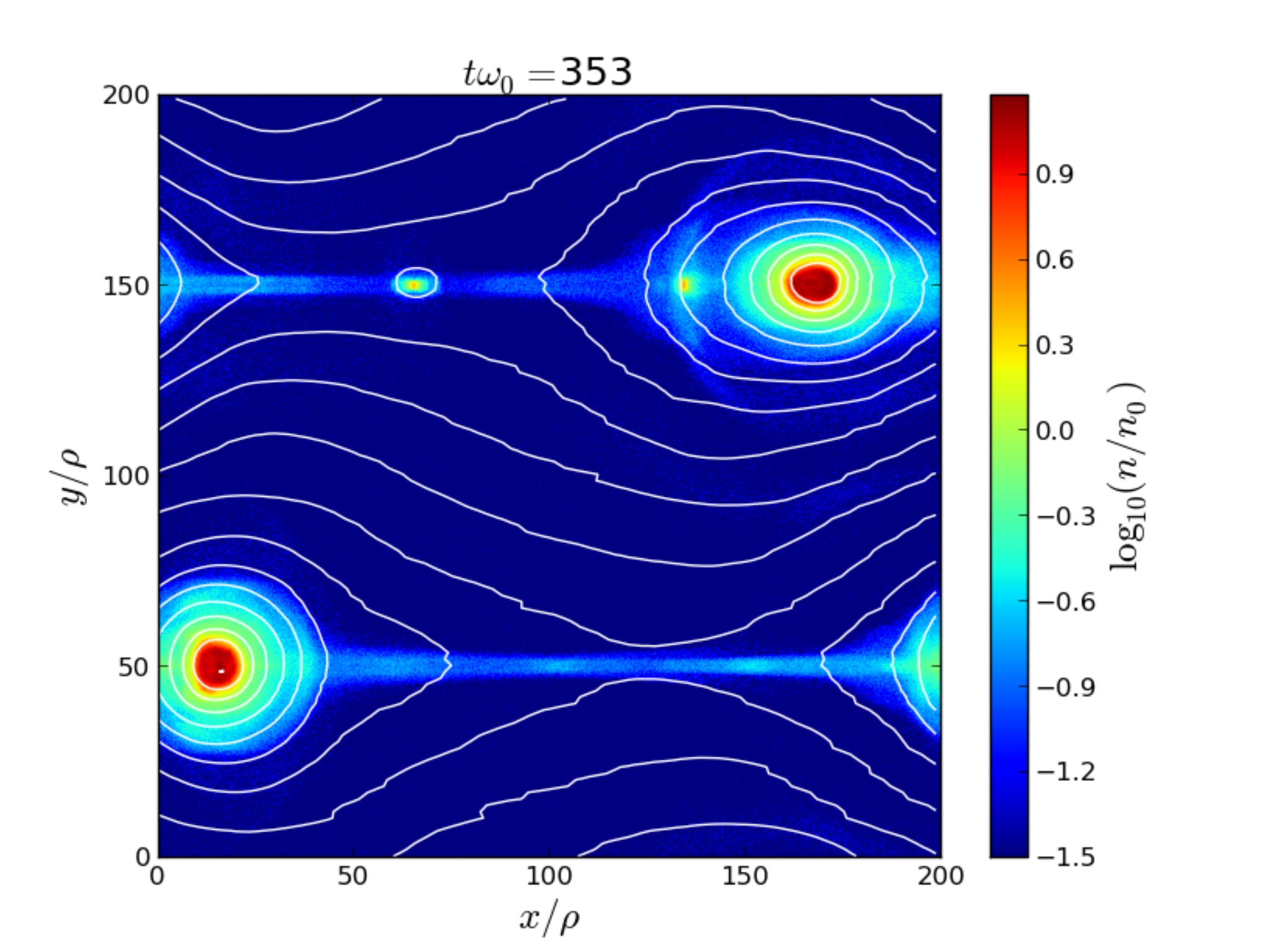}
\caption{Snapshots of the plasma density at $t\omega_0=0$ (top left)$,~132$ (top right)$,~265$ (bottom left)$,~$and $353$ (bottom right) of the 2D simulation {\tt 2DXY0} in the $xy$-plane (with no guide field, $\alpha=0$). Magnetic field lines are represented by solid white lines. In this simulation, the development of the tearing instability forms multiple plasmoids separated by X-points which facilitates fast magnetic reconnection. Reconnection dissipates about $70\%$ of the magnetic energy in the form of energetic particles and radiation (see Figures~\ref{fig_spec_time}-\ref{fig_spec_tearing}).}
\label{fig_xy}
\end{figure*}

The initial electromagnetic field configuration is
\begin{eqnarray}
\mathbf{B}&=&\left\{ \begin{array}{lcl} -B_0\tanh\left(\frac{y-L_{\rm y}/4}{\delta}\right)\mathbf{e_{\rm x}}+\alpha B_0 \mathbf{e_{\rm z}} &\mbox{if} & y<L_{\rm y}/2 \label{eq_mag}\\
~~B_0\tanh\left(\frac{y-3L_{\rm y}/4}{\delta}\right)\mathbf{e_{\rm x}}+\alpha B_0 \mathbf{e_{\rm z}} & \mbox{if} & y>L_{\rm y}/2 \end{array} \right. ,\\
\nonumber \\
\mathbf{E}&=&\mathbf{0}~,
\end{eqnarray}
where $\mathbf{e_{\rm x}},~\mathbf{e_{\rm z}}$ are unit vectors along the $x$- and $z$-directions. $B_0$ is the upstream reconnecting magnetic field and $\alpha$ is a dimensionless parameter of the simulation that quantifies the strength of the guide field component $B_{\rm z}$ in units of $B_0$ (Figure~\ref{fig_config}). Observations constrain the magnetic field in the emitting region to about $1~$mG, which is much higher than the expected average quiescent field of order $100$-$200~\mu$G \citep{2010A&A...523A...2M}. In this work, we choose $B_0=5~$mG to be consistent with our previous studies of the flares \citep{2012ApJ...746..148C, 2013ApJ...770..147C}. Hence, the energy scale at which the radiation reaction force equals the electric force, assuming that $E=B_0=5$~mG, is
\begin{equation}
\gamma_{\rm rad}m_{\rm e}c^2=\sqrt{\frac{3e m^2_{\rm e} c^4}{2r_{\rm e}^2 B_0}}\approx 1.3\times 10^9 m_{\rm e}c^2,
\label{grad}
\end{equation}
where $e$ is the fundamental electric charge. Below, we express lengths in units of the typical initial Larmor radius of the particles in the simulations, i.e., $\rho_0=\theta_0 m_{\rm e}c^2/eB_0\approx 3.4\times 10^{13}~$cm. In all the simulations, the layer half-thickness is then $\delta/\rho_0\approx 2.7$ and the relativistic collisionless electron skin-depth $d_{\rm e}\equiv\sqrt{\theta_0 m_{\rm e}c^2/4\pi n_0 e^2}\approx 1.8\rho_0$. Similarly, timescales are given in units of the gyration time of the bulk of the particles in the plasma, i.e., $\omega_0^{-1}\equiv\rho_0/c\approx 1140~$s. The initial distribution of fields and plasma results in a low plasma-$\beta$ or high magnetization of the upstream plasma (i.e., outside the layers). Here, the magnetization parameter is $\sigma\equiv B_0^2/4\pi (0.1 n_0)\theta_0 m_{\rm e} c^2\approx 16$.

The system is initially set at an equilibrium, i.e., there is a force balance across the reconnection layers between the magnetic pressure and the drifting particle pressure. This equilibrium is unstable to two competing instabilities, namely the relativistic tearing and kink instabilities, as well as oblique modes that combine tearing and kink modes \citep{2005PhRvL..95i5001Z, 2008ApJ...677..530Z, 2011NatPh...7..539D, 2013ApJ...774...41K}. In contrast to \citet{2013ApJ...770..147C}, we choose here not to apply any initial perturbation in order to avoid any artificial enhancement of one type of instability over the other. Instabilities are seeded with the numerical noise only. This choice has a direct computational cost because the lack of perturbation significantly delays the onset of reconnection (See Sections~\ref{evol},~\ref{evol3d}), but it enables a fair comparison between the growth rates of both instabilities (Sections~\ref{fou},~\ref{fou3d}). Another important consequence of this choice specific to this study is the significant radiative cooling of the particles before reconnection can accelerate them (Section~\ref{spec2d}).

\begin{figure*}
\centering
\includegraphics[width=8.5cm]{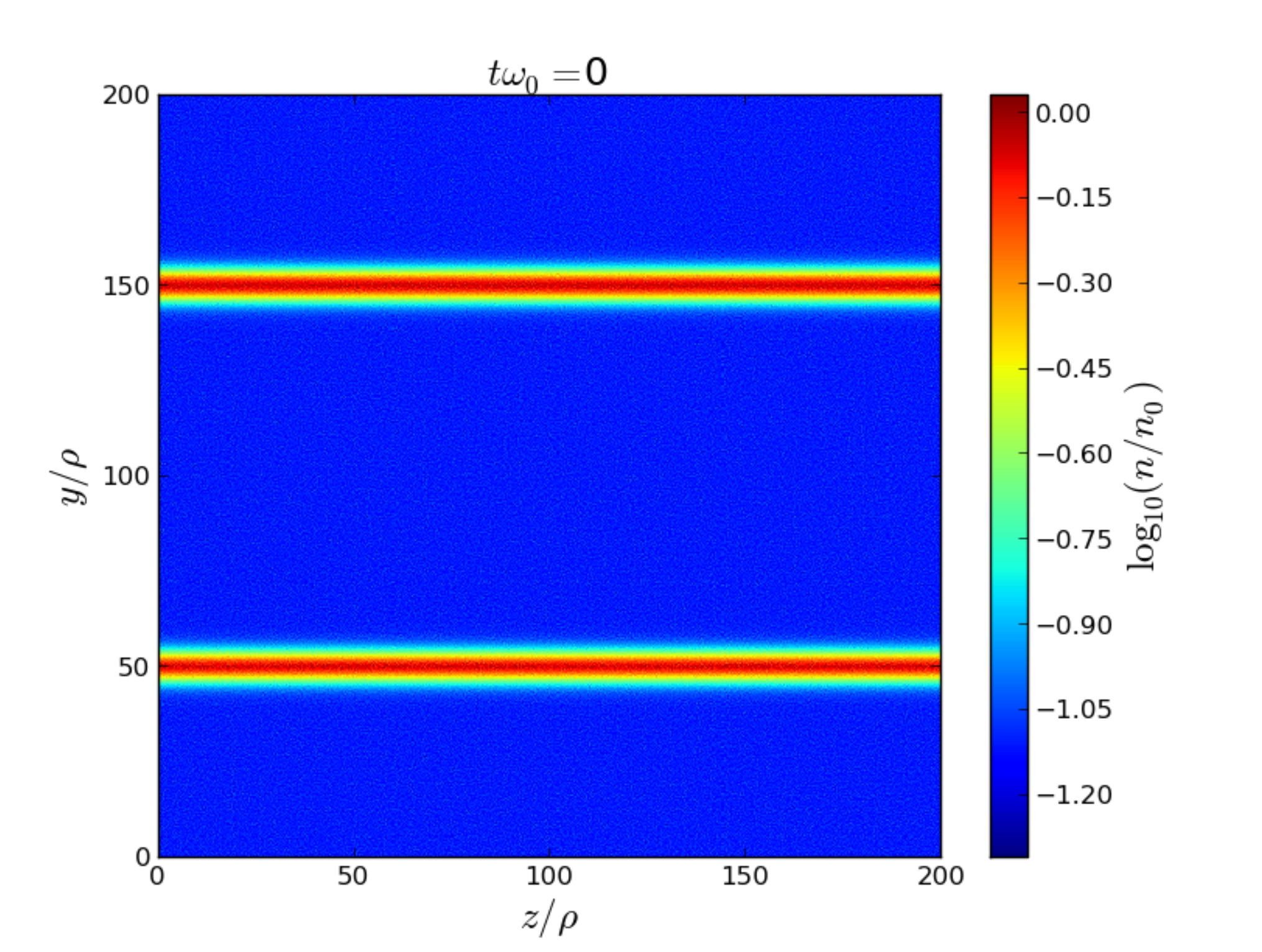}
\includegraphics[width=8.5cm]{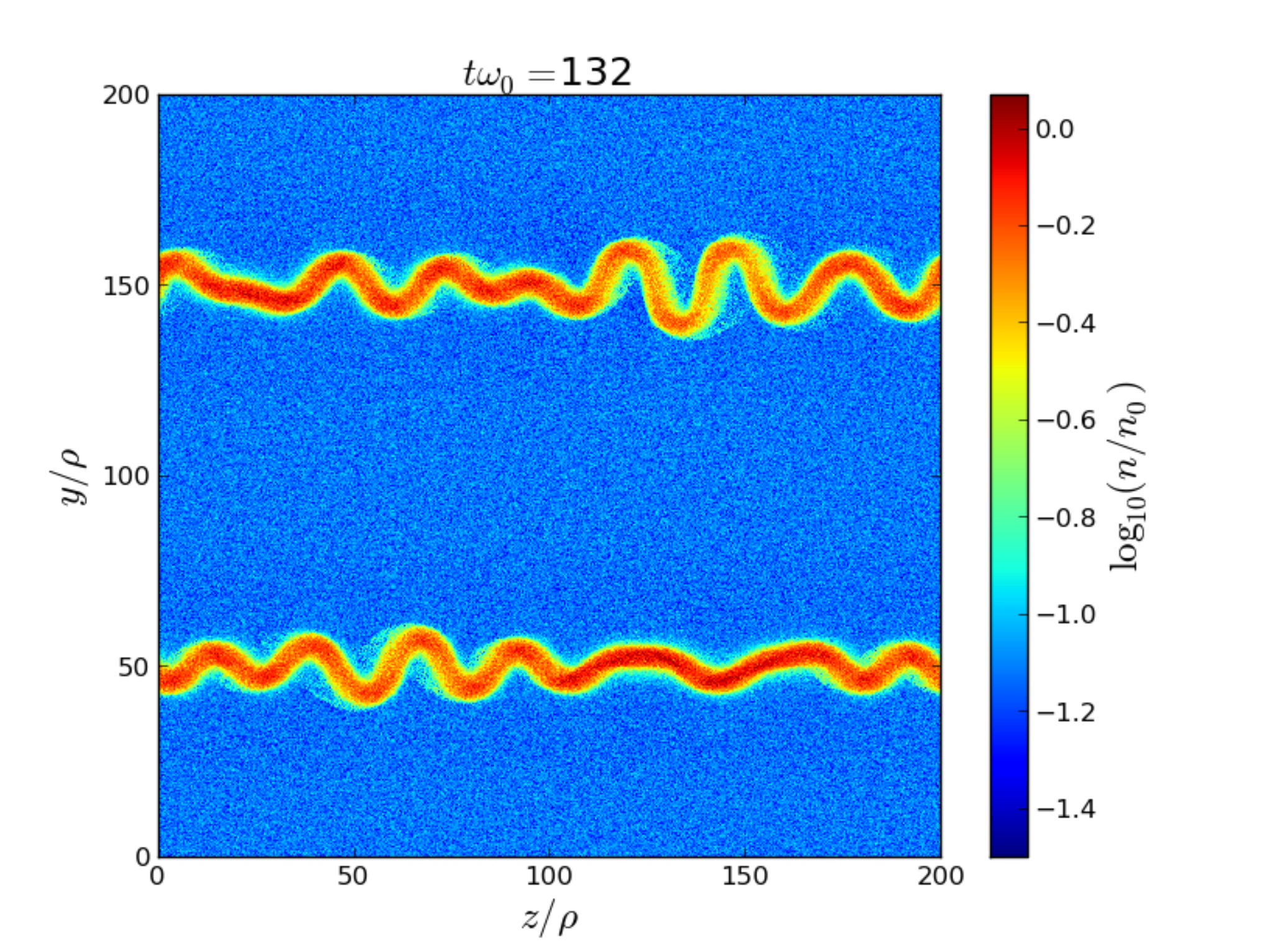}
\includegraphics[width=8.5cm]{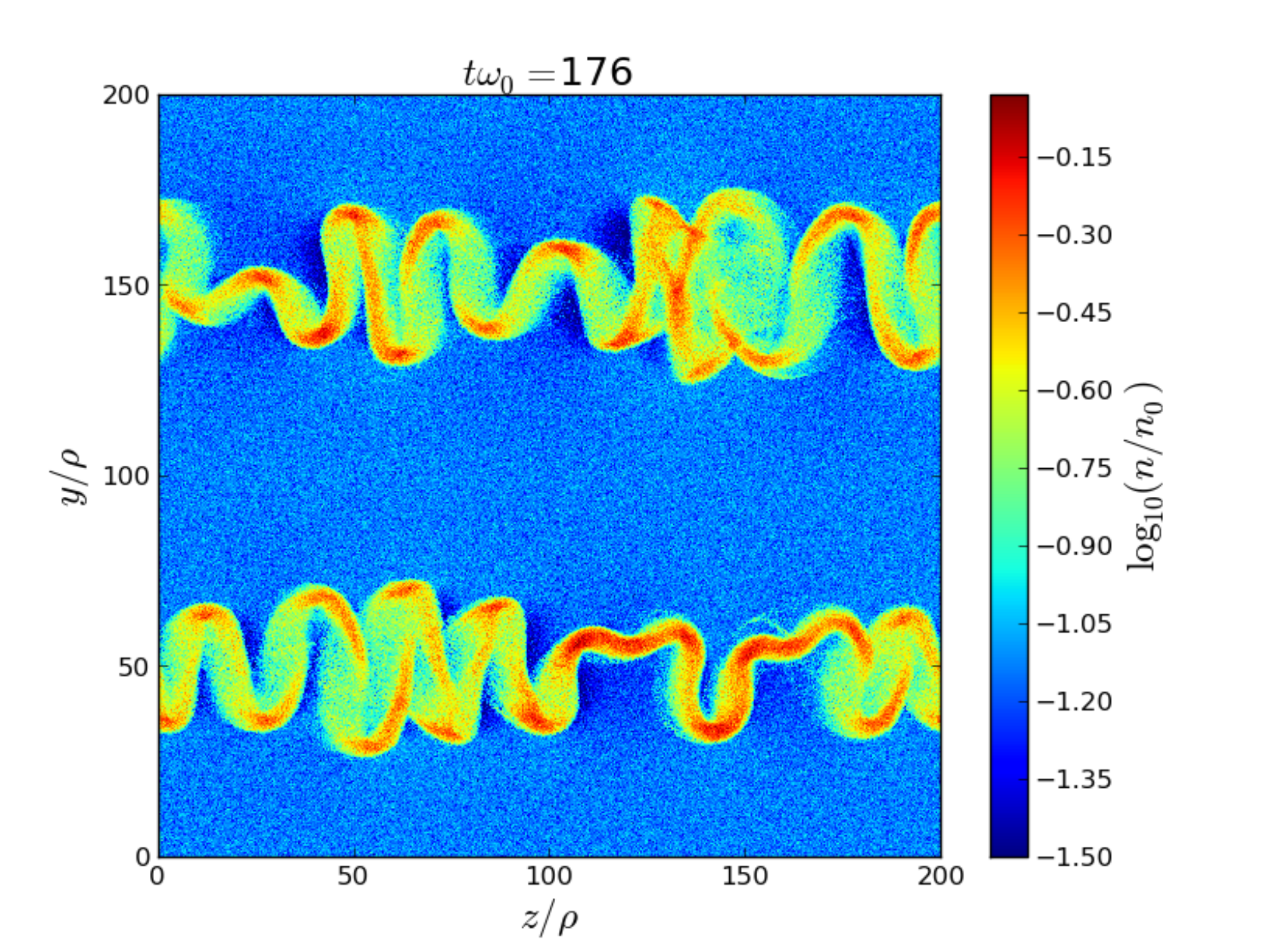}
\includegraphics[width=8.5cm]{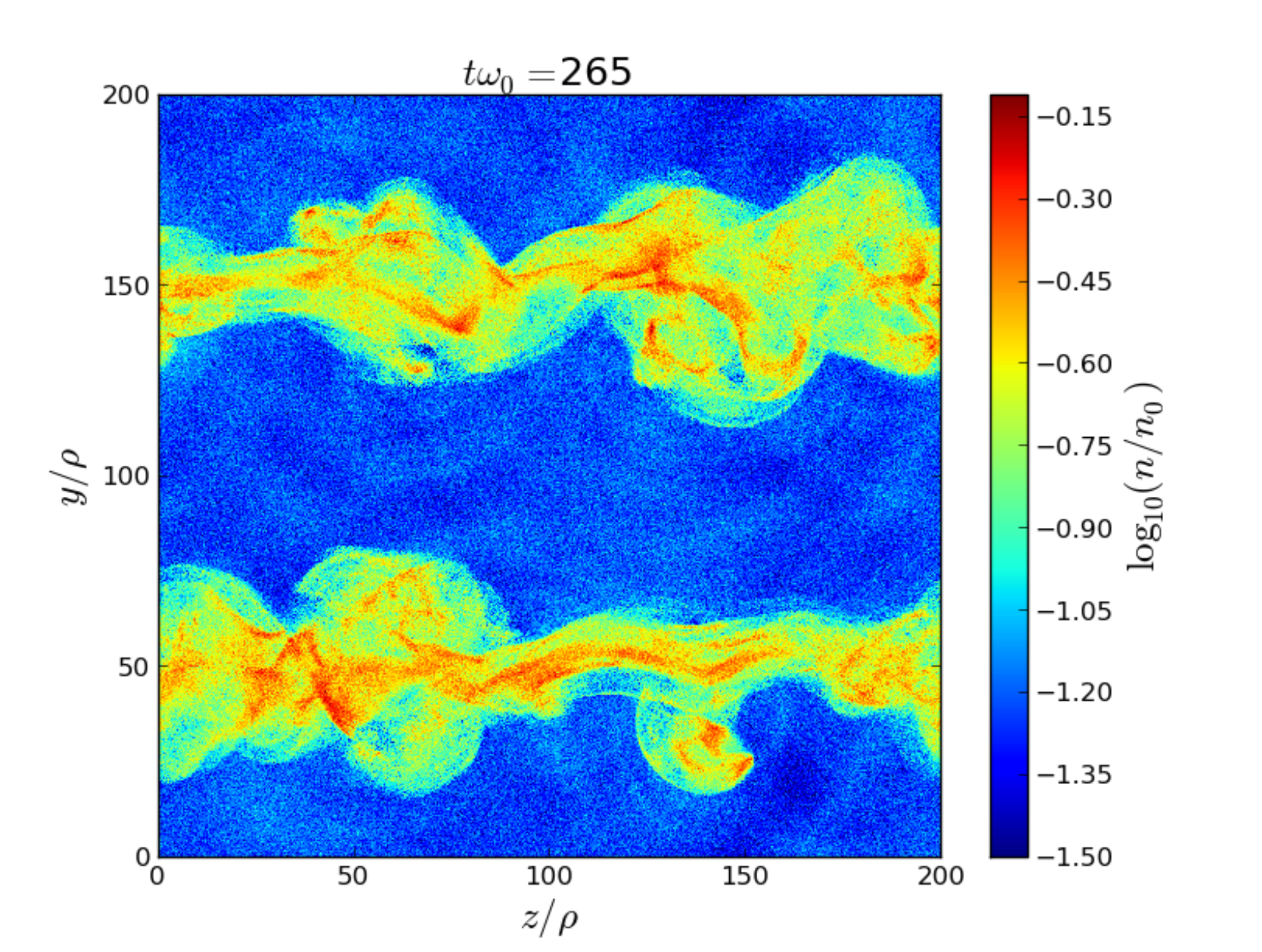}
\caption{Snapshot of the plasma density at $t\omega_0=0$ (top left)$,~132$ (top right)$,~176$ (bottom left)$,~$and $265$ (bottom right) of the 2D simulation {\tt 2DYZ0} in the $yz$-plane (with no guide field, $\alpha=0$). Although this simulation cannot capture magnetic reconnection that proceeds in the $xy$-plane, it shows that the layers rapidly destabilize along the $z$-direction due to the kink instability. The layers are deformed and eventually completely disrupted, leading to efficient dissipation of the magnetic energy (about $55\%$), mostly in the form of heat (see Figures~\ref{fig_spec_time}-\ref{fig_spec_kink}).}
\label{fig_zy}
\end{figure*}

\subsection{Set of simulations}

In this work, we performed a series of $14$ simulations. This set includes $12$ 2D simulations, and $2$ 3D simulations. The 2D simulations are designed to study the effect of the guide field strength, $\alpha=0,~0.25,~0.5,~0.75$ and $1$ on the developments of instabilities (kink and tearing) and on particle acceleration/emission. To analyze the developments both instabilities separately, we follow the same approach as \citet{2007ApJ...670..702Z, 2008ApJ...677..530Z}, i.e., we consider the dynamics of the Harris current layers in the $xy$-plane where the tearing modes alone develop similarly to our previous study in \citet{2013ApJ...770..147C}, and in the $yz$-plane where the kink modes alone develop (the kink and the tearing modes are perpendicular to each other). The box is square of size $L_{\rm x}\times L_{\rm y}=(200\rho_0)^2$ and $L_{\rm y}\times L_{\rm z}=(200\rho_0)^2$ with $1440^2$ cells and 16 particles per cell (all species together). The spatial resolution is $\rho_0/\Delta x\approx 7.2$, where $\Delta x$ is the grid spacing in the $x$-direction ($\Delta x=\Delta y=\Delta z$), which ensures the conservation of the total energy to within $\lesssim 1\%$ error throughout the simulation. From this 2D scan, we identify the best/worst conditions for efficient particle acceleration and emission above the radiation reaction limit in 3D. The 3D box is cubical of size $L_{\rm x}\times L_{\rm y}\times L_{\rm z}=(200\rho_0)^3$ with $1440^3$ grid cells and $16$ particles per cell (all species together). The simulation time step is set at $0.3$ times the critical Courant-Friedrichs-Lewy time step, $\Delta t=0.3\Delta t_{\rm CFL}\approx 0.029\omega_0^{-1}$ in 2D and $\approx 0.024\omega_0^{-1}$ in 3D, in order to maintain satisfactory total energy conservation in the presence of strong radiative damping. Table~\ref{tab_simu} enumerates all the simulations presented here.

\section{RESULTS OF THE 2D RUNS}\label{results2d}

In this section, we present and discuss the results of the 2D runs listed in Table~\ref{tab_simu}. After describing the overall time evolution of the reconnection layers in the $xy$- and $yz$-planes (Section~\ref{evol}), we present a Fourier analysis of the tearing and kink instabilities as a function of the guide field strength (Section~\ref{fou}). Then, we deduce from the particle and photon spectra the most/least favorable conditions for particle acceleration beyond $\gamma_{\rm rad}$ and synchrotron emission $>160~$MeV in 3D (Section~\ref{spec2d}).

\subsection{Description of the time evolution}\label{evol}

Figure~\ref{fig_xy} shows the time evolution of the total plasma density and field lines at four characteristic stages of 2D magnetic reconnection in the $xy$-plane with $\alpha=0$ (run {\tt 2DXY0}). Because there is no initial perturbation, the layers remain static until $t\omega_0\approx 120$ when the layer tears apart into about 7 plasmoids per layer separated by X-points where field lines reconnect. The noise of the macro-particles in the PIC code is sufficient to seed the tearing instability. The reconnection electric field $E_{\rm z}$ is maximum at X-points and is responsible for most of particle acceleration. The high magnetic tension of freshly reconnected field lines pushes the plasma towards the $\pm x$-directions and drives the large scale reconnection outflow that forces magnetic islands to merge with each other. Reconnection proceeds until there is only one big island per layer remaining in the box. At the end of the simulation ($t\omega_0=353$), about $70\%$ of the initial magnetic energy is dissipated in the form of particle kinetic energy. All the energy gained by the particles is then lost via the emission of synchrotron radiation. Adding a guide field does not suppress the tearing instability, but it creates a charge separation across the layer that induces a strong $E_{\rm y}$ electric field (see also \citealt{2008ApJ...677..530Z,2013ApJ...770..147C}).

Figure~\ref{fig_zy} presents the time evolution of the 2D simulation in the $yz$-plane with no guide field (run {\tt 2DYZ0}). The initial setup of fields and particles is identical to run {\tt 2DXY0}, except that the reconnecting field ($B_{\rm x}$) is now perpendicular to the simulation plane. Hence, reconnection and tearing modes cannot be captured by this simulation. Instead, we observe the development of the kink instability as early as $t\omega_0\approx 100$ in the form of a small sinusoidal deformation of the current sheets with respect to the initial layer mid-plane. The sinusoidal deformation proceeds along the $z$-direction, with the deformation amplitude in the $y$-direction increasing rapidly up to about a quarter of the simulation box size (about $50\rho_0$). At this stage, the folded current layers are disrupted, leading to fast and efficient magnetic dissipation. About $55\%$ of the total magnetic energy is dissipated by the end of the simulation. The guide field has a dramatic influence on the stability of the layers. The amplitude of the deformation as well as the magnetic energy dissipated decreases with increasing guide field. For $\alpha\gtrsim 0.75$, the layers remain flat during the entire duration of the simulation and no magnetic energy is dissipated. In this case, the only noticeable time evolution is a slight decrease of the layer thickness due to synchrotron cooling. To maintain pressure balance across the layers with the unchanged upstream magnetic field, the layer must compress to compensate for the radiative energy losses \citep{2011PhPl...18d2105U}. We observed also a compression of the reconnection layer in the $xy$-reconnection simulations.

\subsection{Fourier analysis of unstable modes}\label{fou}

To compare the relative strength of the tearing instability versus the kink instability, we perform a spectral analysis of the fastest growing modes that develop in the simulations. To study the kink instability, we do a fast Fourier transform (FFT) along the $z$-direction of the small variations of the reconnecting magnetic field in the bottom layer mid-plane, $\delta B_{\rm x}(z,t)=B_{\rm x}(y=L_{\rm y}/4,z,t)-B_{\rm x}(y=L_{\rm y}/4,z,0)$, during the early phase of the 2D simulations in the $yz$-plane. For the tearing modes, we follow the same procedure for the fluctuations in the reconnected field along the $x$-direction, $\delta B_{\rm y}(x,t)=B_{\rm y}(x,y=L_{\rm y}/4,t)-B_{\rm x}(y=L_{\rm y}/4,z,0)$, in the 2D simulations in the $xy$-plane.

\begin{figure}
\centering
\includegraphics[width=8.5cm]{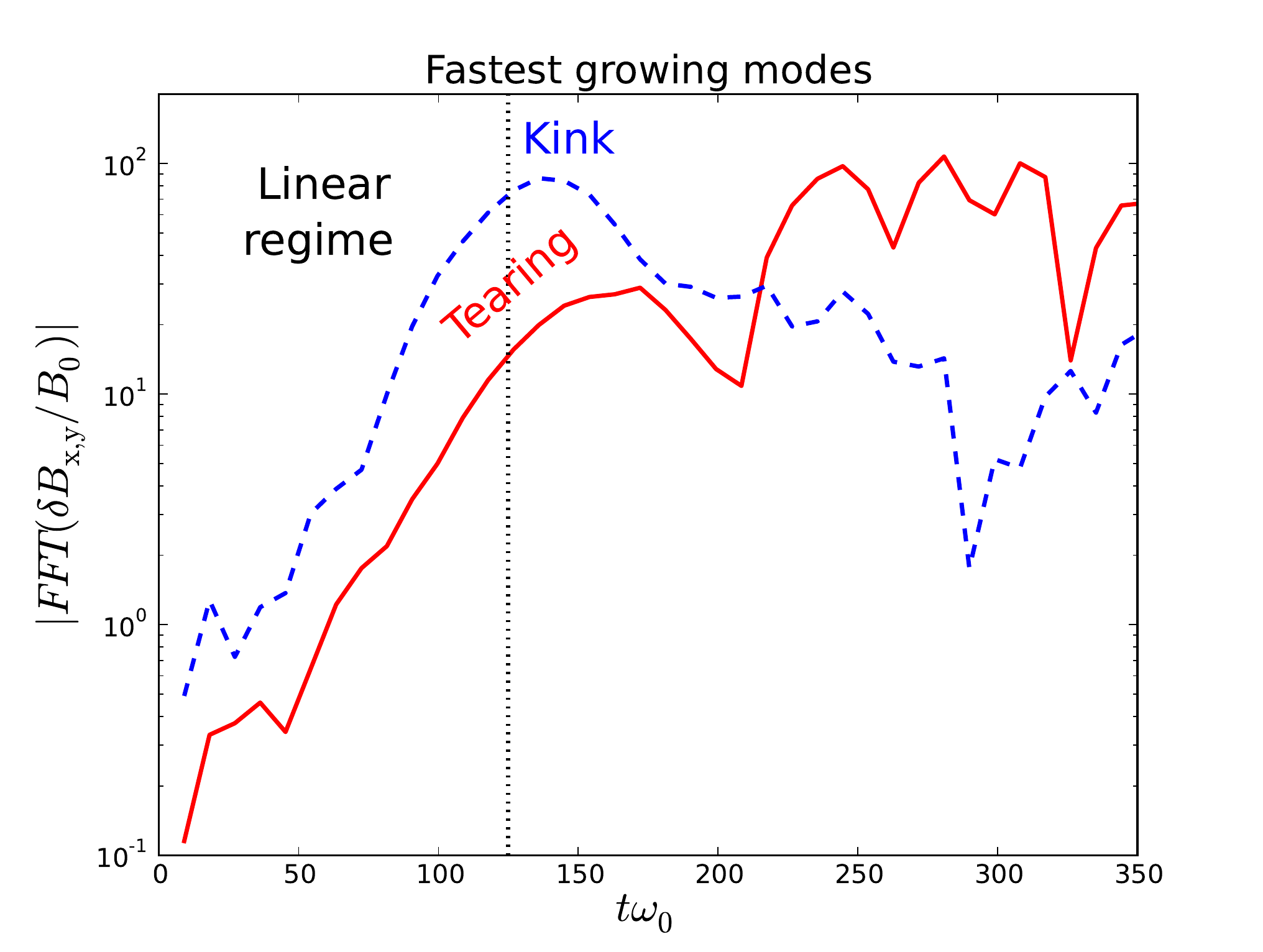}
\includegraphics[width=8.5cm]{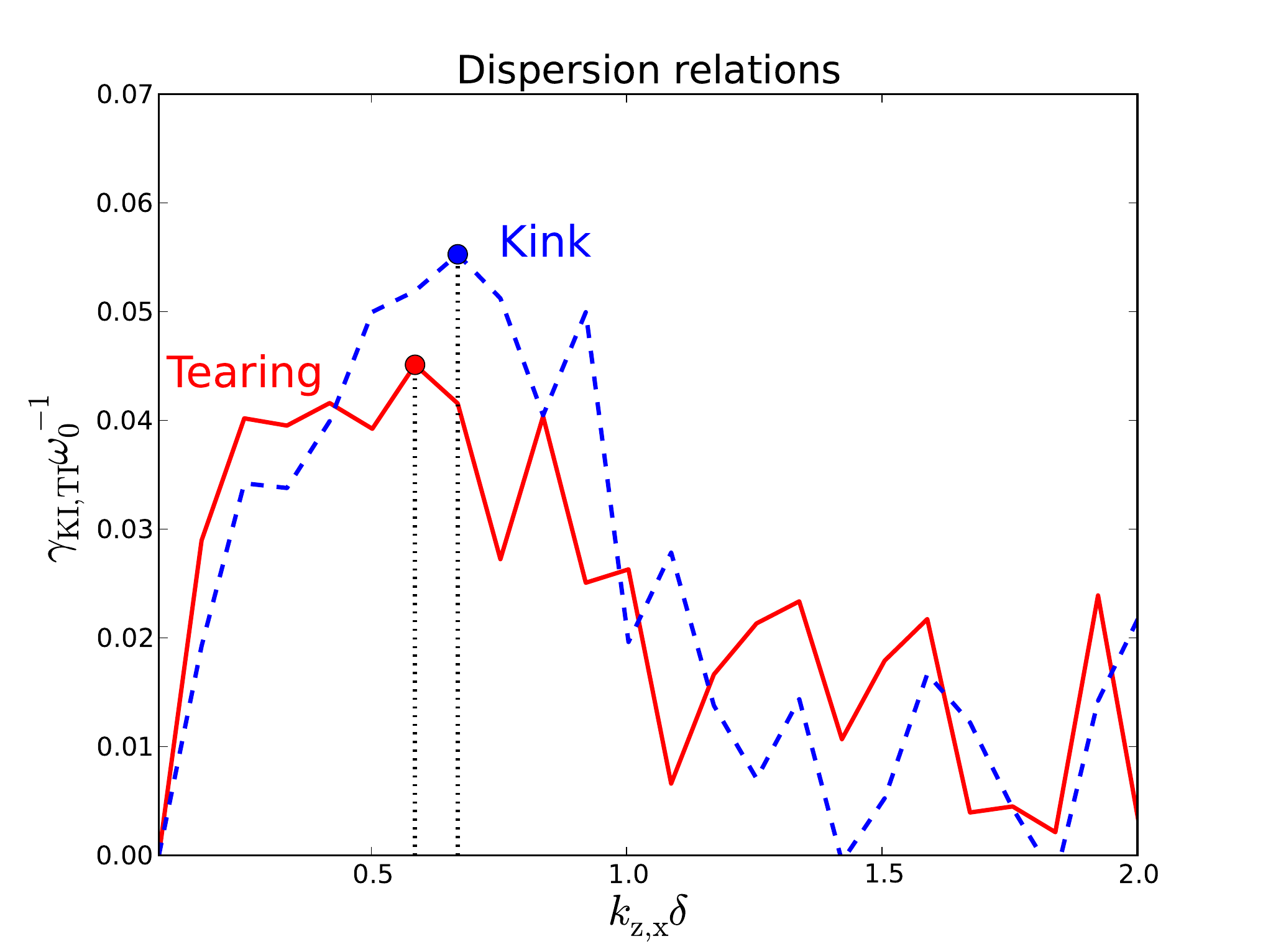}
\caption{Top: Time evolution of the fastest growing tearing (red solid line, $k_{\rm x}\delta\approx 0.58$) and kink (blue dashed line, $k_{\rm z}\delta\approx 0.67$) modes, in the simulation {\tt 2DXY0} and {\tt 2DYZ0} where $\alpha=0$. The duration of the linear phase is $t\omega_0\approx 125$ (delimited by the vertical dotted line) and is about the same in both simulations. Bottom: Dispersion relations of the tearing (red solid line) and kink (blue dashed line) instabilities during the linear stage. This plot shows only the region of small wavenumber $k_{\rm z,x}$ where the most unstable modes are found. The vertical black dotted lines mark the fastest growing modes.}
\label{fig_disp}
\end{figure}

We present in Figure~\ref{fig_disp} the time evolution of the fastest growing modes as well as the dispersion relations for the tearing and kink modes, with no guide field. In the linear regime ($t\omega_0\lesssim 125$), we infer the growth rates by fitting the amplitude of each mode with $\left|\rm{FFT}(\delta B_{\rm x,y}/B_0)\right|\propto\exp\left(\gamma_{\rm gr}(k)t\right)$, where $\gamma_{\rm gr}$ is the growth rate of the mode of wave-number $k$. We find that the fastest growing tearing mode is at $k_{\rm x}\delta\approx 0.58$, which coincides with the analytical expectation of $k_{\rm x}\delta=1/\sqrt{3}$ \citep{1979SvA....23..460Z} as found by \citet{2007ApJ...670..702Z}. The wavelength of this mode is $L_{\rm x}/\lambda_{\rm x}=L_{\rm x}/2\pi\sqrt(3)\delta\approx 7$; this explains the number of plasmoids formed in the early stages of reconnection (see top right panel in Figure~\ref{fig_xy}). The fastest growing kink mode has a wavelength $L_{\rm z}/\lambda_{\rm z}\approx 8$ (or $k_{\rm z}\delta\approx 0.67$) which is consistent with the deformation of the current layers observed in Figure~\ref{fig_zy}, top right panel. The corresponding growth rate is $\gamma_{\rm KI}\omega_0^{-1}\approx 0.055$, which is comparable with the fastest tearing growth rate, $\gamma_{\rm TI}\omega_0^{-1}\approx 0.045$ (Figure~\ref{fig_disp}, bottom panel). This is expected for an ultra-relativistic plasma ($kT\gg m_{\rm e}c^2$) with a drift velocity $\beta_{\rm drift}=0.6$ \citep{2007ApJ...670..702Z}. It is worth noting that the dispersion relations for both the kink and the tearing instabilities are not sharply peaked around the fastest growing modes; a broad range of low-frequency modes is almost equally unstable (i.e., for $0< k_{\rm x,z}\delta\lesssim 1$).

\begin{figure}
\centering
\includegraphics[width=8.5cm]{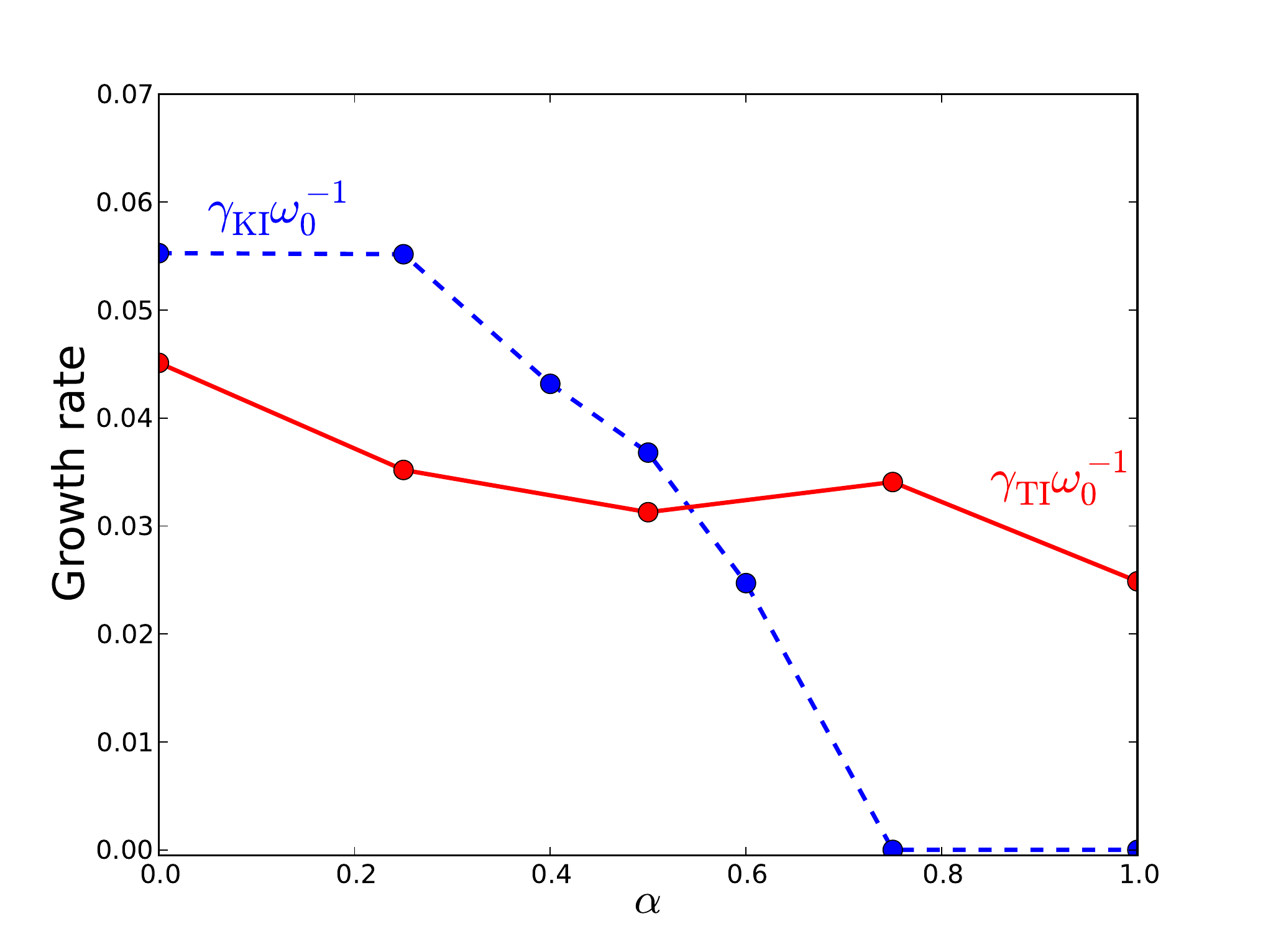}
\caption{Linear growth rates of the fastest growing modes for the tearing (for $\gamma_{\rm TI}(k_{\rm x}=0.58)$, red solid line) and the kink (for $\gamma_{\rm KI}(k_{\rm z}=0.67)$, blue dashed line) instabilities multiplied by $\omega_0^{-1}$, as a function of the guide field strength $\alpha$. Each dot represents one simulation. The analysis of the kink/tearing instability was performed using the set of 2D simulations in the $yz$-/$xy$-plane.}
\label{fig_rate}
\end{figure}

In agreement with \citet{2008ApJ...677..530Z} and as pointed out in Section~\ref{evol}, we find that the kink instability depends sensitively on the guide field strength. Figure~\ref{fig_rate} shows that the fastest growth rate decreases rapidly between $\alpha=0.25$ and $\alpha=0.75$ from $\gamma_{\rm KI}\omega_0^{-1}\approx 0.055$ to undetectable levels. Thus, the guide field stabilizes the layer along the $z$-direction. In contrast, as mentioned earlier in Section~\ref{evol}, the tearing growth rate depends only mildly on $\alpha$; we note a decrease from $\gamma_{\rm TI}\omega_0^{-1}\approx 0.045$ for $\alpha=0$ to $\gamma_{\rm TI}\omega_0^{-1}\approx 0.025$ for $\alpha=1$. For $\alpha\gtrsim 0.5$, the tearing instability dominates over the kink.

\subsection{Particle and photon spectra}\label{spec2d}

\begin{figure}
\centering
\includegraphics[width=8.5cm]{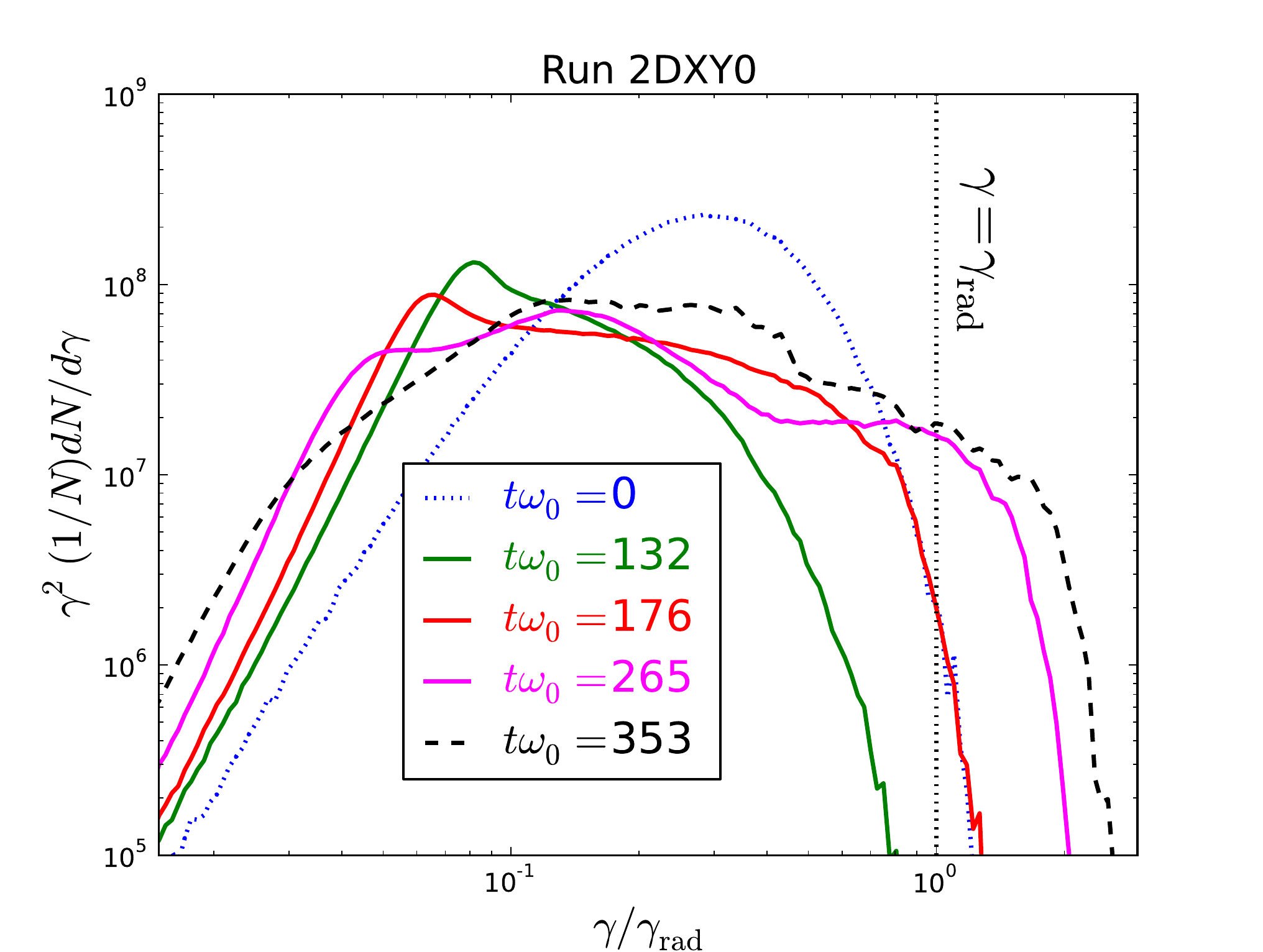}
\includegraphics[width=8.5cm]{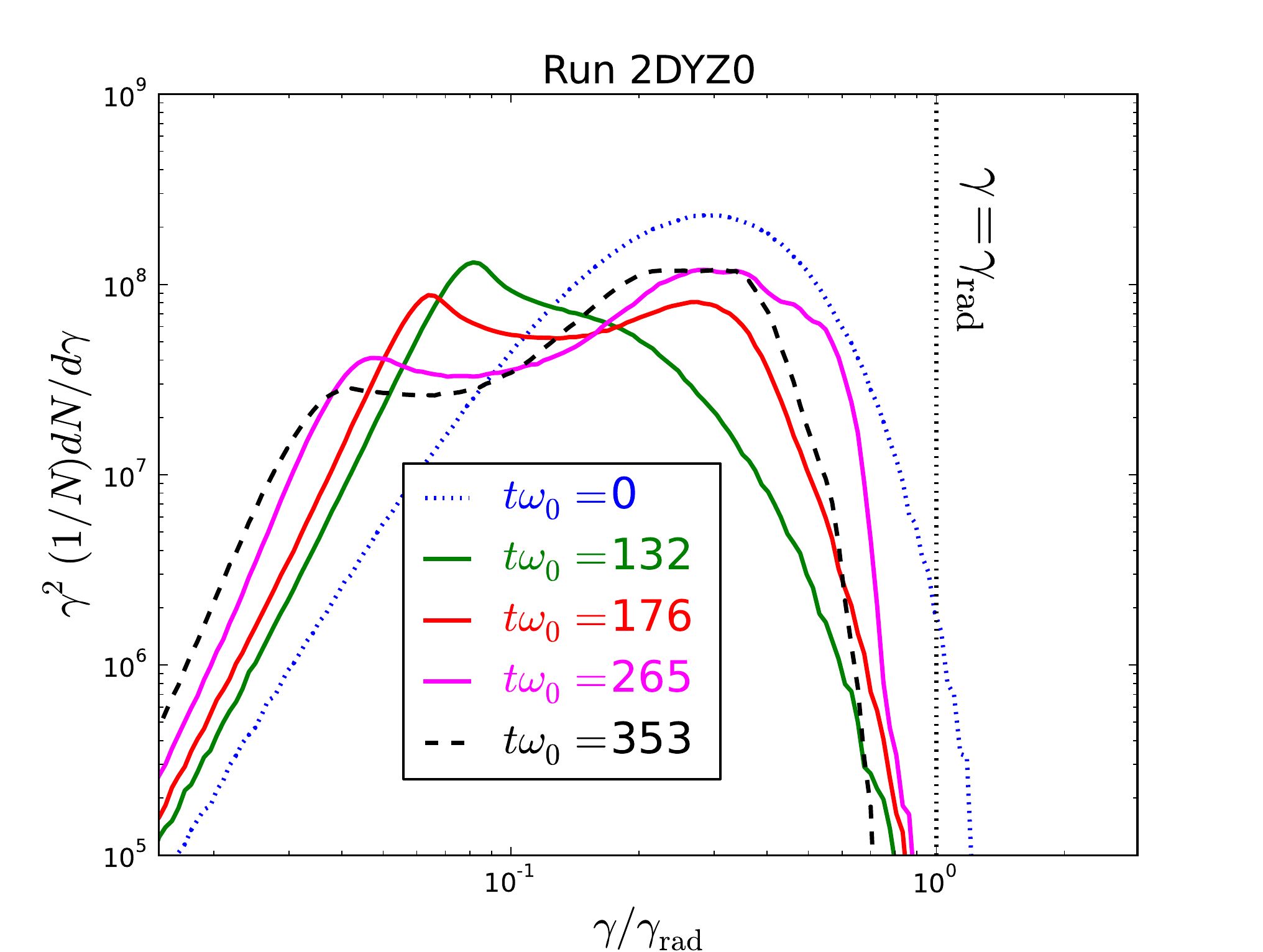}
\caption{Particle energy distribution normalized to the total number of particles ($\gamma^2 (1/{\rm N}) d{\rm N}/d\gamma$) of the 2D simulations in the $xy$-plane (top, run {\tt 2DXY0}) and $yz$-plane (bottom, run {\tt 2DYZ0}) for $\alpha=0$. The spectra are obtained at time $t\omega_0=0$ (dotted line)$,~132,~176,~265$ and $353$ (dashed line) and are averaged over all directions. The particle Lorentz factor is normalized to the nominal radiation reaction limit $\gamma_{\rm rad}\approx 1.3\times 10^9$.}
\label{fig_spec_time}
\end{figure}

The critical quantities of interest here are the particle energy distributions, $\gamma^2 d{\rm N}/d\gamma$, and the instantaneous optically thin synchrotron radiation spectral energy distribution (SED) emitted by the particles, $\nu F_{\nu}\equiv E^2 d{\rm N_{\rm ph}}/dt dE$, where $E$ is the photon energy. Figure~\ref{fig_spec_time} shows the time evolution of the total particle spectra at different times with no guide field in the $xy$-plane (top panel) and in the $yz$-plane (bottom panel). In the early stage ($t\omega_0\lesssim 132$), both simulations are subject to pure synchrotron cooling (i.e., with no acceleration or heating) of the plasma that results in a decrease of the typical Lorentz factor of the particles from $\gamma/\gamma_{\rm rad}\approx 0.3$ at $t\omega_0=0$ to $\gamma/\gamma_{\rm rad}\approx 0.08$ at $t\omega_0=132$. The decrease of the mean particle energy within the layer explains the shrinking of the layer thickness described in Section~\ref{evol}. 

\begin{figure}
\centering
\includegraphics[width=8.5cm]{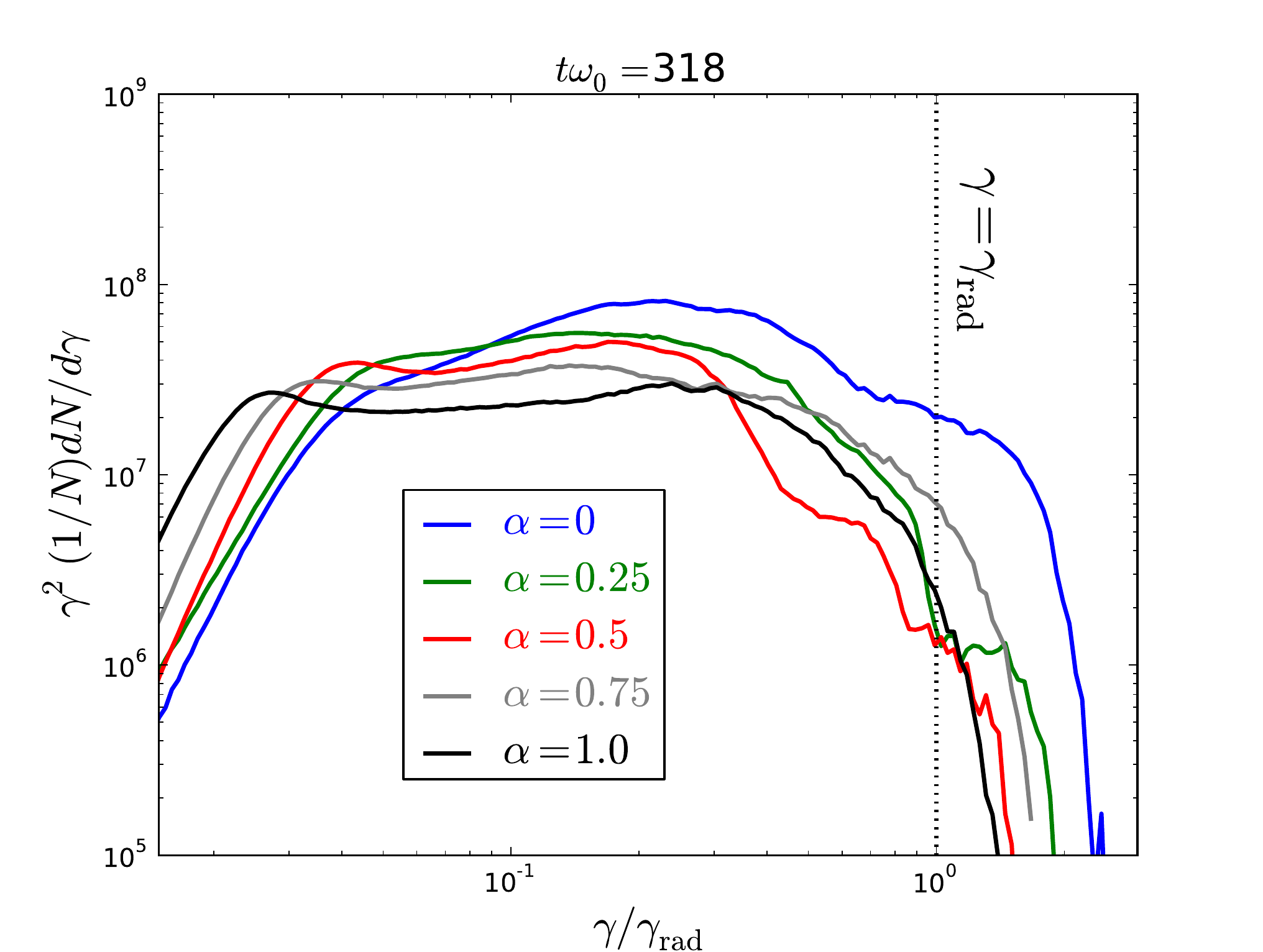}
\includegraphics[width=8.5cm]{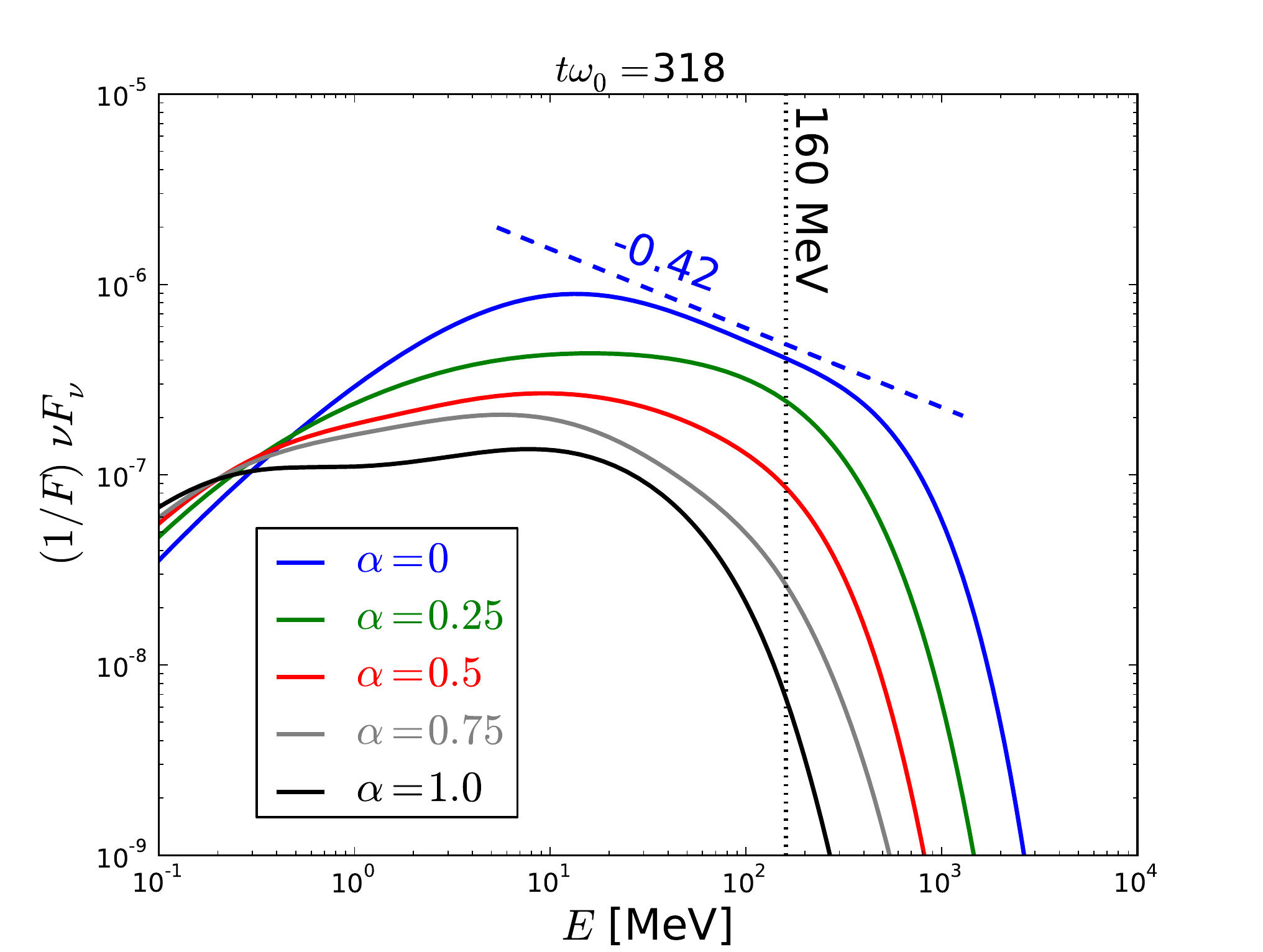}
\caption{Particle energy distribution normalized to the total number of particles ($\gamma^2 (1/{\rm N}) d{\rm N}/d\gamma$, top) and synchrotron radiation spectral energy distribution normalized by the total (frequency-integrated) photon flux ($(1/F)\nu F_{\nu}$, bottom) of the 2D simulations in the $xy$-plane at $t\omega_0=318$, averaged over all directions. The spectra are obtained for $\alpha=0,~0.25,~0.5,~0.75$ and $1$. The particle Lorentz factor in the top panel is normalized to the nominal radiation reaction limit $\gamma_{\rm rad}\approx 1.3\times 10^9$. In the bottom panel, the blue dashed line is a power-law fit of index $\approx -0.42$ of the $\alpha=0$ SED between $E=20$~MeV and $E=350$~MeV.}
\label{fig_spec_tearing}
\end{figure}

At $t\omega_0\gtrsim 132$, the instabilities trigger magnetic dissipation and particles are energized, but the particle spectra differ significantly in both cases. In run {\tt 2DXY0}, where the tearing instability drives reconnection, the particle spectrum extends to higher and higher energy with time until the end of the simulation, where the maximum energy reaches $\gamma_{\rm max}/\gamma_{\rm rad}\approx 2.5$, i.e., well above the nominal radiation reaction limit. The spectrum above $\gamma/\gamma_{\rm rad}=0.1$ cannot be simply modeled with a single power-law, but it is well contained between two steep power laws of index $-2$ and $-3$. We know from our previous study that the high-energy particles are accelerated via the reconnection electric field at X-points and follow relativistic Speiser orbits \citep{2013ApJ...770..147C}. The maximum energy is then given by the electric potential drop along the $z$-direction (neglecting radiative losses), i.e.,
\begin{equation}
\gamma_{\rm max}\sim \frac{e E_{\rm z} L_{\rm x}}{m_{\rm e}c^2}=\frac{e \beta_{\rm rec}B_0 L_{\rm x}}{m_{\rm e}c^2}\approx 3\gamma_{\rm rad},
\label{gmax_tearing}
\end{equation}
for a dimensionless reconnection rate $\beta_{\rm rec}\approx 0.2$. Particles above the radiation reaction limit ($\gamma>\gamma_{\rm rad}$) account for about $5\%$ of the total energy of the plasma at $t\omega_0=318$ (Figure~\ref{fig_spec_tearing}, top panel), and are responsible for the emission of synchrotron radiation above $160~$MeV. Figure~\ref{fig_spec_tearing} (bottom panel) shows the resulting isotropic synchrotron radiation SED at $t\omega_0=318$, where about $11\%$ of the radiative power is $>160~$MeV. The SED peaks at $E=10~$MeV and extends with a power-law of index $-0.42$ up to about $300$-$400$~MeV before cutting off exponentially.

\begin{figure}
\centering
\includegraphics[width=8.5cm]{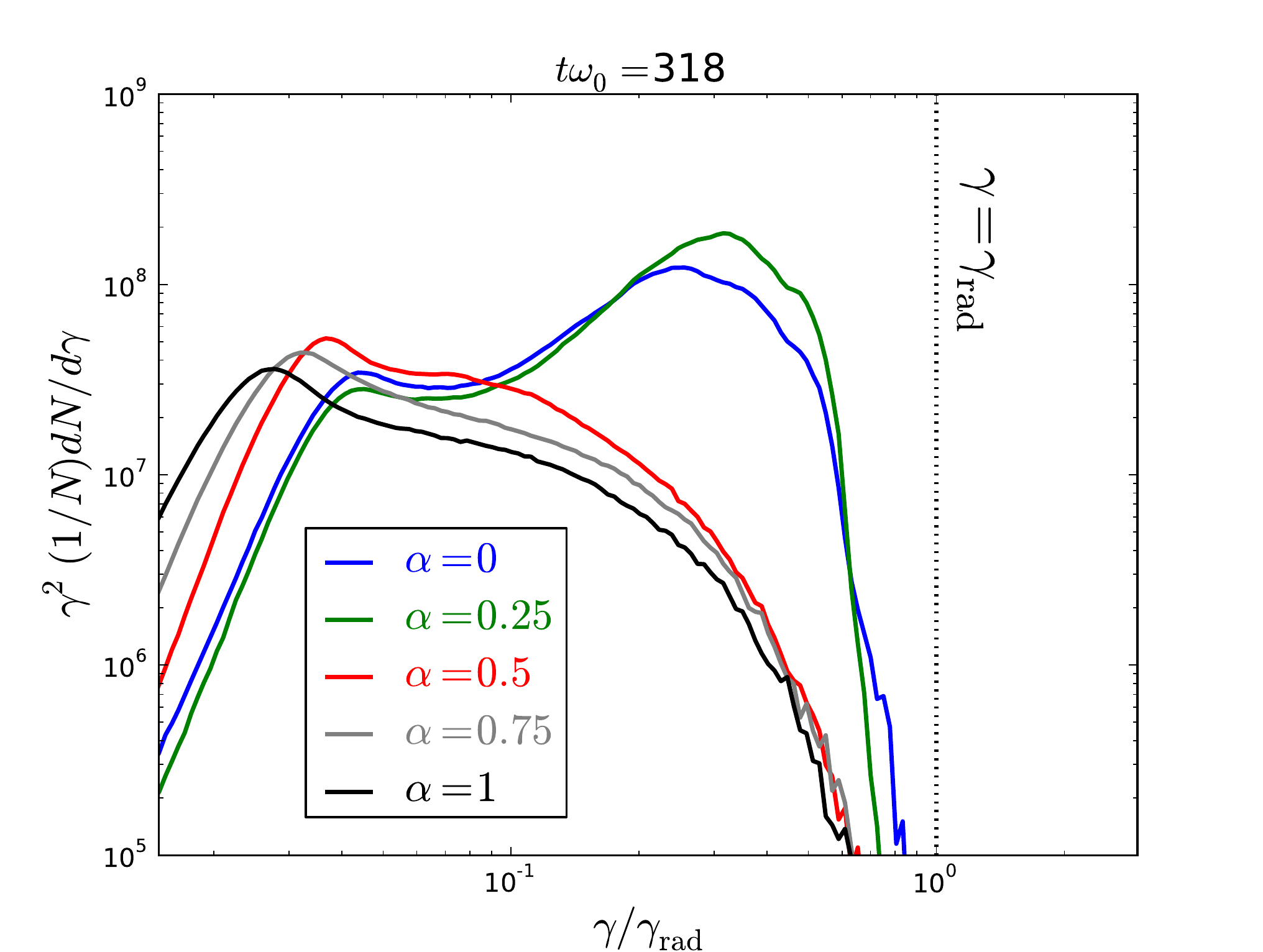}
\includegraphics[width=8.5cm]{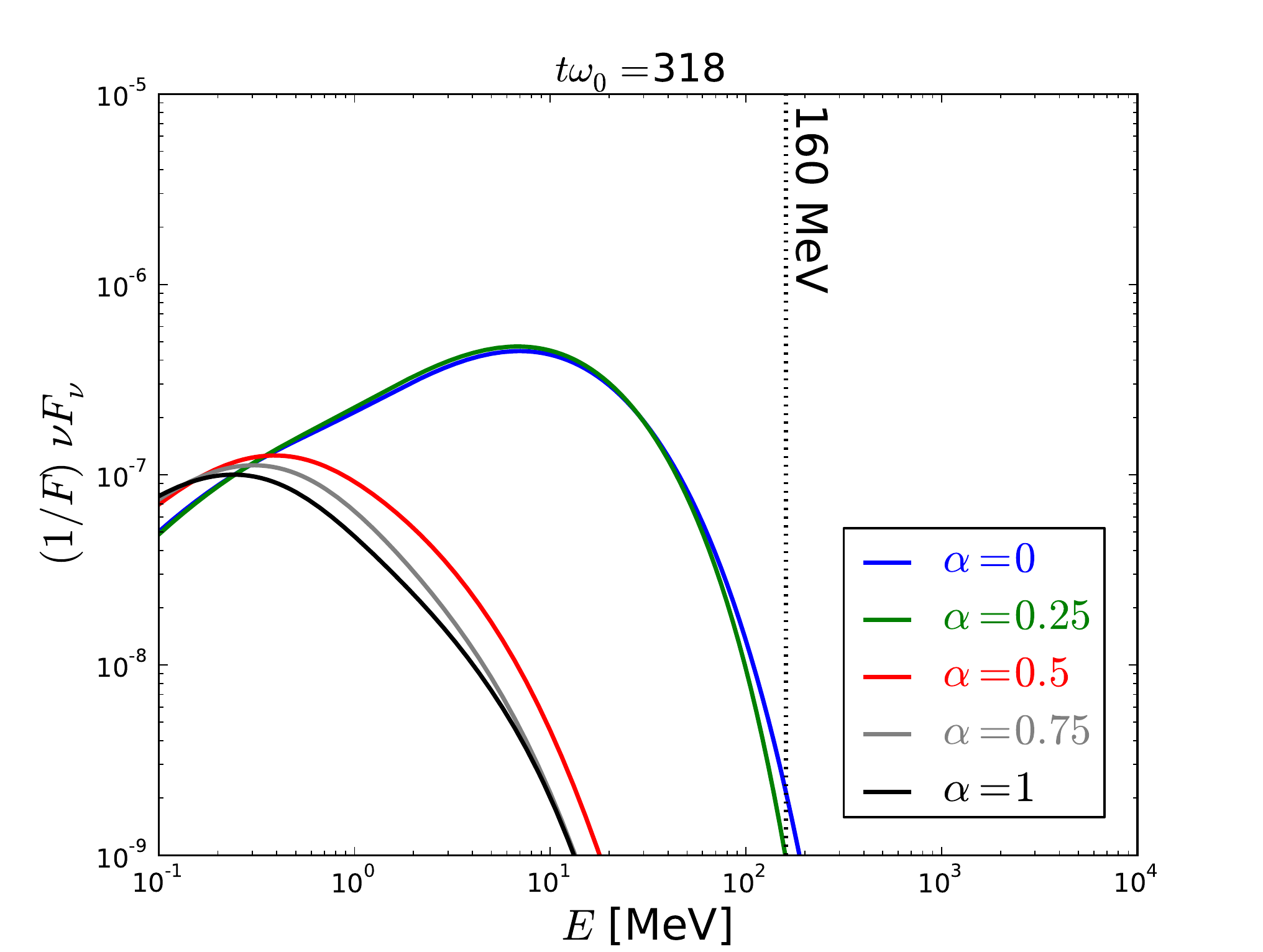}
\caption{Same as in Figure~\ref{fig_spec_tearing}, but for the 2D simulations in the $yz$-plane.}
\label{fig_spec_kink}
\end{figure}

In contrast, in run {\tt 2DYZ0}, where the kink instability drives the annihilation of the magnetic field, the particles are heated up to a typical energy $\gamma/\gamma_{\rm rad}\approx 0.3$. The particle spectrum is composed of a Maxwellian-like distribution on top of a cooled distribution of particles formed at $t\omega_0\lesssim 132$ (Figures~\ref{fig_spec_time},~\ref{fig_spec_kink}). The mean energy of the hot particles corresponds to a nearly uniform redistribution of the total dissipated magnetic energy to kinetic energy of background particles, i.e.,
\begin{equation}
\langle\gamma\rangle \sim 0.55\times \frac{B_0^2/8\pi}{(0.1n_0)m_{\rm e}c^2}=0.55\times\frac{\sigma\theta_0}{2}\approx 0.34\gamma_{\rm rad},
\label{gmax_kink}
\end{equation}
where the numerical factor $0.55$ accounts for the fraction of the total magnetic energy dissipated at the end of the simulation. Hence, the development of the kink prevents the acceleration of particles above $\gamma_{\rm rad}$ and the emission of synchrotron photons above $160~$MeV. Figure~\ref{fig_spec_kink} (bottom panel) shows that the total synchrotron radiation SED peaks and cuts off at $E=10~$MeV, far below the desired energies $>160~$MeV.

Because a moderate guide field suppresses the effect of the kink instability, hence magnetic dissipation, the particles are not heated for $\alpha\gtrsim 0.5$, and the initial spectrum continues cooling until the end of the $yz$-plane simulation where the particles radiate low-energy ($\sim 1$~MeV) synchrotron radiation (Figure~\ref{fig_spec_kink}). In the $xy$-plane reconnection simulations, the guide field tends to decrease the maximum energy of the particles and of the emitted radiation (Figure~\ref{fig_spec_tearing}, see also \citealt{2013ApJ...770..147C}). The guide field deflects the particles outside the layer, reducing the time spent by the particle within the accelerating region.

\section{RESULTS OF THE 3D RUNS}\label{results3d}

From the previous section, we find that the tearing and kink modes grow at a similar rate and wavelength in our setup. Both instabilities lead to fast dissipation of the magnetic energy in the form of thermal and non-thermal particles. The kink instability tends to disrupt the layer, which prevents non-thermal particle acceleration and emission above the standard radiation reaction limit. It is desirable to impose a moderate guide field to diminish the negative effect of the kink on particle acceleration, but too strong a guide field is not advantageous either, as it decreases the maximum energy reached by the particles and radiation. Hence, we decided to run a 3D simulation with an $\alpha=0.5$ guide field (run {\tt 3D050}, see Table~\ref{tab_simu}), which appears to be a good compromise. For comparison, we also performed a 3D simulation without guide field (run {\tt 3D0}). In this section, we first describe the time evolution of 3D reconnection in the two runs (Section~\ref{evol3d}). Then, we provide a quantitative analysis of the most unstable modes in the $(k_{\rm x}\times k_{\rm z})$-plane in the linear regime (Section~\ref{fou3d}). In addition, we address below the question of particle acceleration, emission (Section~\ref{spec3d}), particle and radiation anisotropies (Section~\ref{anis3d}), the expected radiative signatures (i.e., spectra and lightcurves) and comparison with the {\em Fermi}-LAT observations of the Crab flares (Sections~\ref{comp_obs},~\ref{light}).

\subsection{Plasma time evolution}\label{evol3d}

\begin{figure*}
\centering
\includegraphics[width=8.5cm]{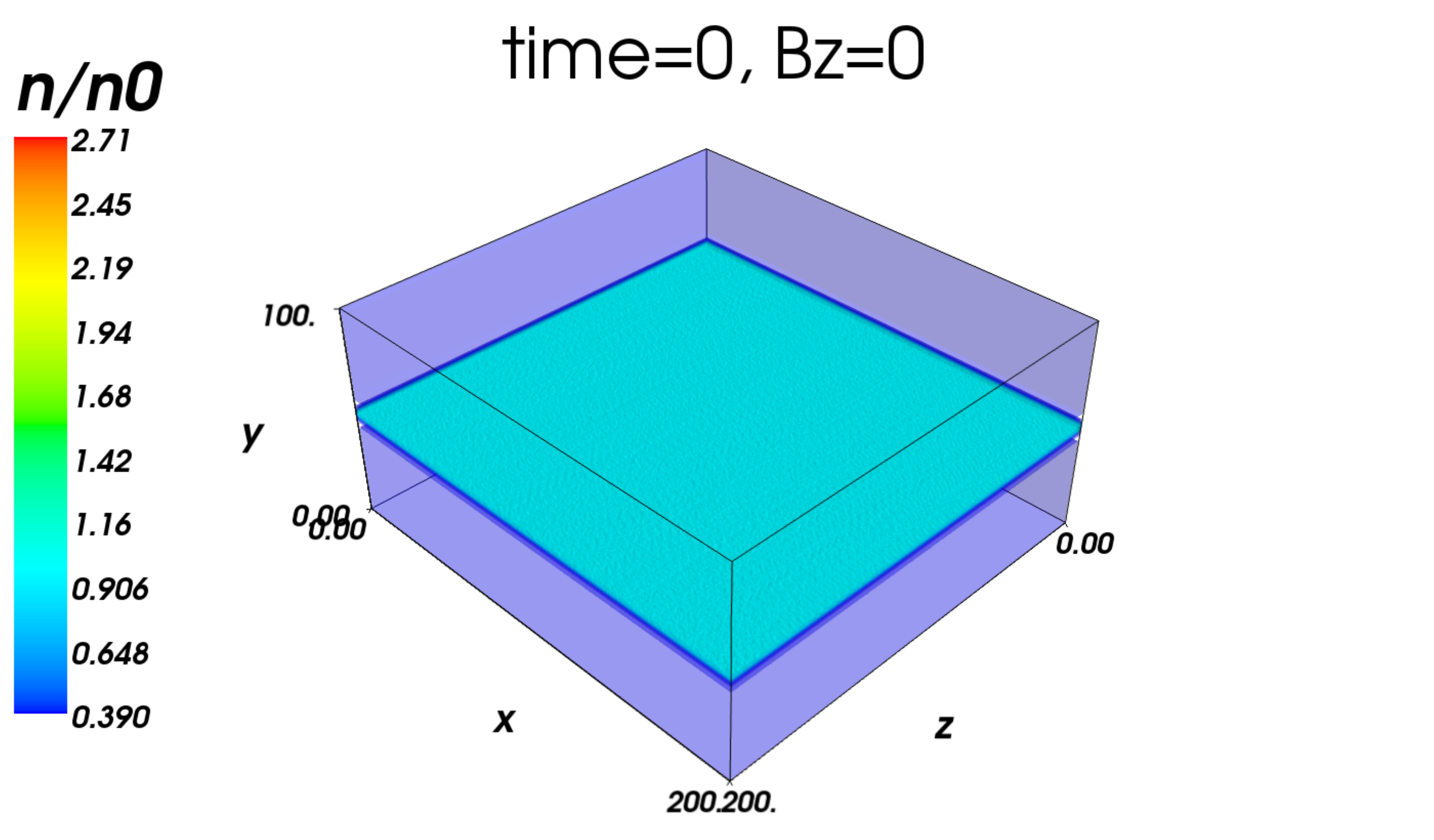}
\includegraphics[width=8.5cm]{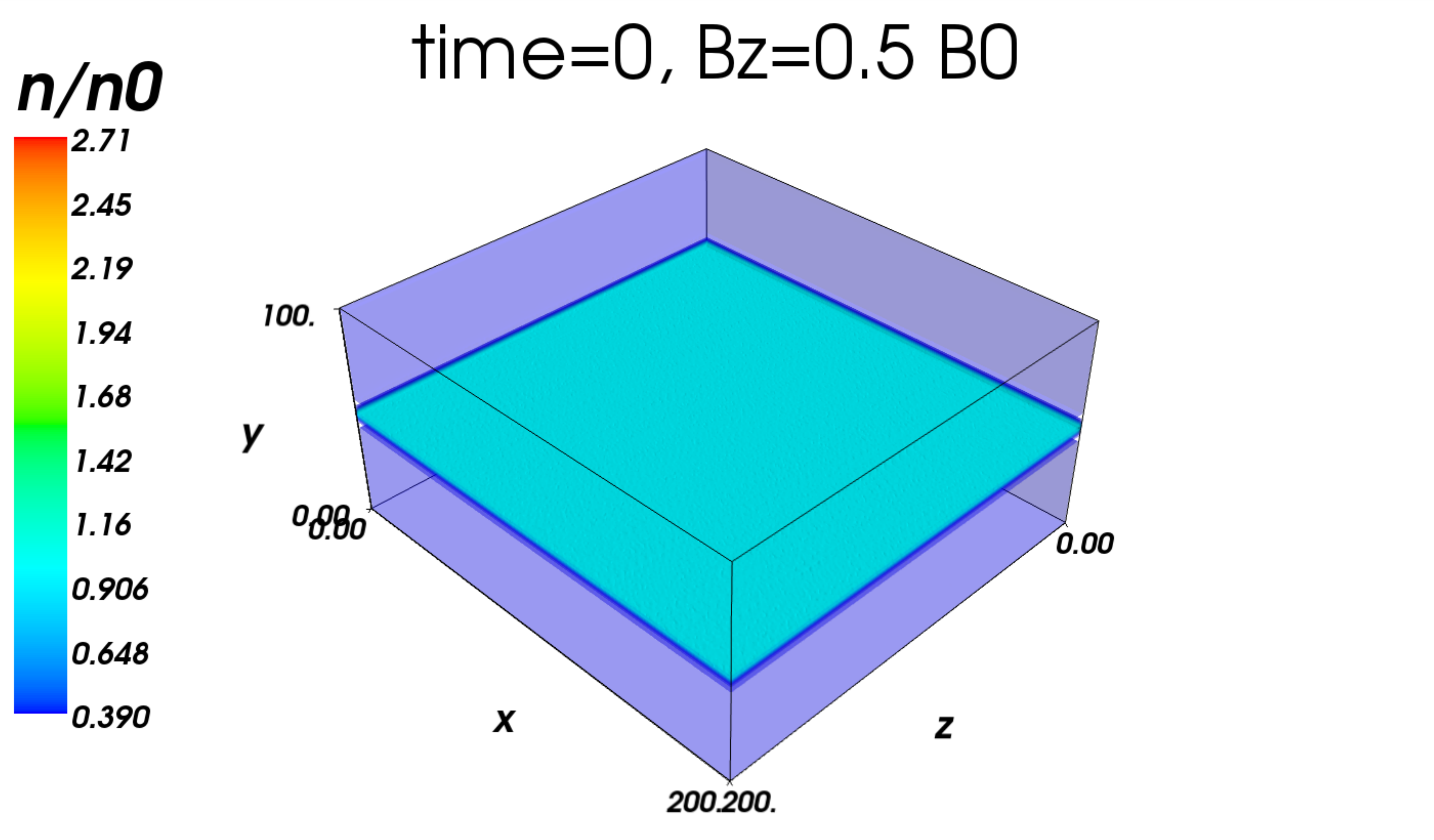}
\includegraphics[width=8.5cm]{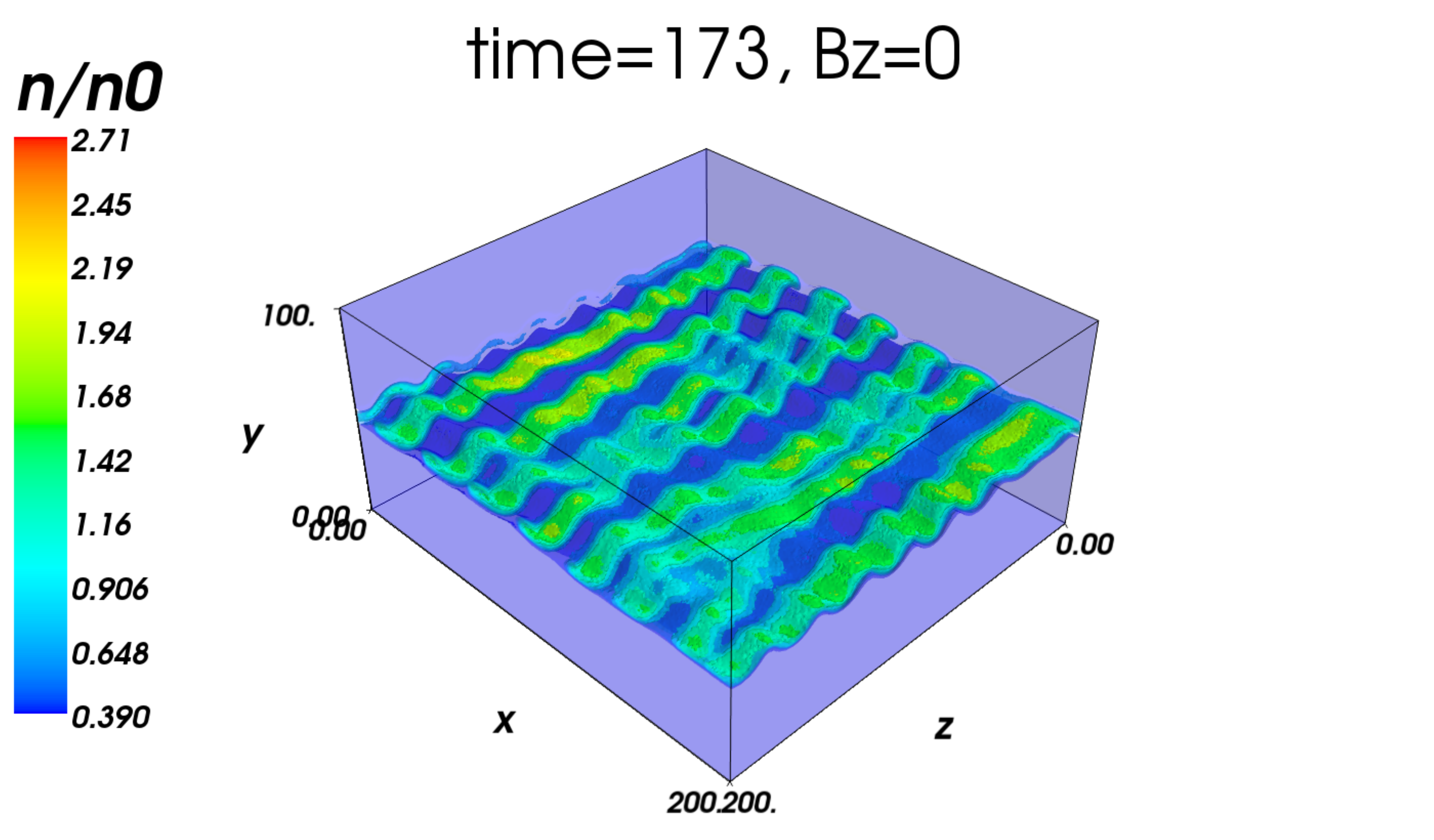}
\includegraphics[width=8.5cm]{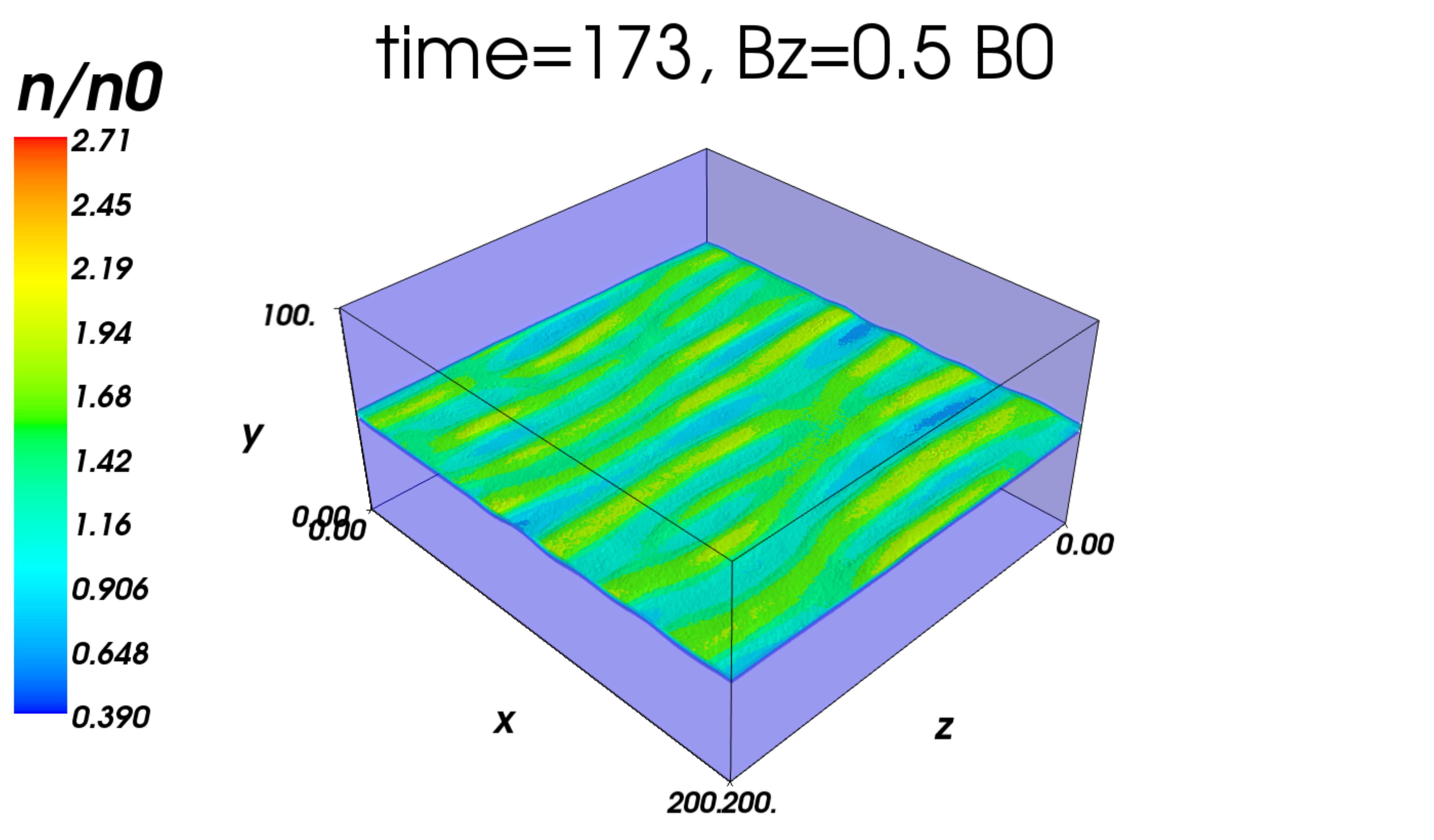}
\includegraphics[width=8.5cm]{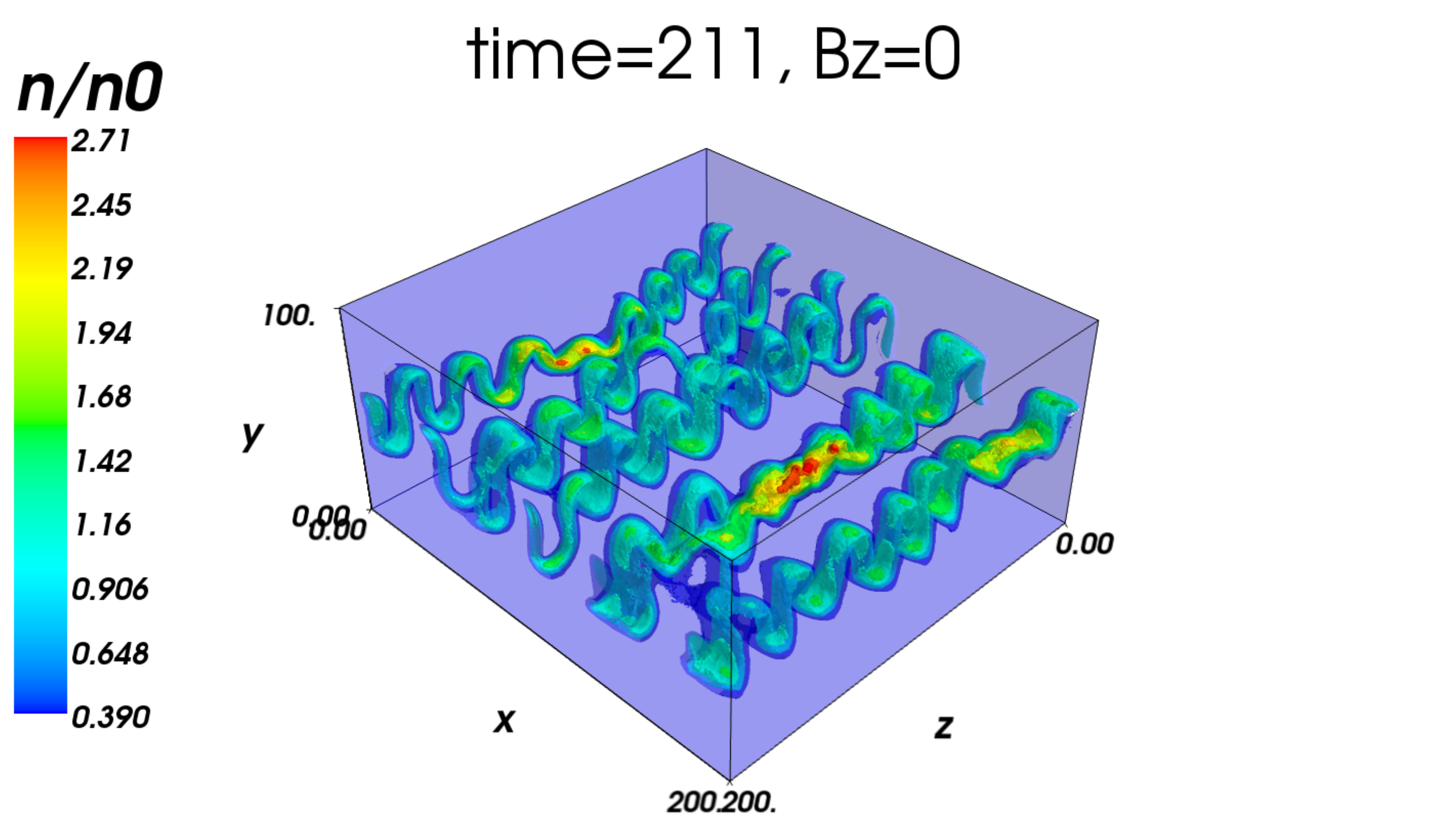}
\includegraphics[width=8.5cm]{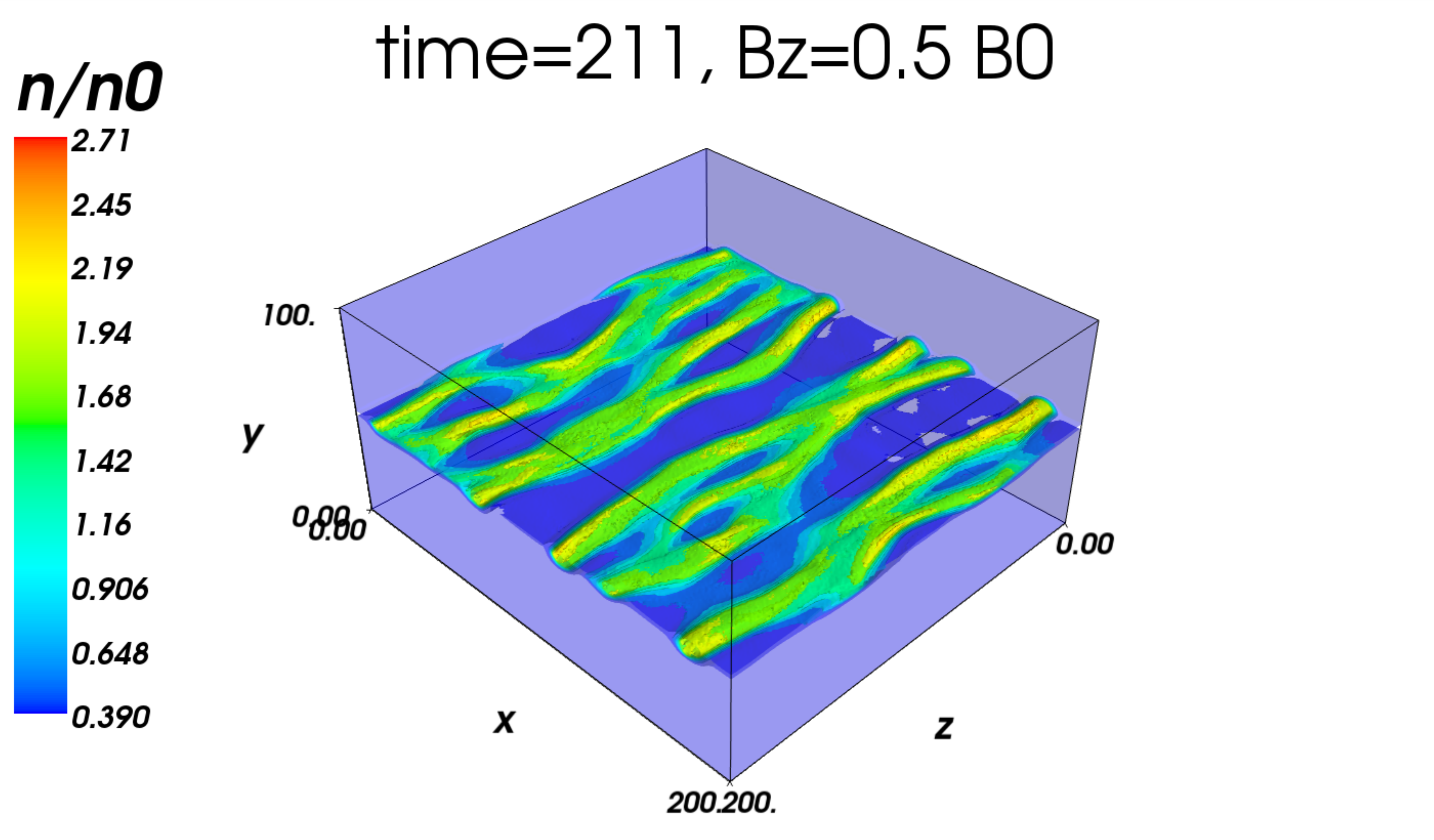}
\includegraphics[width=8.5cm]{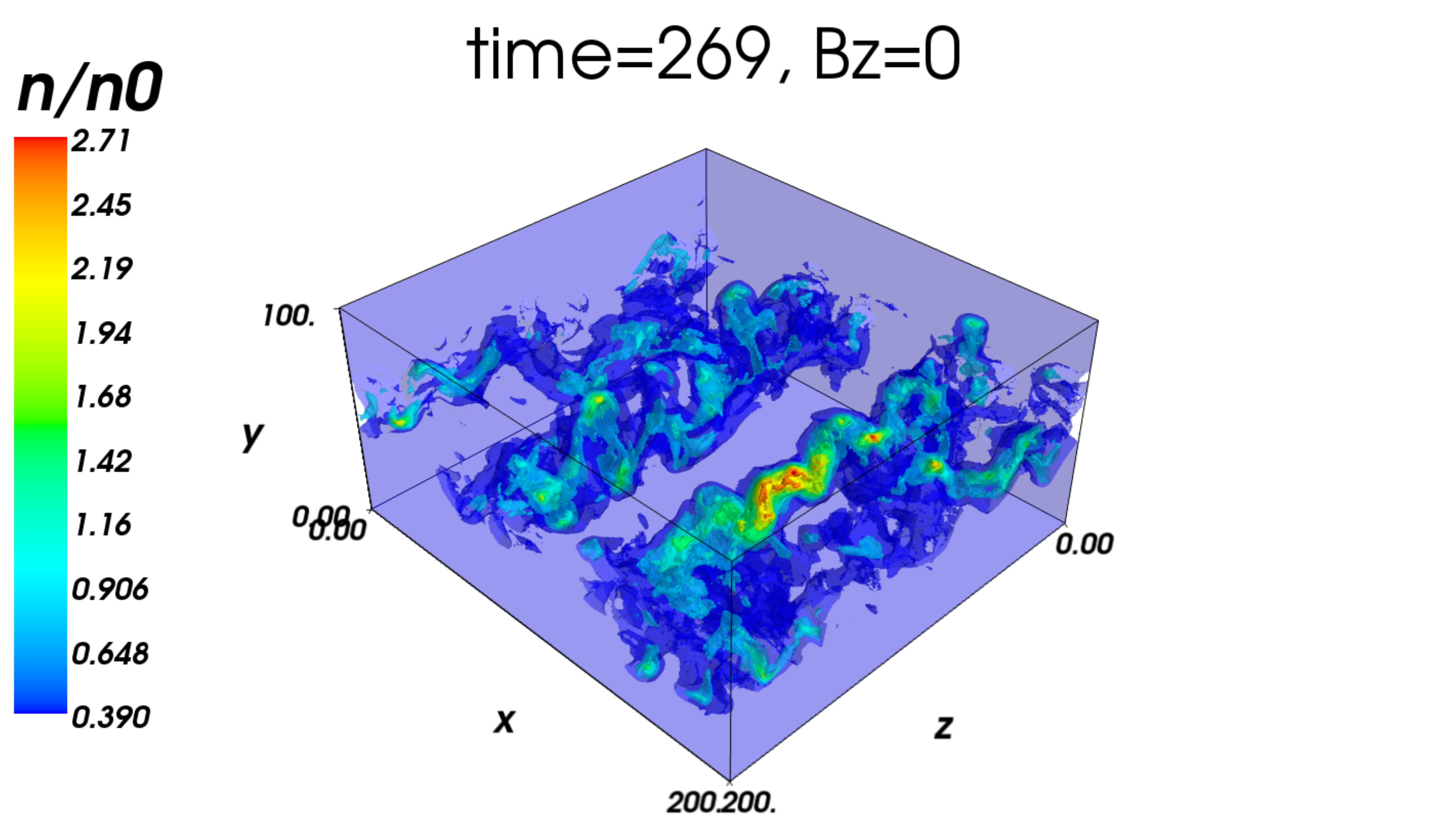}
\includegraphics[width=8.5cm]{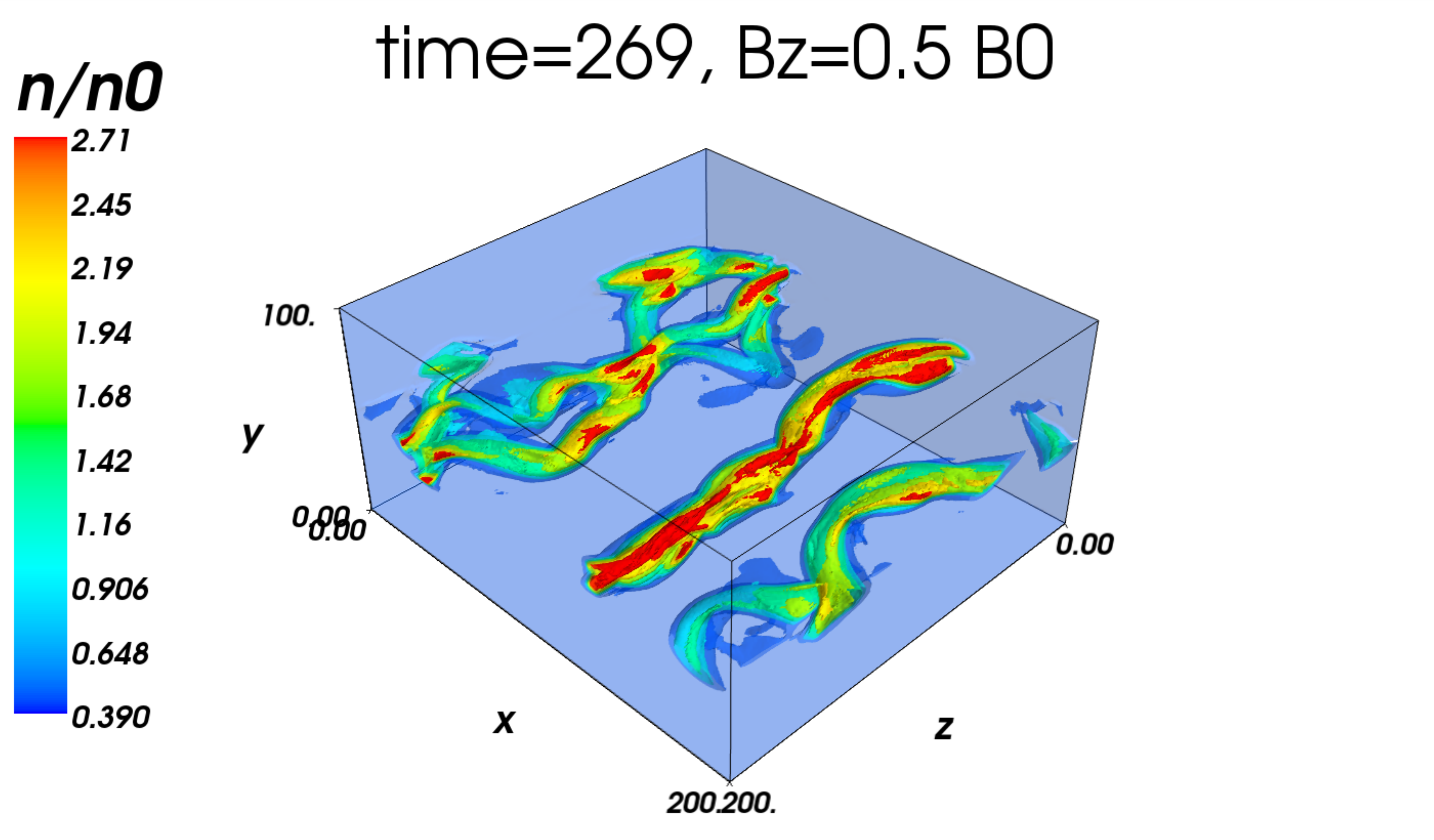}
\caption{Time evolution of the plasma density (color-coded isosurfaces) in the bottom half of the simulation box at $t\omega_0=0,~173,~211,~$and $269$ (from top to bottom), for $\alpha=0$ (left panels, run {\tt 3D0}) and $\alpha=0.5$ (right panels, run {\tt 3D050}). Low-density isosurfaces (blue) are transparent in order to see the high-density regions (red) nested in the flux tubes. The time is given in units of $\omega_0^{-1}$, and spatial coordinates are in units of $\rho_0$.}
\label{fig_3d}
\end{figure*}

\begin{figure*}
\includegraphics[width=8.5cm]{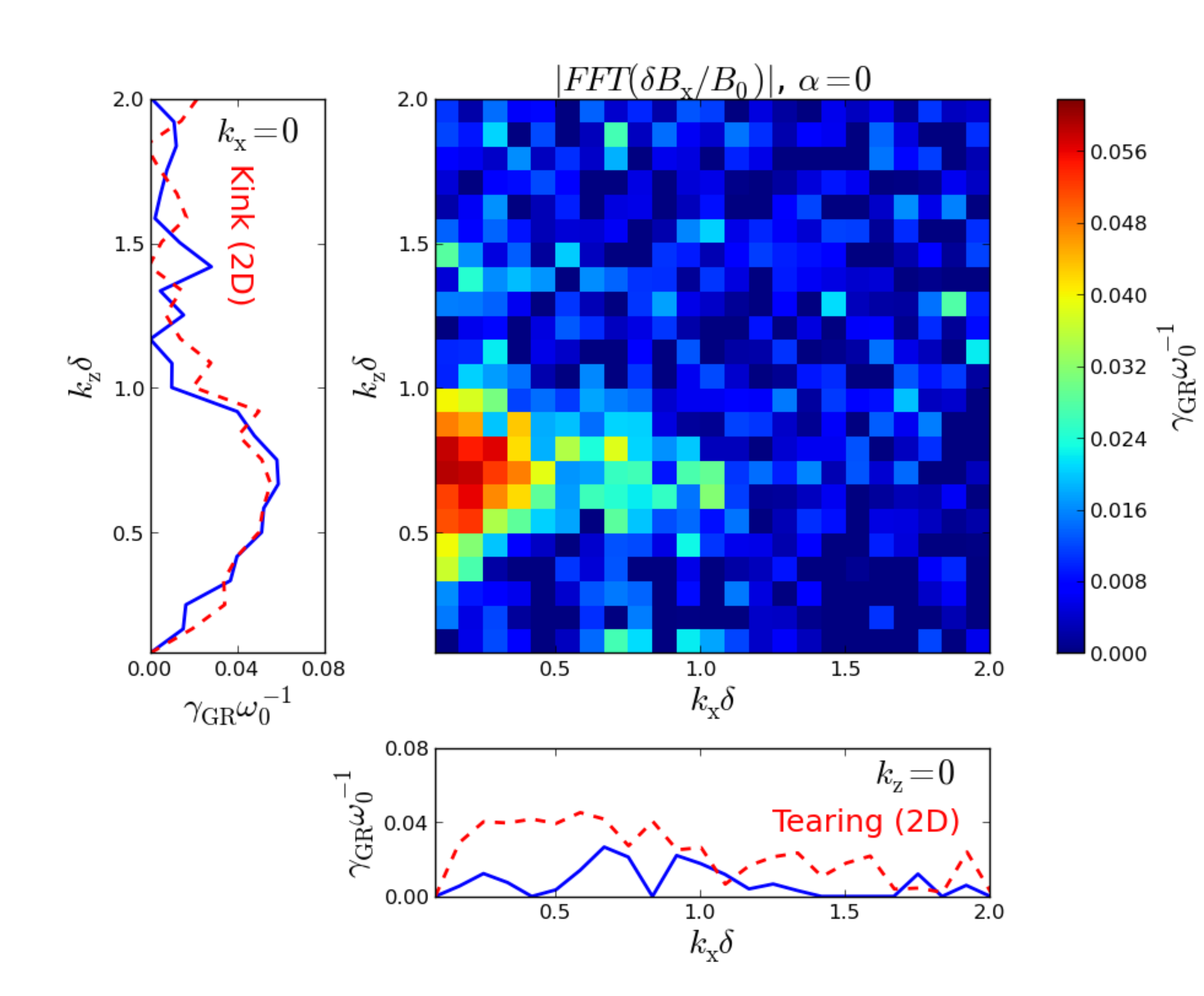}
\includegraphics[width=8.5cm]{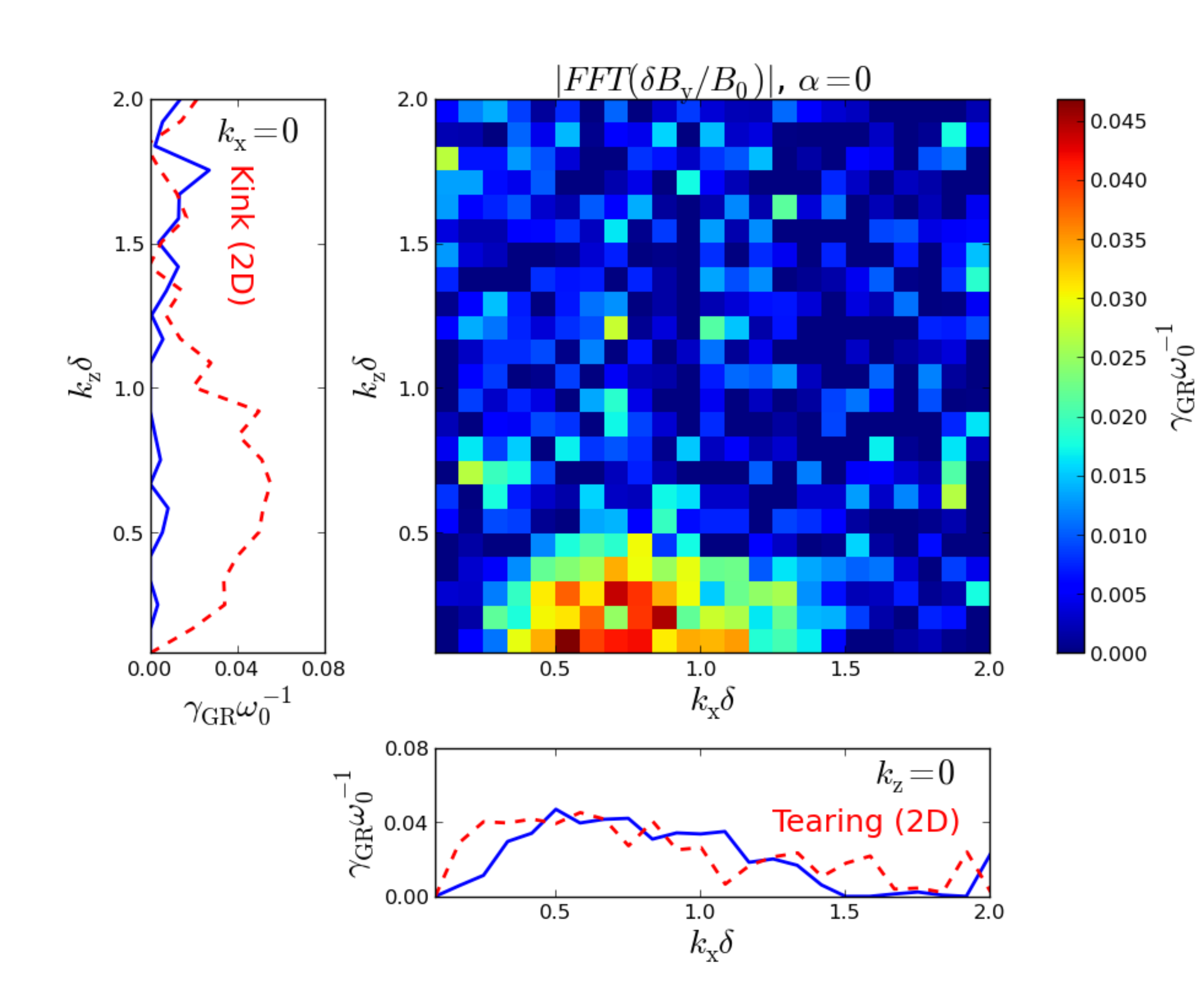}
\caption{Linear growth rates in run {\tt 3D0} $\gamma_{\rm GR}$ times $\omega_0^{-1}$ in the $(k_{\rm x}\times k_{\rm z})$-plane (color-coded plots) using the fluctuations in $B_{\rm x}$ (left panel) which are most sensitive to kink-like modes, and in $B_{\rm y}$ (right panel) which are most sensitive to tearing-like modes. The blue solid lines in each subplots give the growth rates along the $k_{\rm x}$-axis for $k_{\rm z}=0$ (bottom subplots) and along the $k_{\rm z}$-axis for $k_{\rm x}=0$ (left subplots). The red dashed lines show the dispersion relation for the pure kink and tearing modes obtained in Section~\ref{fou} for comparison.}
\label{fig_disp2d}
\end{figure*}

\begin{figure*}
\includegraphics[width=8.5cm]{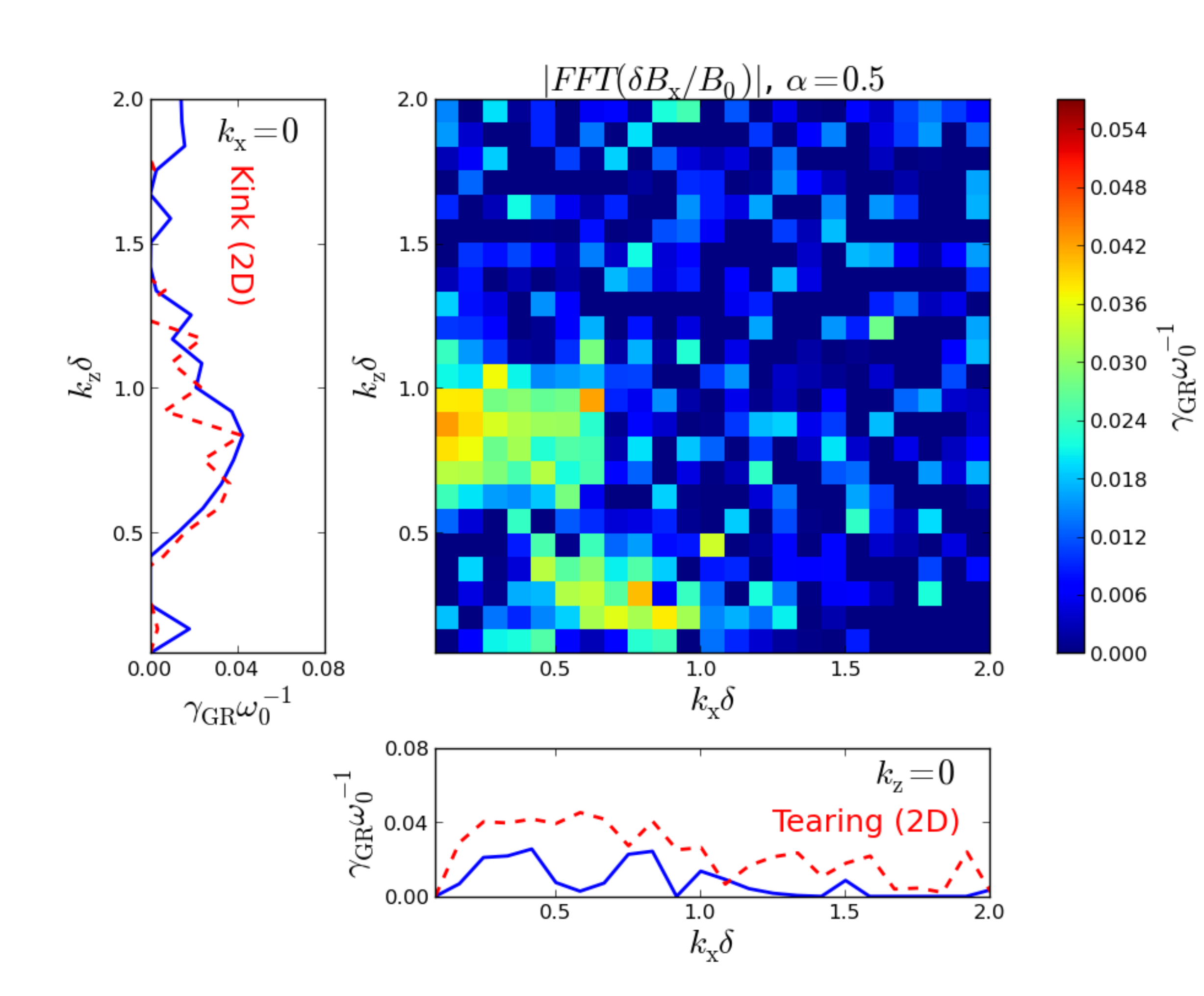}
\includegraphics[width=8.5cm]{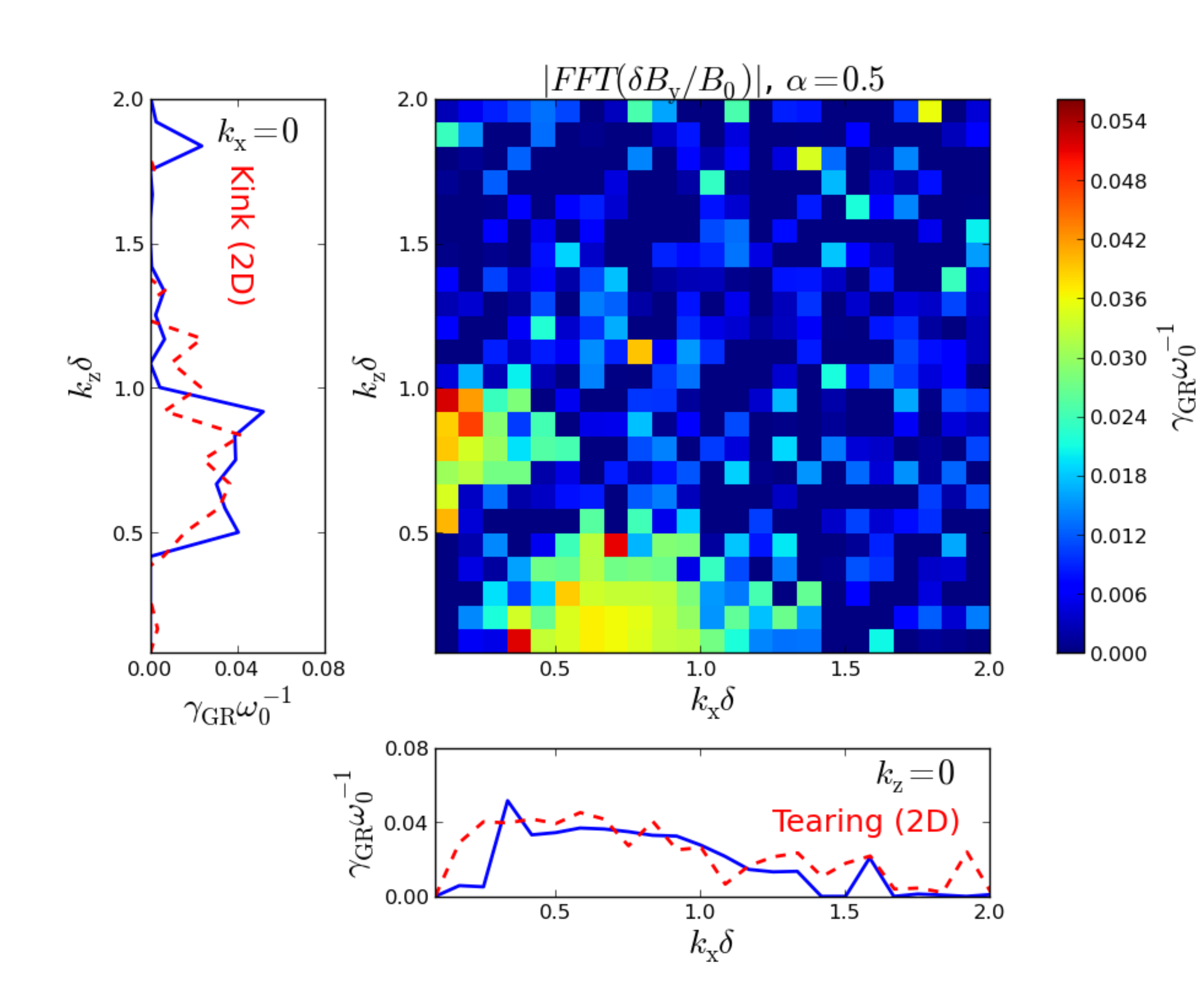}
\caption{Same as in Fig.~\ref{fig_disp2d} with an $\alpha=0.5$ guide field (run {\tt 3D050}).}
\label{fig_disp2d_guide}
\end{figure*}

Figure~\ref{fig_3d} (left panels) shows the time evolution of the plasma density\footnote{Movies are available at this URL: \url{http://benoit.cerutti.free.fr/movies/Reconnection_Crab3D/}.} in the zero-guide field simulation at $t\omega_0=0,~173,~211$ and $269$. The initial stage where the layer remains apparently static lasts for about $t\omega_0=144$, i.e., half of the whole simulation time. At $t\omega_0\gtrsim 144$, overdensities appear in the layers in the form of 7-8 tubes (flux ropes) elongated along the $z$-direction. These structures are generated by the tearing instability and are the 3D generalization of the magnetic islands observed in 2D reconnection. As the simulation proceeds into the non-linear regime, the flux ropes merge with each other creating bigger ones, as magnetic islands do in 2D reconnection. However, in 3D this process does not happen at the same time everywhere along the $z$-direction, which results in the formation of a network of interconnected flux ropes at intermediate times ($173\lesssim t\omega_0\lesssim 211$).

In parallel to this process, the kink instability deforms the two layers along the $z$-direction in the form of sine-like translation of the layers' mid-planes in the $\pm y$-directions. During the most active period of reconnection ($t\omega_0\gtrsim 173$), the kink instability takes over and eventually destroys the flux ropes formed by the tearing modes (see left bottom panel in Figure~\ref{fig_3d}). Only a few coherent structures survive at the end of the simulation ($t\omega_0=269$). In particular the reconnection electric field, which is strongest along the X-lines between two flux ropes, loses its initial coherence. This results in efficient particle heating but poor particle acceleration (see below, Section~\ref{spec3d}). At the end of this run about $52\%$ of the total magnetic energy is dissipated, although the simulation does not reach the fully saturated state. 

The right panels in Figure~\ref{fig_3d} shows the time evolution of the plasma density for $\alpha=0.5$ guide field. One sees immediately that the guide field effectively suppresses the kink deformations of the layers in the $\pm y-$directions, as expected from the 2D simulations in the $yz$-plane (See Section~\ref{results2d}) and from \citet{2008ApJ...677..530Z}. In contrast, the tearing instability seems undisturbed and breaks the layer into a network of 8 flux tubes. Towards the end of the simulation, there are about 3 well-defined flux ropes containing almost all the plasma that went through reconnection. At this point in time, $20\%$ of the total magnetic energy (i.e., including the reconnecting and the guide field energy) has dissipated, in agreement with the 2D run {\tt 2DXY050}.

\begin{figure}
\centering
\includegraphics[width=8.5cm]{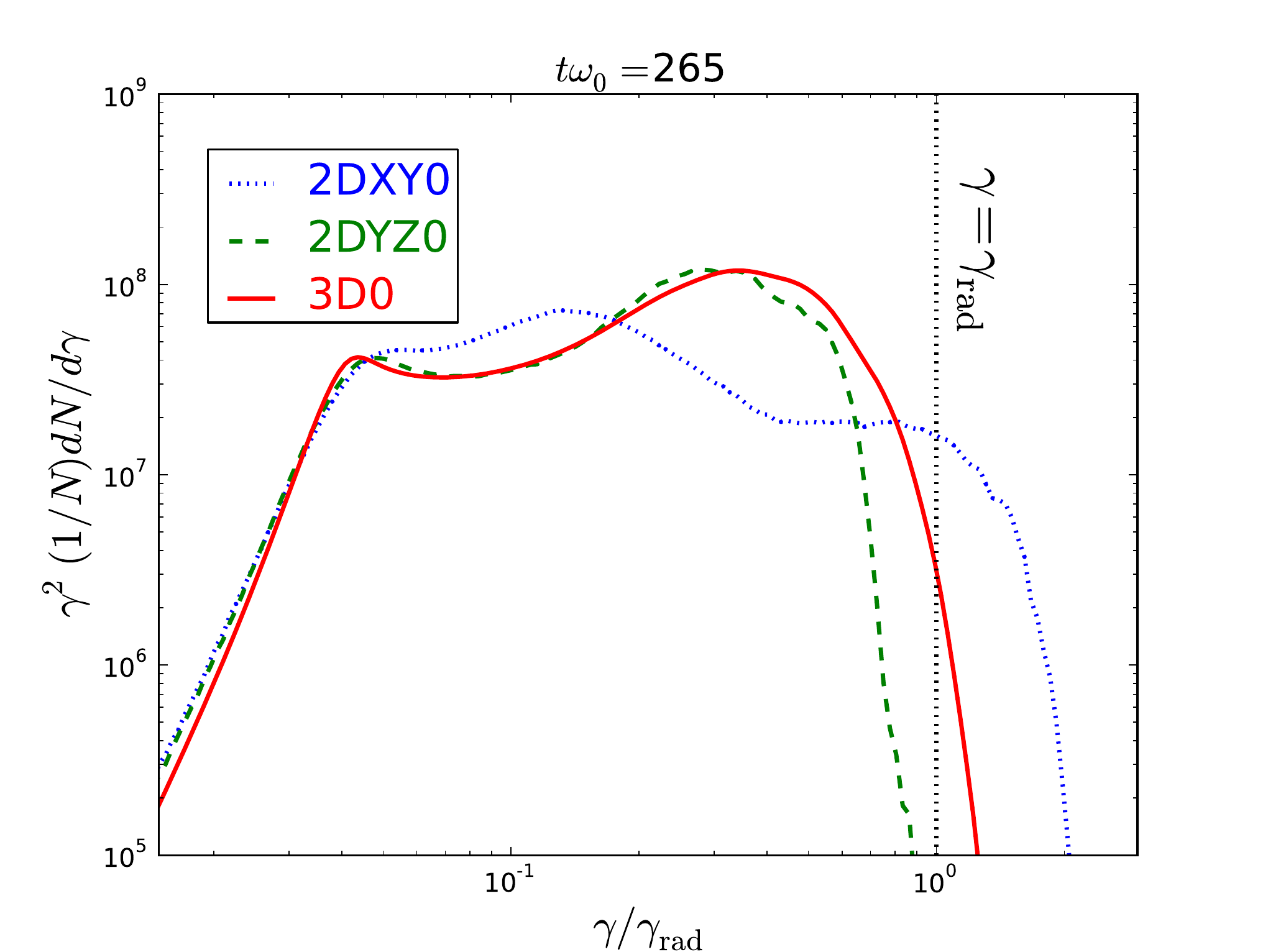}
\includegraphics[width=8.5cm]{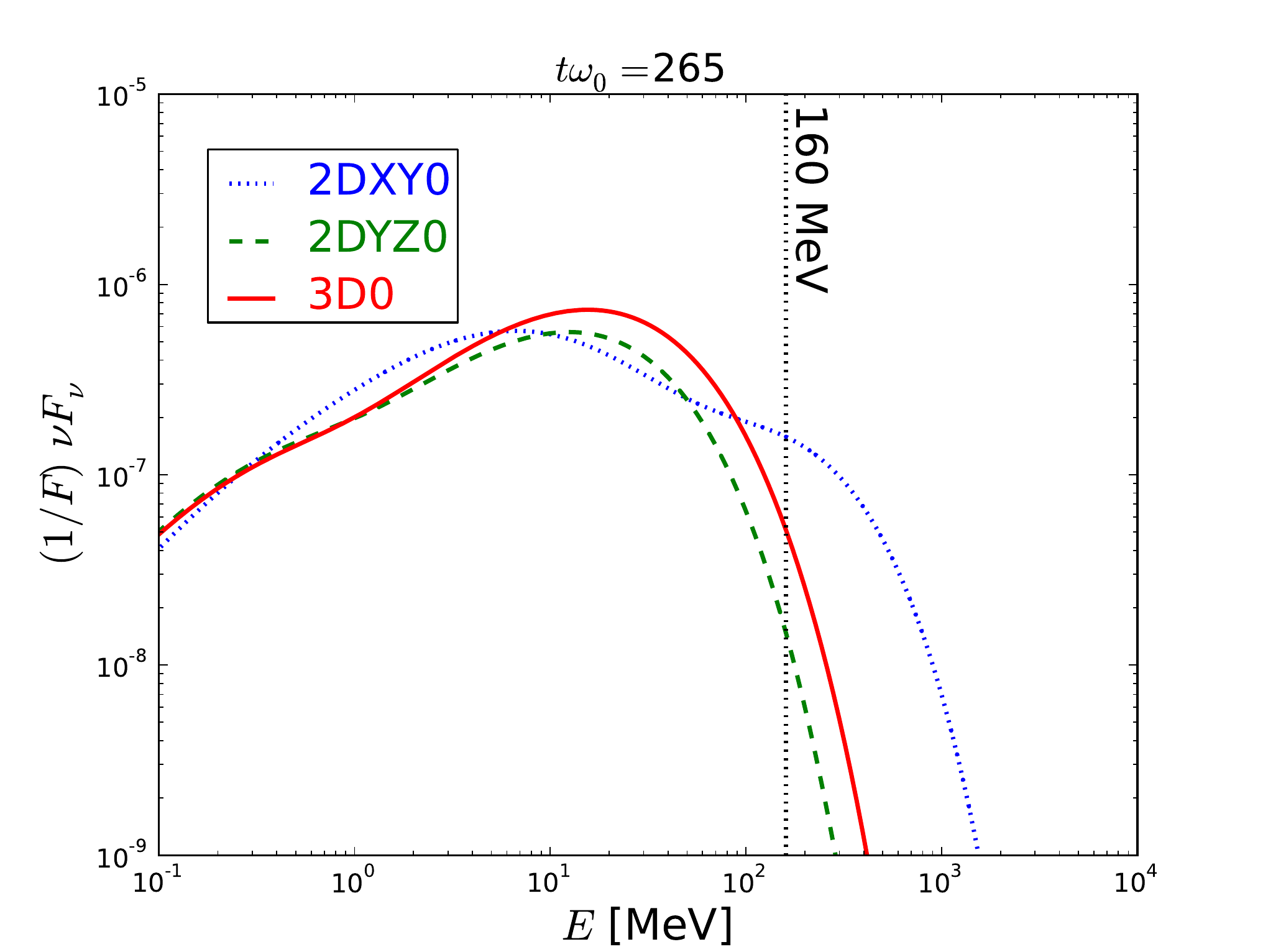}
\caption{Isotropically averaged particle energy distribution (top panel) and SED (bottom panel) obtained in 2D ($xy$-plane, blue dotted line, and in the $yz$-plane, green dashed line) and 3D (red solid line) with no guide field.}
\label{fig_spec_3d}
\end{figure}

\begin{figure}
\centering
\includegraphics[width=8.5cm]{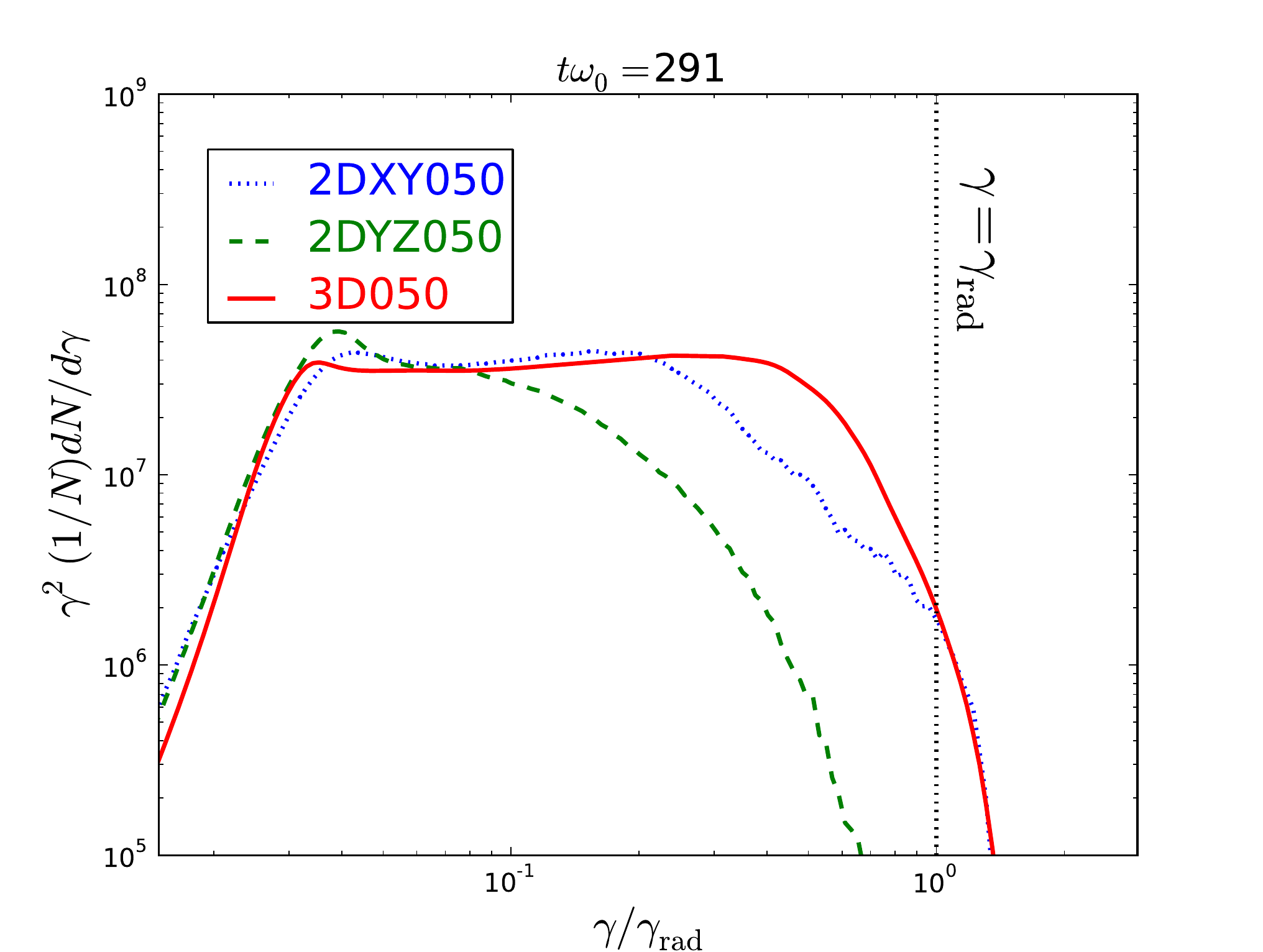}
\includegraphics[width=8.5cm]{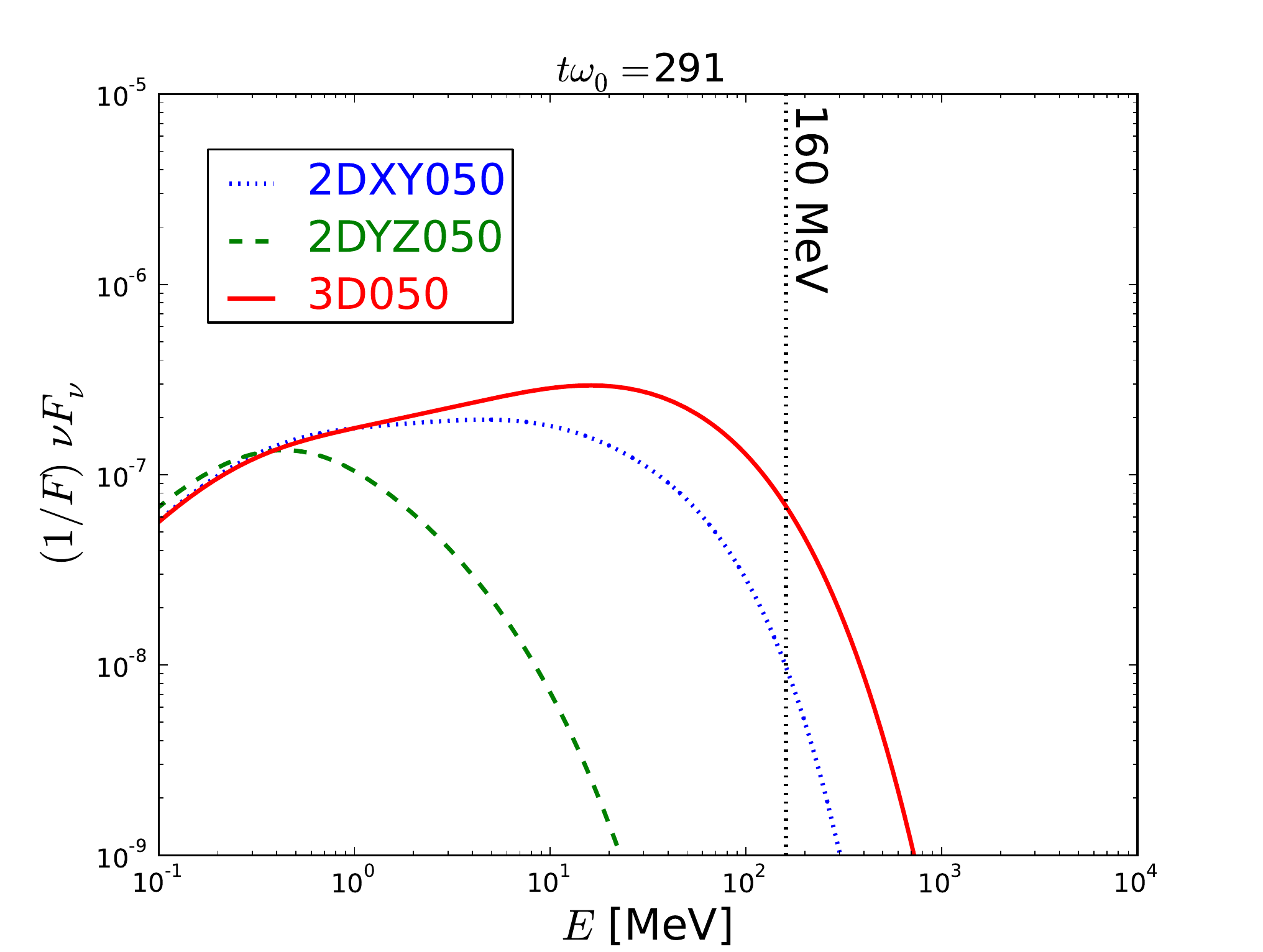}
\caption{Same as in Figure~\ref{fig_spec_3d}, but with an $\alpha=0.5$ guide field.}
\label{fig_spec_3d_guide}
\end{figure}

\subsection{Fourier analysis of unstable modes}\label{fou3d}

Following the analysis presented in Section~\ref{fou}, we perform a Fourier decomposition of the magnetic fluctuations in the bottom layer mid-plane, $(x,y=L_{\rm y}/4,z)$, to study the most unstable modes that develop in the 3D simulations. Figure~\ref{fig_disp2d} presents the growth rate of each modes in the $(k_{\rm x}\times k_{\rm z})$-plane estimated from the variations of $B_{\rm x}$ (left panel) and $B_{\rm y}$ (right panel), for $\alpha=0$. As pointed out in Section~\ref{fou} and by \citet{2008ApJ...677..530Z} and \citet{2013ApJ...774...41K}, we find that the reconnecting field $B_{\rm x}$ effectively captures the kink-like modes along $k_{\rm z}$ whereas the reconnected field $B_{\rm y}$ is most sensitive to tearing-like modes along $k_{\rm x}$. The dispersion relations show that pure kink (along $k_{\rm z}$ for $k_{\rm x}=0$) and pure tearing (along $k_{\rm x}$ for $k_{\rm z}=0$) modes grow at rates in very good agreement with the corresponding 2D simulations. With a growth rate $\gamma_{\rm GR}\approx 0.06\omega_0$, the fastest growing mode in the simulation is a pure kink mode of wavenumber $k_{\rm z}\delta\approx 0.7$, or $L_{\rm z}/\lambda_{\rm z}\approx 8$ consistent with the deformation of the layer observed in the earlier stage of reconnection (Figure~\ref{fig_3d}, left panels) and with the 2D run {\tt 2DYZ0}. The fastest tearing mode has a growth rate $\gamma_{\rm GR}\approx 0.045\omega_0$ at $k_{\rm x}\delta\approx 0.5$ and generates the $\approx 7$ initial flux ropes obtained in the simulation. The $(k_{\rm x}\times k_{\rm z})$-plane is also filled with oblique modes, i.e., waves with a non-zero $k_{\rm x}$- and $k_{\rm z}$-component, with growth rates comparable to the fastest tearing and kink modes. The existence of these modes is reflected by the flux ropes being slightly tilted in the $xz$-plane. Adding an $\alpha=0.5$ guide field decreases the amplitude of the low-frequency ($k_{\rm z}\delta\lesssim 1$) growth rates of the kink modes (Figure~\ref{fig_disp2d_guide}). In particular, the growth rate of the fastest mode for $\alpha=0$, $k_{\rm z}\delta=0.7$, decreases from $0.06\omega_0$ to $0.03\omega_0$. As a result, the fastest growing kink mode is now at $k_{\rm z}\delta=0.8$ with a rate $\approx 0.04\omega_0$, while the fastest growing tearing modes is approximatively unchanged, in excellent agreement with the 2D runs (Figure~\ref{fig_disp2d_guide}).

\begin{figure*}[htp]
\centering
\includegraphics[width=8.5cm]{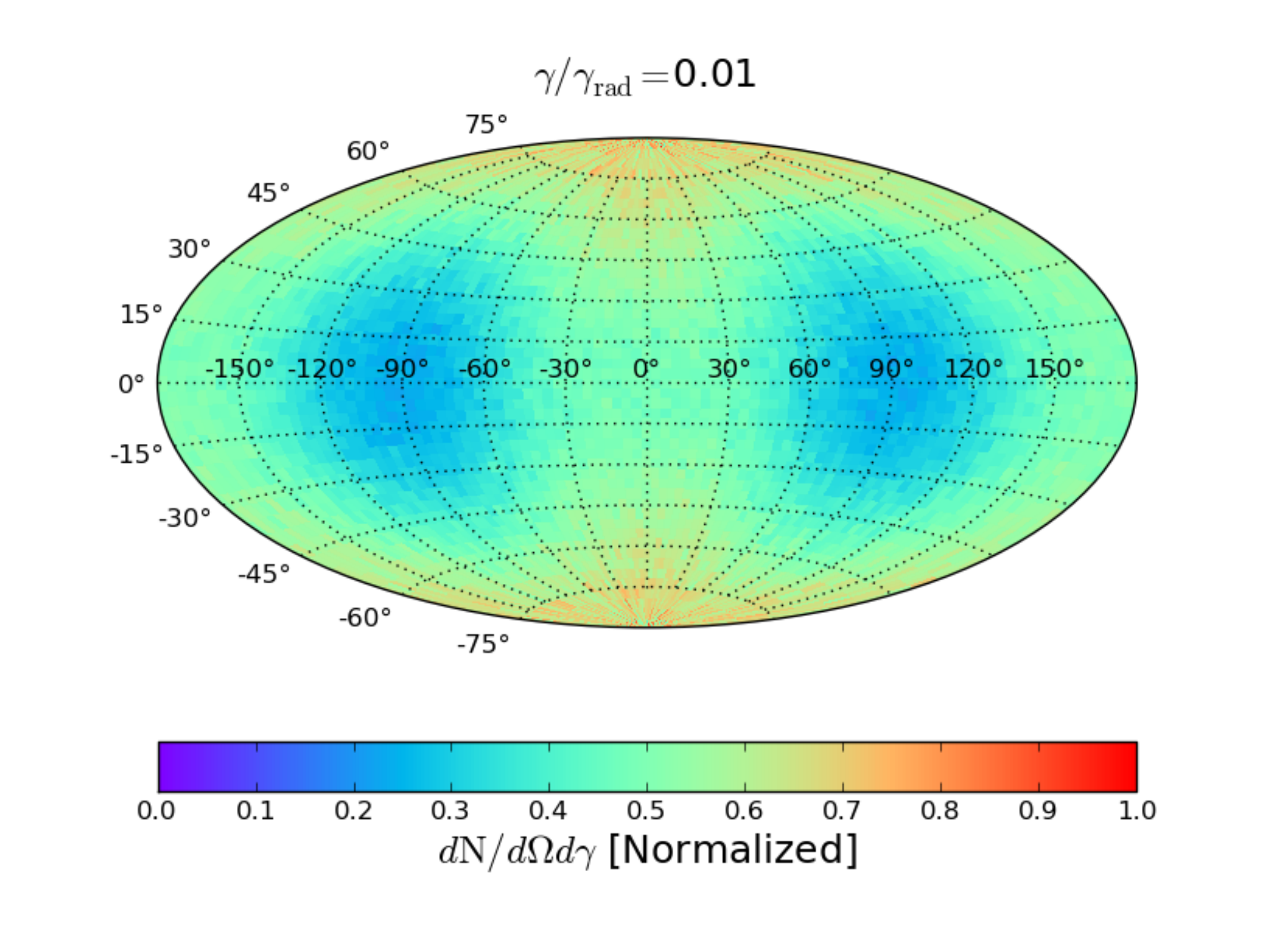}
\includegraphics[width=8.5cm]{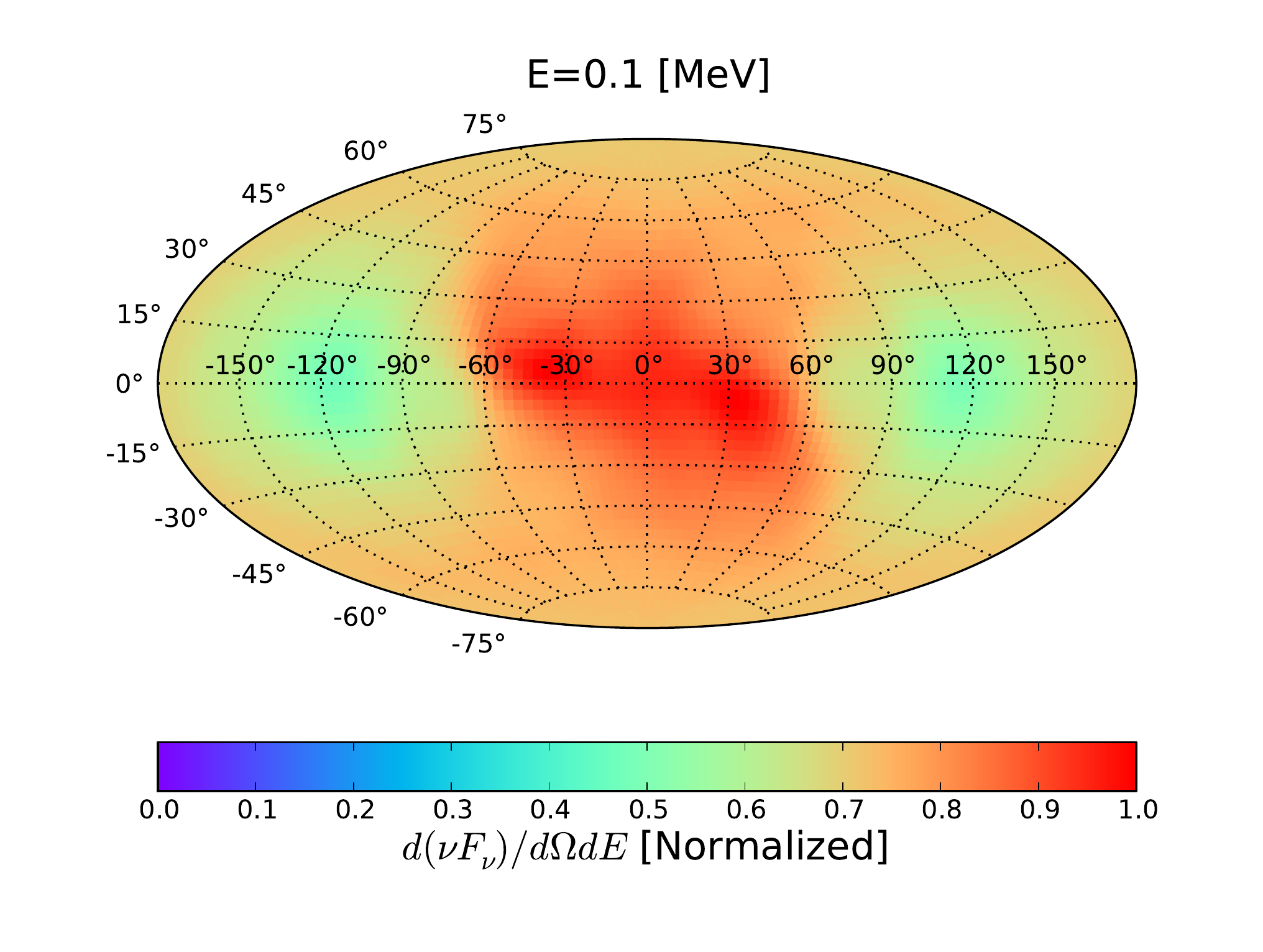}
\includegraphics[width=8.5cm]{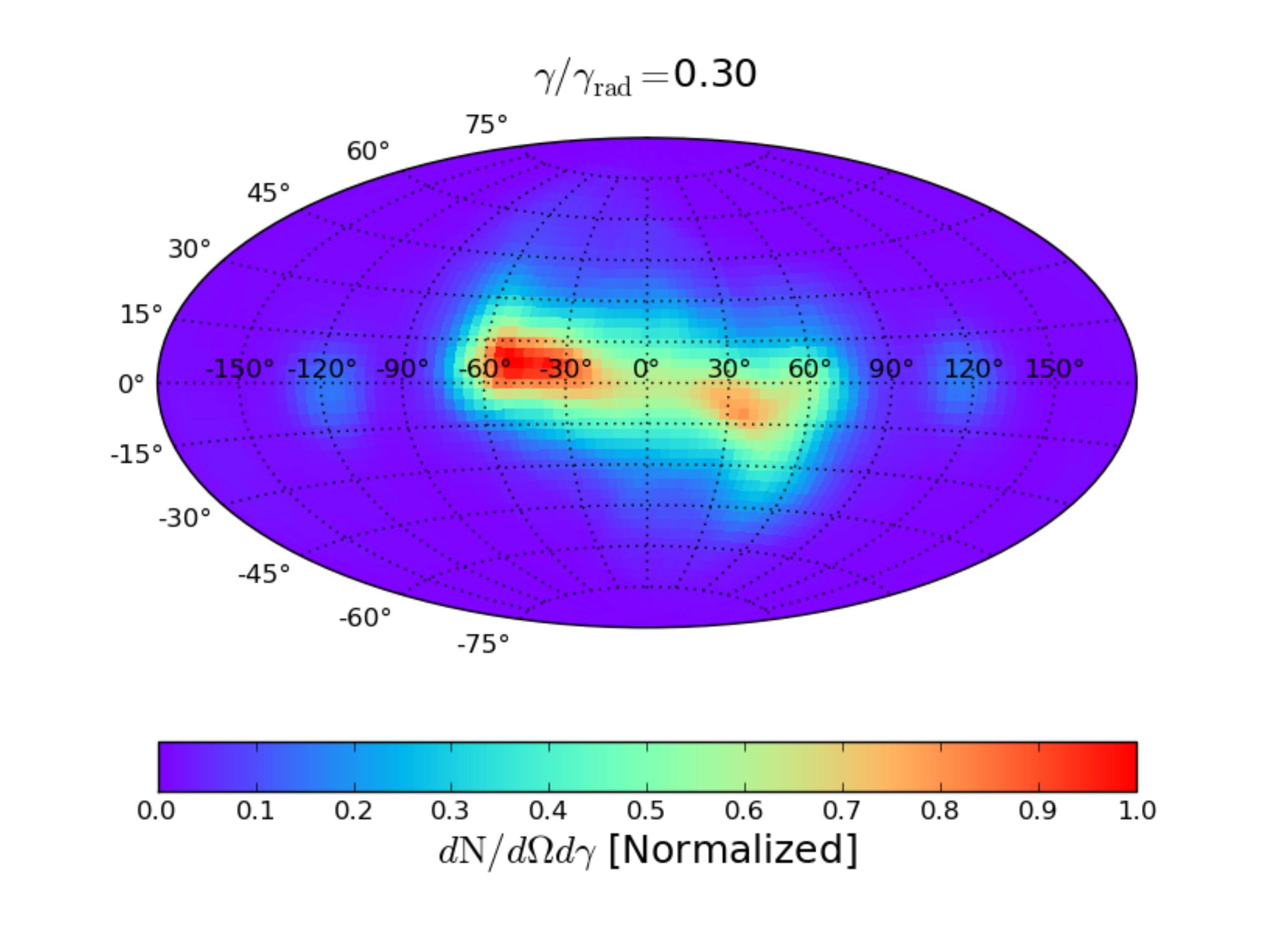}
\includegraphics[width=8.5cm]{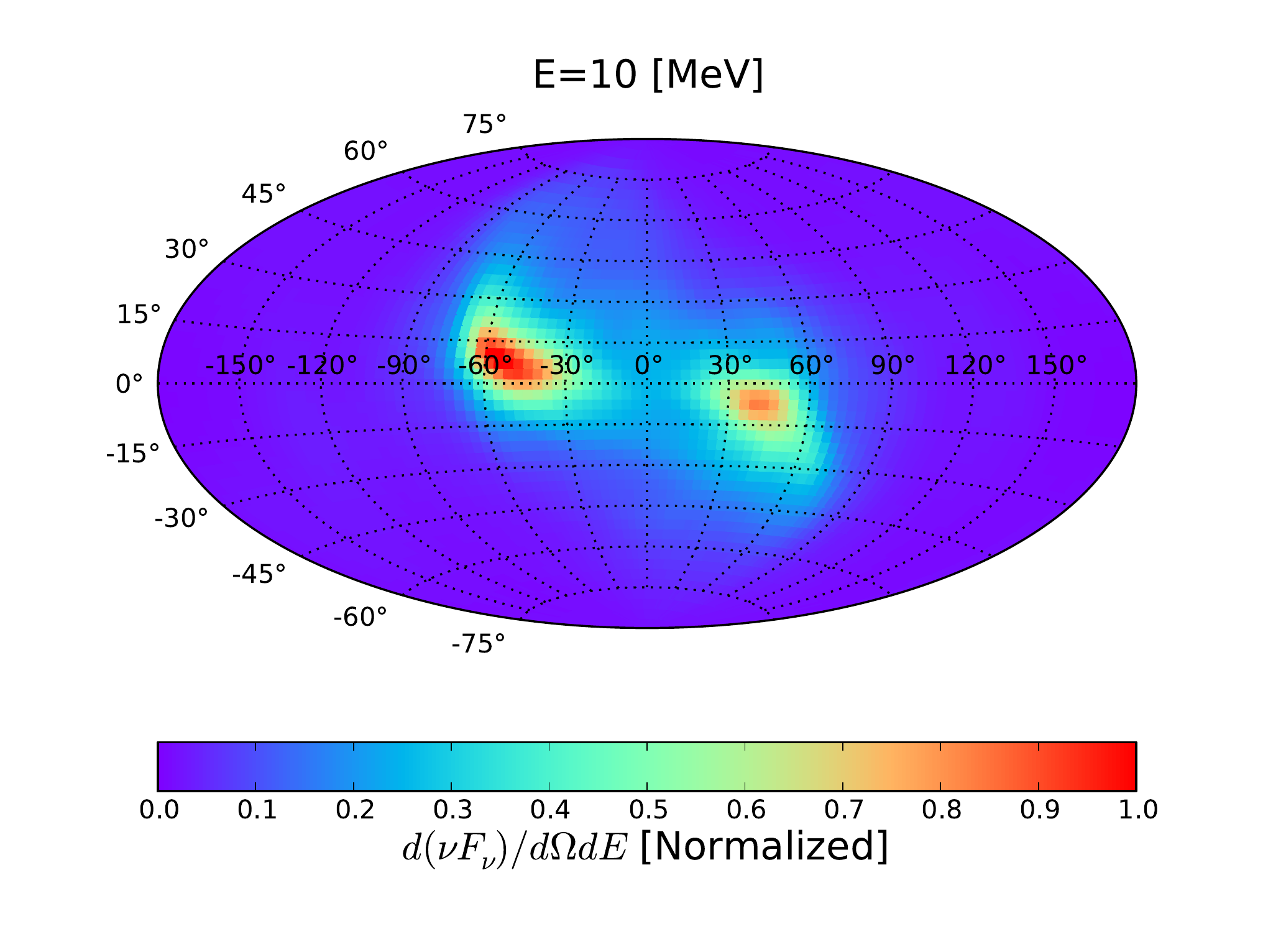}
\includegraphics[width=8.5cm]{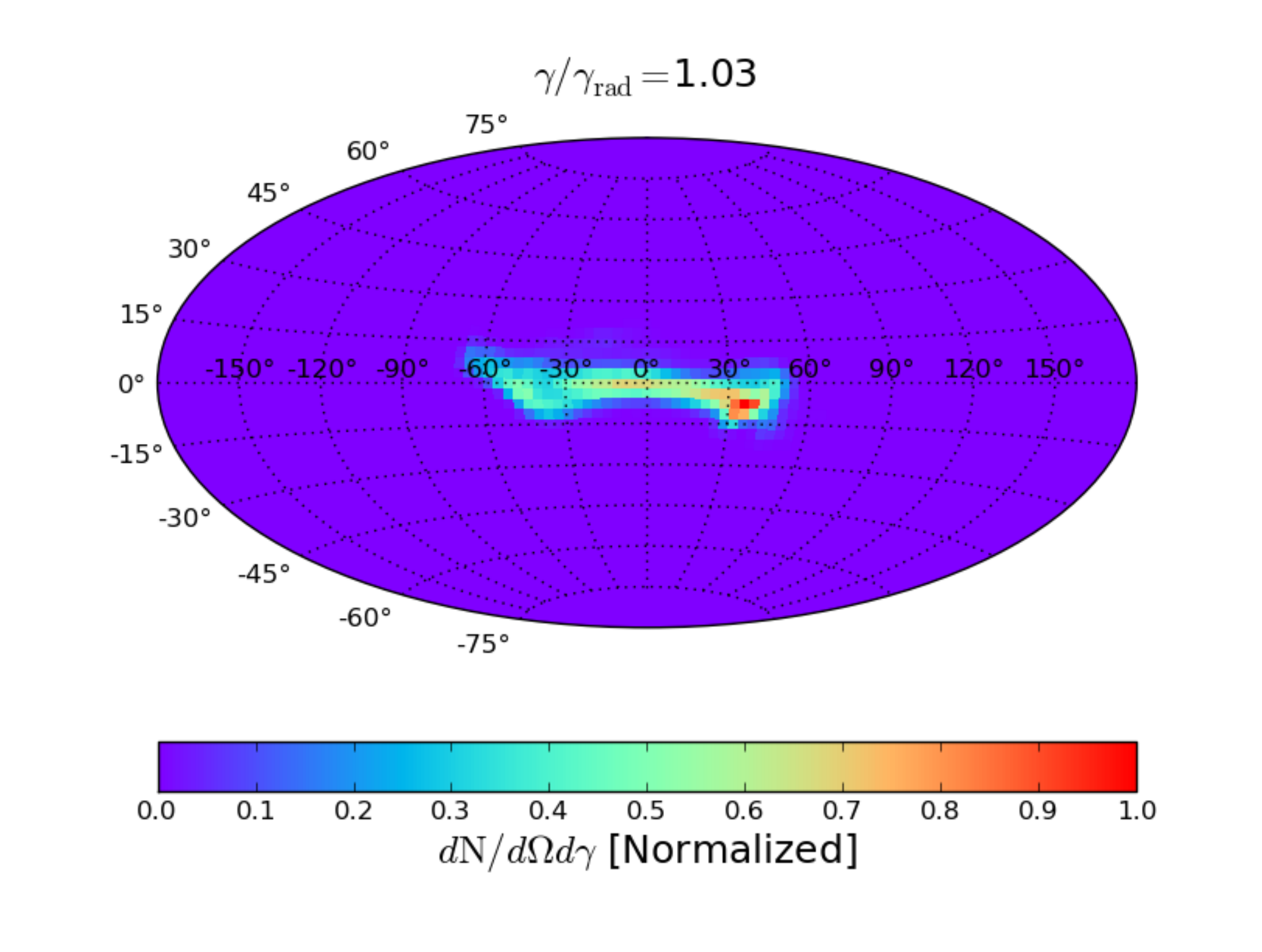}
\includegraphics[width=8.5cm]{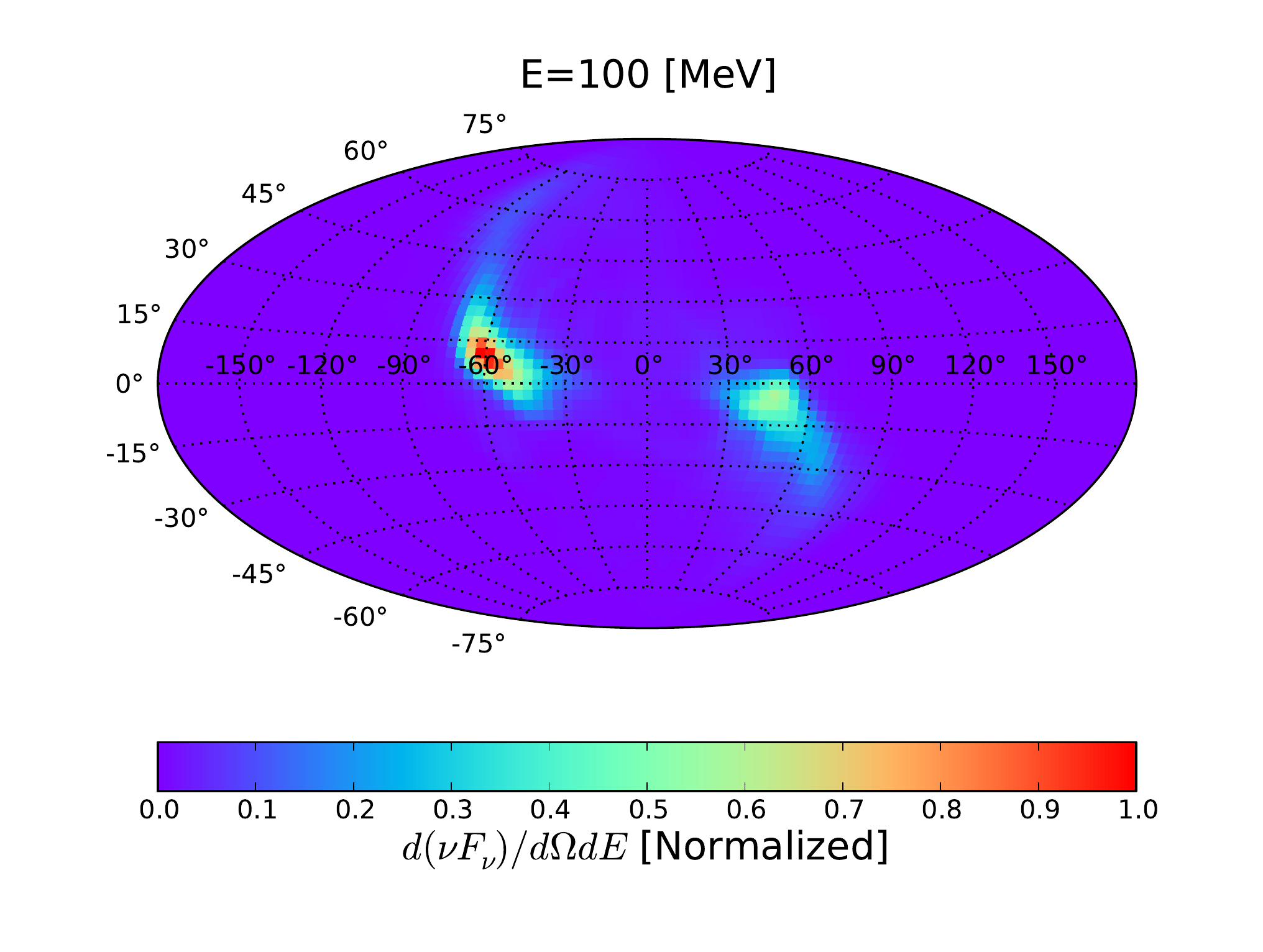}
\caption{Positron angular distributions ($d{\rm N}/d\Omega d\gamma$, left panels) and their synchrotron radiation angular distribution ($d(\nu F_{\nu})/d\Omega dE$, right panels) in run {\tt 3D050} ($\alpha=0.5$) at $t\omega_0=291$. Each panel is at a different energy bin: $\gamma/\gamma_{\rm rad}=0.01$ (top left), $0.3$ (middle left) and $1$ (bottom left) for the particles and $E=0.1~$MeV (top right) $10~$MeV (middle right) and $100~$MeV (bottom right) for the photons. The color-coded scale is linear and normalized to the maximum value in each energy bin. The angular distribution is shown in the Aitoff projection, where the horizontal-axis is the longitude, $\lambda$, varying between $\pm180\degr$ ($-z$-axis) and the vertical axis is the latitude, $\phi$, varying between $-90\degr$ ($-y$-direction) to $+90\degr$ ($+y$-direction). The origin of the plot corresponds to the $+z$-direction.}
\label{fig_anis}
\end{figure*}

\subsection{Particle and photon spectra}\label{spec3d}

Figure~\ref{fig_spec_3d} presents the particle and photon energy distributions averaged over all directions at $t\omega_0=265$, with no guide field. The distributions are remarkably similar to the 2D run {\tt 2DYZ0} ones, and differ significantly from run {\tt 2DXY0}. The high-energy part of the particle spectrum peaks at $\gamma/\gamma_{\rm rad}=0.3$ which is the signature of particle heating via magnetic dissipation rather than particle acceleration through tearing-dominated reconnection (Section~\ref{spec2d}, Eq.~\ref{gmax_kink}). We note that the spectrum extends to higher energy than the pure magnetic dissipation scenario, slightly above $\gamma_{\rm rad}$, suggesting that there is a non-thermal component as well. On the contrary, in the $\alpha=0.5$ guide field case (run {\tt 3DG050}, Figure~\ref{fig_spec_3d_guide}), things are closer to the pure tearing reconnection case of run {\tt 2DXY050}. The particle energy distribution is almost flat in the $0.04\lesssim\gamma/\gamma_{\rm rad}\lesssim 0.4$ range, but barely reaches above $\gamma_{\rm rad}$ as in the zero-guide field case. Nevertheless, the synchrotron emission $>160~$MeV is more intense than in the zero guide field case. Even though there is clear evidence for particle acceleration above the radiation reaction limit, the effect remains slightly weaker than in 2D with no guide field. A bigger box size would help to improve the significance of this result.

\subsection{Particle and photon anisotropies}\label{anis3d}

The angular distribution of the particles is of critical interest for determining the apparent isotropic radiation flux seen by a distant observer who probes one direction only. In 2D reconnection, we expect a pronounced beaming of the particles that increases rapidly with their energy \citep{2012ApJ...754L..33C, 2013ApJ...770..147C}. We confirm here that this phenomenon exists also in 3D, even with a finite guide field. Figure~\ref{fig_anis} presents energy-resolved maps of the angular distribution of the positrons (left panels) and their optically thin synchrotron radiation (right panel) in run {\tt 3D050}. The direction of motion of the particles is measured with two angles: the latitude, $\phi$, varying between $-90\degr$ and $90\degr$, defined as
\begin{equation}
\phi=\sin^{-1}\left(\frac{u_{\rm y}}{\sqrt{u_{\rm x}^2+u_{\rm y}^2+u_{\rm z}^2}}\right),
\label{phi}
\end{equation}
and the longitude, $\lambda$, defined between $-180\degr$ and $180\degr$ given by
\begin{equation}
\lambda = \left\{ \begin{array}{lcl} \cos^{-1}\left(\frac{u_{\rm z}}{\sqrt{u_{\rm x}^2+u_{\rm z}^2}}\right) & \mbox{if} & \sin\lambda>0\\
-\cos^{-1}\left(\frac{u_{\rm z}}{\sqrt{u_{\rm x}^2+u_{\rm z}^2}}\right) & \mbox{if} & \sin\lambda<0 \end{array} \right. ,
\label{lambda}
\end{equation}
where $u_{\rm x},~u_{\rm y}$, and $u_{\rm z}$ are the components of the particle 4-velocity vector. 

We find that the low-energy particles ($\gamma/\gamma_{\rm rad}\lesssim 0.1$) nearly conserve the initially imposed isotropy, because they are still upstream and have not been energized by reconnection. In contrast, the high-energy particles ($\gamma/\gamma_{\rm rad}\gtrsim 0.1$) are significantly beamed along the reconnection plane (at X-lines and with flux ropes) within $\phi=\pm 15\degr$ and $\lambda=\pm 60\degr$. The $\lambda=\pm 60\degr$ angle is of special interest here because it coincides with the direction of the undisturbed magnetic field lines outside the reconnection layers for a $\alpha=0.5$ guide field ($\lambda_0=\pm\tan^{-1}(1/\alpha)\approx\pm 63\degr$). The particles are accelerated along the $z$-direction by the reconnection electric field, and move back and forth across the layer mid-plane following relativistic Speiser orbits \citep{2013ApJ...770..147C}. At the same time, the particles are deflected away by the reconnected field and the guide field creating a characteristic ``S'' shape in the angular maps. To a lesser extent, the zero-guide field case also presents some degree of anisotropy, but the deformation and then the disruption of the layer by the kink instability effectively broaden the beams.

The synchrotron angular distribution closely follows the particle one, essentially because relativistic particles radiate along their direction of motion within a cone of semi-aperture angle $\sim 1/\gamma\ll 1$. However, there is a noticeable offset between the distribution of the highest-energy particles with $\gamma\gtrsim\gamma_{\rm rad}$ and the radiation above $100$~MeV. This discrepancy is due to the different zones where particles accelerate and where particles radiate. In the accelerating zone, the electric field is intense and leads to linear particle acceleration along the $z$-axis, whereas the perpendicular magnetic field, $B_{\perp}$, is weak deep inside the reconnection layers, yielding little synchrotron radiation. These high-energy particles then radiate $\gtrsim 100$~MeV emission abruptly, i.e., within a fraction of a Larmor gyration, only when they are deflected outside the layer where $B_{\perp}\sim B_0$. The beam dump is well localized at $\lambda=\pm 60\degr$, i.e., along the upstream magnetic field lines (see hot-spots in Figure~\ref{fig_anis}, bottom-right panel).

\begin{figure}
\centering
\includegraphics[width=8.5cm]{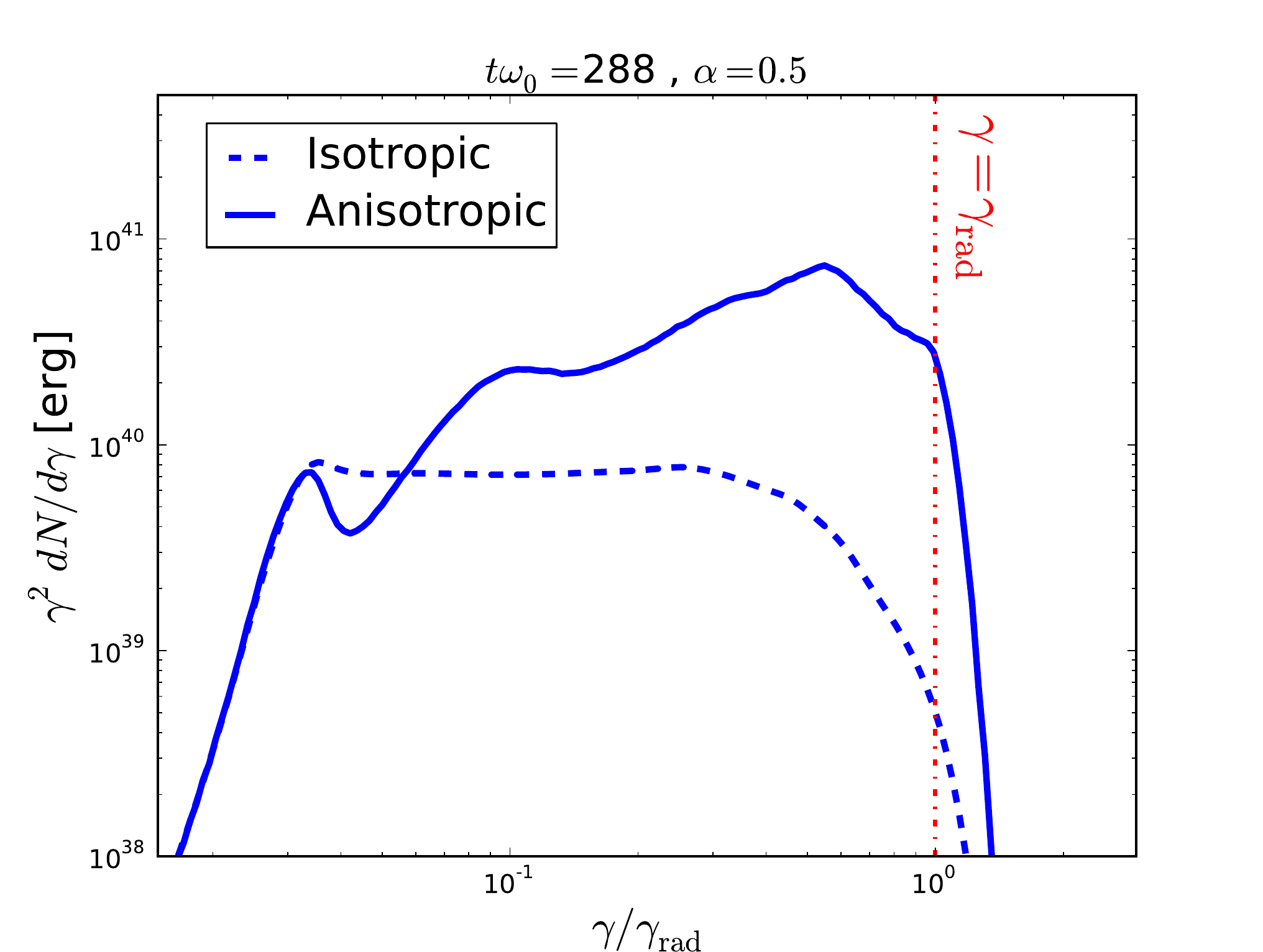}
\includegraphics[width=8.5cm]{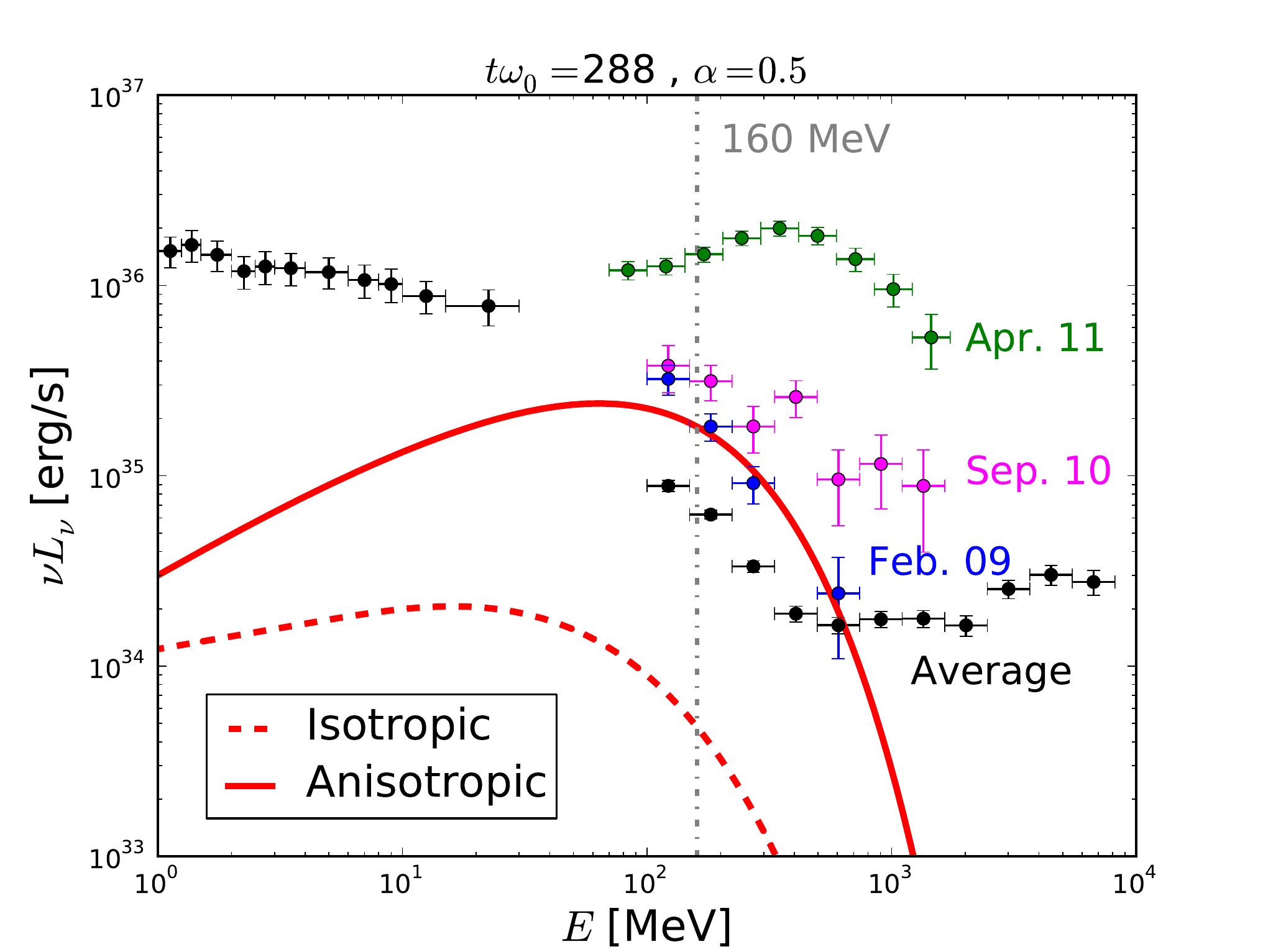}
\caption{Top: Comparison between the isotropically averaged particle energy distribution (dashed line) and the apparent isotropic distribution in the $\phi=-9.2\degr,~\lambda=34.5\degr$ direction (solid line). Bottom: Comparison between the isotropically averaged synchrotron SED (dashed line), the apparent isotropic SED in the $\phi=5.5\degr,~\lambda=-63.6\degr$ direction at $t\omega_0=288$, for $\alpha=0.5$ (solid line), and the observed {\em Fermi}-LAT spectra (data points) during the flares in February 2009, September 2010, and in April 2011, as well as the average quiescent spectrum from 1~MeV to 10~GeV \citep{2011Sci...331..739A, 2012ApJ...749...26B}. Observed fluxes are converted into isotropic luminosities, assuming that the nebula is at 2~kpc from Earth.}
\label{fig_spec_anis}
\end{figure}

\subsection{Apparent spectra and comparison with observations}\label{comp_obs}

As a consequence of this strong anisotropy, the observed spectra of particles and radiation depend sensitively on the viewing angle. Figure~\ref{fig_spec_anis} compares the isotropic particle (top panel) and photon energy distributions (apparent intrinsic isotropic luminosities, $\nu L_{\nu}$, bottom panel) with the distributions along one of the directions dominated by the highest energy particles ($\phi=-9.2\degr,~\lambda=34.5\degr$) and radiation ($\phi=5.5\degr,~\lambda=-63.6\degr$) as it appears to a distant observer at $t\omega_0=288$, for $\alpha=0.5$ (see Section~\ref{anis3d} and Figure~\ref{fig_anis}). The bottom panel in Figure~\ref{fig_spec_anis} also compares the {\em Fermi}-LAT measurements of the gamma-ray spectra of the February 2009, September 2010 and April 2011 flares and the average quiescent spectrum of the Crab Nebula \citep{2011Sci...331..739A, 2012ApJ...749...26B} with the simulated gamma-ray spectra. The particle spectrum in the ($\phi=-9.2\degr,~\lambda=34.5\degr$) direction is very hard, and peaks at $\gamma/\gamma_{\rm rad}\approx 0.6$ with a total apparent isotropic energy $\approx 7\times 10^{40}~$erg, which corresponds to about half the total magnetic energy dissipated by the end of the simulation, or about 10 times more energy than in the isotropic distribution. The photon spectrum in the ($\phi=5.5\degr,~\lambda=-63.6\degr$) direction peaks at about $100~$MeV but extends up to $\sim1~$GeV. The resulting gamma-ray luminosity above 100~MeV is $L_{\gamma}\approx 3\times 10^{35}~$erg/s, which is about 40 times brighter than the isotropically averaged flux from the layer, and about 3 times brighter than the observed average (quiescent) luminosity of the Crab Nebula ($L_{\gamma}^{\rm Crab}\approx 10^{35}~$erg/s, assuming a distance of 2~kpc between the nebula and the observer). Thanks to beaming, the level of gamma rays expected from the model is consistent with the moderately bright flares, such as the February 2009 or September 2010 ones (see Figure~\ref{fig_spec_anis}, bottom panel), but the simulation cannot reproduce the flux of the brightest flares, such as the April 2011 event. Presumably, increasing the box size would be enough to increase the density of the emitting particles $>100~$MeV and account for the most intense flares.

\begin{figure}
\centering
\includegraphics[width=8.5cm]{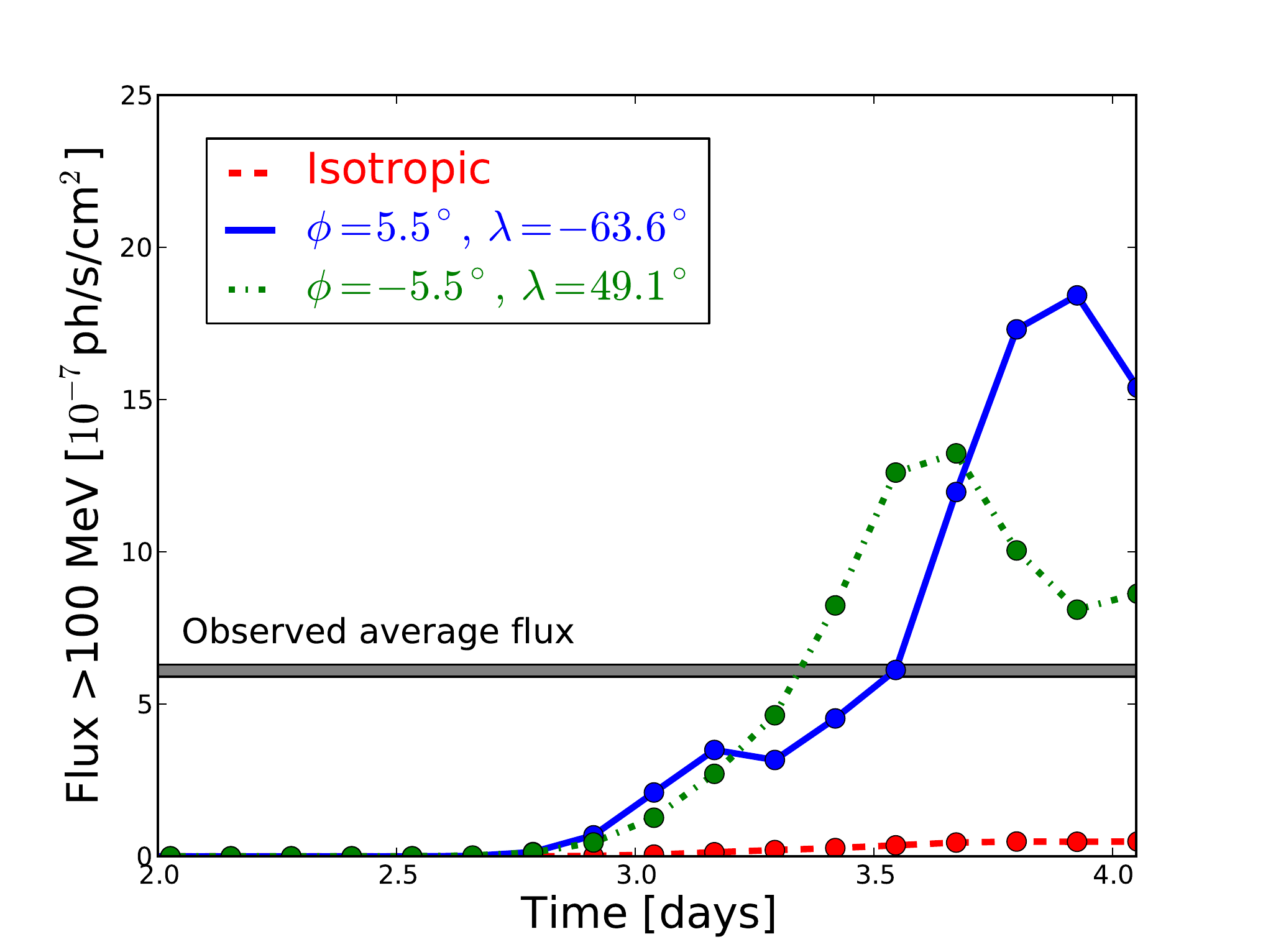}
\caption{Instantaneous synchrotron photon flux integrated above $100~$MeV as a function of time in the directions $\phi=5.5\degr,~\lambda=-63.6\degr$ (blue solid line), $\phi=-5.5\degr,~\lambda=49.1\degr$ (green dot-dashed line), and isotropically averaged (red dotted line) in run {\tt 3D050}. Fluxes are given in ph/s/cm$^{2}$, assuming a distance of $2~$kpc between the layer and the observer. The horizontal gray band shows the average observed flux of the Crab Nebula above 100~MeV measured by the {\em Fermi}-LAT, equal to $6.1\pm 0.2\times 10^{-7}$ ph/s/cm$^{2}$ \citep{2012ApJ...749...26B}.}
\label{fig_lightcurve}
\end{figure}

\subsection{Lightcurves}\label{light}

The beam of high-energy radiation is also time variable, both in direction and in intensity. Figure~\ref{fig_lightcurve} presents the computed time evolution of the synchrotron photon flux integrated above $100$~MeV in the directions defined by $\phi=5.5\degr,~\lambda=-63.6\degr$ and $\phi=-5.5\degr,~\lambda=49.1\degr$, as well as the computed time evolution of the isotropically-averaged flux and the observed average flux in the Crab Nebula measured by the {\em Fermi}-LAT \citep{2012ApJ...749...26B} for comparison. This calculation assumes that all the photons emitted at a given instant throughout the box reach the observer at the same time, i.e., it ignores the time delay between photons emitted in different regions with respect to the observer. Along the directions probed here, the $>100$~MeV flux doubling timescale is of order $10-20\omega_0^{-1}$ or $3$-$6$ hours, for both the rising and the decaying time, which is compatible with the observations of the Crab flares \citep{2012ApJ...749...26B, 2013ApJ...775L...37S}, as well as with our previous 2D simulations in \citet{2013ApJ...770..147C}. Shorter variability timescales may still exist in the simulation but our measurement is limited by the data dumping period set at $T_{\rm dump}\approx 10\omega_0^{-1}$. These synthetic lightcurves also clearly illustrate the effect of the particle beaming on the observed flux discussed in Section~\ref{comp_obs}. Along the direction of the beam, the $>100~$MeV flux can be $\gtrsim 10$ times more intense than the isotropically-averaged one, and even exceeds the measured quiescent gamma-ray flux of the Crab Nebula by a factor $\approx 2$-$3$ at the peak of the lightcurves 3.5 to 4~days after the beginning of the simulation. Each time the beam crosses our line of sight, we see a rapid bright flare of the most energetic radiation emitted in the simulation.

The high-energy particles are strongly bunched within the magnetic flux ropes (within magnetic islands in 2D, see \citealt{2012ApJ...754L..33C, 2013ApJ...770..147C}). As a result, the typical size of the emitting regions is comparable to the dimensions of the flux ropes, i.e., of order $L_{\rm x}/10\approx 20\rho_0$ along the $x$- and $y$-directions (Figure~\ref{fig_3d}), which corresponds to about $6$~light-hours. We conclude that particle bunching is at the origin of the ultra-short time variability ($3$-$6$ hours) found in the reconstructed lightcurve (Figure~\ref{fig_lightcurve}). Particle bunching and anisotropy help to alleviate the severe energetic constraints imposed by the Crab flares.

\section{CONCLUSION}\label{ccl}

We found that, unlike classical models of particle acceleration, 3D relativistic pair plasma reconnection can accelerate particles above the standard radiation reaction limit in the Crab Nebula. We also confirm the existence of a strong energy-dependent anisotropy of the particles and their radiation, resulting in an apparent boosting of the high-energy radiation observed when the beam crosses our line of sight. In this case, the simulated gamma-ray flux $>100~$MeV exceeds the measured quiescent flux from the nebula by a factor 2-3, and reproduces well the flux of moderately bright flares, such as the February 2009 or the September 2010 events. Simulating brighter flares (e.g., the April 2011 flare) may be achieved with a larger box size. In addition, the bunching of the energetic particles within the magnetic flux ropes results in rapid time variations of the observed gamma-ray flux ($\lesssim 6$~hours). The results are consistent with observations of the Crab flares and with our previous 2D simulations \citep{2013ApJ...770..147C}, although this extreme acceleration is less pronounced than in 2D due to the deformation of the layer by the kink instability in 3D. If there is no guide field, we found that the kink instability grows faster than the tearing instability, resulting in the disruption of the reconnection layers and significant particle heating rather than reconnection and non-thermal particle acceleration. In agreement with \citet{2007ApJ...670..702Z, 2008ApJ...677..530Z}, we observe that a moderate guide field ($\alpha\sim 0.5$) is enough to reduce the negative effect of the kink on the acceleration of particles. However, a strong guide field (i.e., $\alpha\gtrsim 1$) quenches particle acceleration and the emission of high-energy emission because it deflects the particles away from the X-lines too rapidly.

Applying a guide field is probably not the only way to suppress kink instability. In the Harris configuration, an initial ultra-relativistic drifting particle flow (with $\beta_{\rm drift}\gtrsim 0.6$) is expected to foster tearing-dominated reconnection \citep{2007ApJ...670..702Z} as observed by \citet{2011PhPl...18e2105L}. Alternatively, starting with an out-of-equilibrium layer could also drive a fast onset of reconnection \citep{2013ApJ...774...41K}. In real systems, the reconnection layer is not likely to be smooth, flat, undisturbed, and in equilibrium. A small perturbation in the field lines, like a pre-existing X-point, can favor fast reconnection before the kink instability has time to grow. Hence, while we have shown one set of conditions emitting $>160~$MeV radiation, there may be other conditions allowing reconnection to produce similar results.

The reconnection model could be refined if future multi-wavelength observations can pin down the location of the flares in the Crab Nebula (so far there is nothing obvious, see e.g., \citealt{2013ApJ...765...56W}). One promising location for reconnection-powered flares could be within the jets in the polar regions where the plasma is expected to be highly magnetized (i.e., $\sigma\gtrsim 1$) with stronger magnetic field close to the pulsar rotational axis \citep{2011ApJ...737L..40U, 2012ApJ...746..148C, 2012MNRAS.427.1497L, 2013MNRAS.428.2459K, 2013arXiv1309.0375M, 2013arXiv1310.2531P}. In addition, theoretical studies \citep{1998ApJ...493..291B}, numerical simulations \citep{2011ApJ...728...90M, 2013MNRAS.431L..48P, 2013arXiv1310.2531P, 2013arXiv1309.0375M}, and possibly X-ray observations \citep{2011evhe.confE...2W} indicate that the jets are unstable to kink instabilities. The non-linear development of these instabilities could lead to the formation of current sheets, presumably with a non-zero guide field, and then to magnetic dissipation in the Crab Nebula in the form of powerful gamma-ray flares, which may contribute to solving the ``$\sigma$-problem'' in pulsar wind nebulae \citep{1974MNRAS.167....1R, 1984ApJ...283..694K}. 

Relativistic reconnection may also be at the origin of other astrophysical flares. Most notably, TeV gamma-ray flares observed in blazars (e.g., \citealt{2007ApJ...664L..71A, 2007ApJ...669..862A, 2011ApJ...730L...8A}) present challenges similar to the Crab flares (e.g., ultra-short time variability, problematic energetics) that could be solved by invoking relativistic reconnection in a highly magnetized jet \citep{2009MNRAS.395L..29G, 2010MNRAS.402.1649G, 2011MNRAS.413..333N, 2012MNRAS.425.2519N, 2012ApJ...754L..33C, 2013MNRAS.431..355G}. The physical conditions in blazar jets are quite different than in the Crab Nebula (e.g., composition, inverse Compton drag, pair creation), which may change the dynamics of reconnection. The current sheet that forms beyond the light-cylinder in pulsars offers another interesting environment for studying relativistic reconnection subject to strong synchrotron cooling. Pairs energized by reconnection may be at the origin of the GeV pulsed emission in gamma-ray pulsars \citep{1996A&A...311..172L, 2012MNRAS.424.2023P, 2012arXiv1210.3346U, 2013A&A...550A.101A}. We speculate that synchrotron radiation from the particles could even account for the recently reported $>100~$GeV pulsed emission from the Crab (see, e.g., \citealt{2011Sci...334...69V,2012A&A...540A..69A}) if particle acceleration above the radiation reaction limit operates in this context.

\acknowledgements BC thanks G.~Lesur for discussions about the linear analysis of the unstable modes in the simulations. The authors thank the referee for his/her useful comments. BC acknowledges support from the Lyman Spitzer Jr. Fellowship awarded by the Department of Astrophysical Sciences at Princeton University, and the Max-Planck/Princeton Center for Plasma Physics. This work was also supported by NSF grant PHY-0903851, DOE Grants DE-SC0008409 and DE-SC0008655, NASA grant NNX12AP17G through the Fermi Guest Investigator Program. Numerical simulations were performed on the local CIPS computer cluster Verus and on Kraken at the National Institute for Computational Sciences (\url{www.nics.tennessee.edu/}). This work also utilized the Janus supercomputer, which is supported by the National Science Foundation (award number CNS-0821794), the University of Colorado Boulder, the University of Colorado Denver, and the National Center for Atmospheric Research. The Janus supercomputer is operated by the University of Colorado Boulder. The figures published in this work were created with the {\tt matplotlib} library \citep{hunter2007} and the 3D visualization with {\tt Mayavi2} \citep{mayavi2011}.

\end{document}